%% file: ToricPEPS.tex
\documentclass[letterpaper,twocolumn,superscriptaddress,showpacs,nofootinbib,notitlepage,accepted=2021-11-15]{quantumarticle}
\pdfoutput=1
\usepackage[numbers,sort&compress]{natbib}
\usepackage[utf8]{inputenc}
\usepackage{amsmath, amssymb}
\usepackage{verbatim}
\usepackage[usenames,dvipsnames]{xcolor}
\usepackage{graphicx}
\usepackage[pdftex, plainpages=false]{hyperref}
\usepackage{mathtools}
\usepackage{stmaryrd, ifthen}
\usepackage{array}
\usepackage{tabularx,float}
\usepackage[export]{adjustbox}
\usepackage{bm, bbm, yfonts, wasysym}
\usepackage{enumitem}
\usepackage{calc}
\usepackage{etoolbox}
\mathtoolsset{showonlyrefs=false}
\usepackage{footnote}
\usepackage{dcolumn}
\usepackage{microtype}
\usepackage{booktabs}

\usepackage{tikz}
\usetikzlibrary{shapes.geometric,shapes.misc}
\usetikzlibrary{arrows,matrix,calc,scopes,decorations.markings,snakes}
\usetikzlibrary{arrows.meta}
\usetikzlibrary{intersections, patterns,fit} 
\usetikzlibrary{decorations.pathreplacing,angles,quotes}

\hypersetup{%
	bookmarks=true,         
	bookmarksopen=true,
	unicode=true,           
	pdftoolbar=true,        
	pdfmenubar=true,        
	pdffitwindow=false,     
	pdfstartview={FitH},    
	pdftitle={On tensor network representations of the (3+1)d toric code},
	pdfauthor={},
	pdfsubject={},          
	pdfcreator={},          
	pdfproducer={pdfLaTeX}, 
	pdfkeywords={toric code} {topological} {tensor networks} {peps},
	pdfnewwindow=true,      
	colorlinks,
	citecolor=BlueViolet,
	filecolor=BlueViolet,
	linkcolor=BlueViolet,
	urlcolor=BlueViolet,
	linktocpage=true
}

\newcommand{\includeTikz}[3]{
	\includegraphics[scale=1, valign=c, raise=#1 pt]{fig/#2}
}

\newcommand{\snum}[1]{\text{\small $#1$}}
\newcommand{\pepssf}{\text{\small\textsf{PEPS}}}
\newcommand{\mpssf}{\text{\small\textsf{MPS}}}
\newcommand{\tc}{\text{\small TC}}

\newcommand{\la}{\langle}
\newcommand{\ra}{\rangle}
\newcommand{\cX}{{\rm c}X}
\newcommand{\q}{\quad}
\newcommand{\sss}{\scriptstyle}
\newcommand{\ssss}{\scriptscriptstyle}
\newcommand{\nn}{\nonumber}
\newcommand{\diam}{\raisebox{0.4pt}[0pt][0pt]{\scalebox{0.9}{\rotatebox[origin=c]{45}{$\hspace{-0.6pt} {\sss \boxempty}$}}}}
\newcommand{\diamsss}{\raisebox{0.3pt}[0pt][0pt]{\scalebox{0.8}{\rotatebox[origin=c]{45}{$\hspace{-0.6pt} {\ssss \boxempty}$}}}}
\newcommand{\tet}{\includegraphics[scale=1, valign=c]{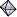}}
\newcommand{\ttet}{\includegraphics[scale=1, valign=c]{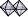}}
\newcommand{\btet}{\includegraphics[scale=1, valign=c]{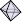}}
\newcommand{\bttet}{\includegraphics[scale=1, valign=c]{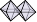}}
\newcommand{\otimesbk}{\;\underline{\otimes}\;}

\DeclareSymbolFont{bbold}{U}{bbold}{m}{n}
\DeclareSymbolFontAlphabet{\mathbbold}{bbold}

\include{tikzPic}

\begin{document}
	
\title{On tensor network representations of the (3+1)d toric code}

\author{Clement Delcamp}
\email{delcamp@pks.mpg.de}
\affiliation{Max-Planck-Institut f{\"u}r Quantenoptik,  Hans-Kopfermann-Stra{\ss}e 1, 85748 Garching, Germany}
\affiliation{Munich Center for Quantum Science and Technology (MCQST), Schellingstra{\ss}e 4, 80799 M{\"u}nchen, Germany}
\author{Norbert Schuch}
\email{norbert.schuch@gmail.com}
\affiliation{Max-Planck-Institut f{\"u}r Quantenoptik,  Hans-Kopfermann-Stra{\ss}e 1, 85748 Garching, Germany}
\affiliation{Munich Center for Quantum Science and Technology (MCQST), Schellingstra{\ss}e 4, 80799 M{\"u}nchen, Germany}
\affiliation{University of Vienna, Faculty of Physics, Boltzmanngasse 5, 1090 Wien, Austria}
\affiliation{University of Vienna, Faculty of Mathematics,\unpenalty~Oskar-Morgenstern-Platz 1, 1090 Wien, Austria}

\begin{abstract}
	\noindent
	We define two dual tensor network representations of the (3+1)d toric code ground state subspace. These two representations, which are obtained by initially imposing either family of stabilizer constraints, are characterized by different virtual symmetries generated by string-like and membrane-like operators, respectively. We discuss the topological properties of the model from the point of view of these virtual symmetries, emphasizing the differences between both representations. In particular, we argue that, depending on the representation, the phase diagram of boundary entanglement degrees of freedom is naturally associated with that of a (2+1)d Hamiltonian displaying either a global or a gauge $\mathbb Z_2$-symmetry.
\end{abstract}

\maketitle

\input{_Introduction.tex}

\input{_ToricCode.tex}

\input{_PEPS2D.tex}

\input{_PEPS3D.tex}

\input{_Discussion.tex}

\acknowledgements

\emph{We would like to thank Markus Hauru, Mohsin Iqbal and David Stephen for stimulating discussions, as well as Frank Verstraete and especially Dominic Williamson for collaboration on the closely related project \cite{WDSV}. This project has received funding from the European Research Council (ERC) under the European Union’s Horizon 2020 research and innovation programme through the ERC Starting Grant WASCOSYS (No.~ 636201)  and the ERC Consolidator Grant SEQUAM (No.~ 863476), as well as the Deutsche Forschungsgemeinschaft (DFG, German Research Foundation) under Germany’s Excellence Strategy -- EXC-2111 -- 390814868.}

\appendix
\input{_Appendix.tex}

\bibliographystyle{unsrtnat}
\bibliography{refs}
	
\end{document}

%% file: tikzPic.tex
\tikzset{->1-/.style={decoration={
			markings,
			mark=at position #1 with {\arrow{Triangle[fill=black, width=3.5pt, length=3.5pt]}}},postaction={decorate}}}

\tikzset{->2-/.style={decoration={
					markings,
					mark=at position #1 with {\arrow{Triangle[fill=white, width=5pt, length=5pt]}}},postaction={decorate}}}

\tikzset{
	every picture/.append style={
		line join=round,
		line cap=round,
		line width = 0.5pt
	}
}

\tikzstyle{dot}=[dotted]
\tikzstyle{op8}=[opacity=0.8]
\tikzstyle{op7}=[opacity=0.7]
\tikzstyle{op6}=[opacity=0.6]
\tikzstyle{op5}=[opacity=0.5]
\tikzstyle{op4}=[opacity=0.4]
\tikzstyle{op3}=[opacity=0.3]
\tikzstyle{op2}=[opacity=0.2]
\tikzstyle{op1}=[opacity=0.1]
\tikzstyle{sha}=[shorten >= 0.3em]
\tikzstyle{shb}=[shorten <= 0.3em]
\tikzstyle{shab}=[shorten >= 0.3em, shorten <= 0.3em]
\tikzstyle{shaMax}=[shorten >= 0.6em]
\tikzstyle{shbMax}=[shorten <= 0.6em]
\tikzstyle{shabMax}=[shorten >= 0.6em, shorten <= 0.6em]
\tikzstyle{peps0}=[draw=none, fill=white,fill opacity=0.55]
\tikzstyle{peps}=[fill=black,fill opacity=0.1]
\tikzstyle{peps2}=[draw=none, fill=BrickRed,fill opacity=0.18]
\tikzstyle{peps2b}=[draw=none, fill=BrickRed,fill opacity=0.18]
\tikzstyle{peps1}=[draw=none, fill=BlueViolet,fill opacity=0.15]
\tikzstyle{peps1spe}=[draw=none, fill=BlueViolet,fill opacity=0.15]
\tikzstyle{peps1b}=[draw=none, fill=BlueViolet,fill opacity=0.15]
\tikzstyle{peps3}=[fill=black,fill opacity=0.1]
\tikzstyle{pauliZ}=[circle, fill=white, inner sep=0.15pt, draw=black]
\tikzstyle{pauliX}=[circle, fill=white, inner sep=-0.16pt, draw=black]
\tikzstyle{ket}=[regular polygon, fill=white, inner sep =-2.1pt]
\tikzstyle{mixLine}=[line width=0.5pt]

\newcommand{\interL}[5]{
	\path[name path=P#1#2] (#1) -- (#2);
	\path[name path=P#3#4] (#3) -- (#4);
	\path [name intersections={of=P#1#2 and P#3#4,by={sp}}];
	\draw[sha,#5] (#1) -- (sp);
	\draw[shb, #5] (sp) -- (#2);
}

\newcommand{\interLL}[7]{
	\path[name path=P#1#2] (#1) -- (#2);
	\path[name path=P#3#4] (#3) -- (#4);
	\path[name path=P#5#6] (#5) -- (#6);	
	\path [name intersections={of=P#1#2 and P#3#4,by={sp1}}];
	\path [name intersections={of=P#1#2 and P#5#6,by={sp2}}];	
	\draw[sha, #7] (#1) -- (sp1);
	\draw[shab, #7] (sp1) -- (sp2);
	\draw[shb, #7] (sp2) -- (#2);
}

\newcommand{\convexpath}[2]{
	[   
	create hullcoords/.code={
		\global\edef\namelist{#1}
		\foreach [count=\counter] \nodename in \namelist {
			\global\edef\numberofnodes{\counter}
			\coordinate (hullcoord\counter) at (\nodename);
		}
		\coordinate (hullcoord0) at (hullcoord\numberofnodes);
		\pgfmathtruncatemacro\lastnumber{\numberofnodes+1}
		\coordinate (hullcoord\lastnumber) at (hullcoord1);
	},
	create hullcoords
	]
	($(hullcoord1)!#2!-90:(hullcoord0)$)
	\foreach [
	evaluate=\currentnode as \previousnode using \currentnode-1,
	evaluate=\currentnode as \nextnode using \currentnode+1
	] \currentnode in {1,...,\numberofnodes} {
		let \p1 = ($(hullcoord\currentnode) - (hullcoord\previousnode)$),
		\n1 = {atan2(\y1,\x1) + 90},
		\p2 = ($(hullcoord\nextnode) - (hullcoord\currentnode)$),
		\n2 = {atan2(\y2,\x2) + 90},
		\n{delta} = {Mod(\n2-\n1,360) - 360}
		in 
		{arc [start angle=\n1, delta angle=\n{delta}, radius=#2]}
		-- ($(hullcoord\nextnode)!#2!-90:(hullcoord\currentnode)$) 
	}
}

\newcommand{\ddott}[1]{
	\node[circle, fill = black, inner sep=1.6pt] at (#1) {};
	\node[circle, fill = white, inner sep=1.1pt] at (#1) {};
	\node[circle, fill = black, inner sep=0.6pt] at (#1) {};
}

\newcommand{\univPEPO}[3]{
	\node at (#1) (a) {};
	\node at (#2) (b) {};
	\fill[peps0] \convexpath{a,b}{7pt};
	\fill[#3] \convexpath{a,b}{7pt};
	\draw[] \convexpath{a,b}{7pt};
	\ddott{#1};
	\ddott{#2};
}

\newcommand{\univPEPS}[3]{
	\node at (#1) (a) {};
	\node at (#2) (b) {};
	\fill[peps0] \convexpath{a,b}{7pt};
	\fill[#3] \convexpath{a,b}{7pt};
	\draw[] \convexpath{a,b}{7pt};
	\node[circle, fill = black, inner sep=1.2pt] at (a) {};
	\node[circle, fill = black, inner sep=1.2pt] at (b) {};
}

\newcommand{\univTriPEPO}[4]{
	\node at (#1) (a) {};
	\node at (#2) (b) {};
	\node at (#3) (c) {};
	\fill[peps0] \convexpath{a,b,c}{7pt};
	\fill[#4] \convexpath{a,b,c}{7pt};
	\draw[] \convexpath{a,b,c}{7pt};
	\ddott{#1};
	\ddott{#2};
	\ddott{#3};
}

\newcommand{\univTensor}[5]{
	\node at (#1) (a) {};
	\node at (#2) (b) {};
	\node at (#3) (c) {};
	\node at (#4) (d) {};
	\fill[peps0] \convexpath{a,b,c,d}{7pt};
	\fill[#5] \convexpath{a,b,c,d}{7pt};
	\draw[] \convexpath{a,b,c,d}{7pt};
	\node[circle, fill = black, inner sep=1.2pt] at (a) {};
	\node[circle, fill = black, inner sep=1.2pt] at (b) {};
	\node[circle, fill = black, inner sep=1.2pt] at (c) {};
	\node[circle, fill = black, inner sep=1.2pt] at (d) {};
}

\newcommand{\plaqOp}[4]{
	\ifthenelse{#3=1}{
		\univPEPO{#1,{#2+1}}{{#1+1},#2}{#4};
		\univPEPO{{#1+1},{#2+2}}{{#1+2},{#2+1}}{#4};
		\draw[] ({#1+0.68},{#2+0.68}) -- ({#1+1.32},{#2+1.32});
	}{}
	\ifthenelse{#3=2}{
		\univPEPO{#1,{#2+1}}{{#1+1},{#2+2}}{#4};
		\univPEPO{{#1+1},#2}{{#1+2},{#2+1}}{#4};
		\draw[] ({#1+0.68},{#2+1.32}) -- ({#1+1.32},{#2+0.68});
	}{}
	\ddott{{#1+1},#2};
	\ddott{#1,{#2+1}};
	\ddott{{#1+1},{#2+2}};
	\ddott{{#1+2},{#2+1}};
}

\newcommand{\tetra}[2]{
	\coordinate (0) at (#1,#2);
	\coordinate (1) at ({#1+5},#2);
	\coordinate (2) at ({#1+1.5},{#2+4});
	\coordinate (3) at ({#1+1.5},{#2-3});
	\coordinate (4) at ({#1-2},{#2+1});
	\coordinate (5) at ({#1+3},{#2+1});
	
	\draw[] (1) -- (5) -- (4);
	\draw[] (2) -- (5) -- (3);
	\fill[peps0] (1) -- (3) -- (4) -- (2) -- (1);
	\fill[peps1] (1) -- (3) -- (4) -- (2) -- (1);
	\draw[] (1) -- (2) -- (4) -- (3) -- (1);
	\draw[] (1) -- (0) -- (4);
	\draw[] (2) -- (0) -- (3);
}

\newcommand{\doubleTetra}[2]{
	\coordinate (0) at (#1,#2);
	\coordinate (1) at ({#1+5},#2);
	\coordinate (2) at ({#1+1.5},{#2+4});
	\coordinate (3) at ({#1+1.5},{#2-3});
	\coordinate (4) at ({#1-2},{#2+1});
	\coordinate (5) at ({#1+3},{#2+1});
	
	\draw[double, double distance between line centers=1.6pt] (1) -- (5) -- (4);
	\draw[double, double distance between line centers=1.6pt] (2) -- (5) -- (3);
	\fill[peps0] (1) -- (3) -- (4) -- (2) -- (1);
	\fill[peps1] (1) -- (3) -- (4) -- (2) -- (1);
	\draw[double, double distance between line centers=1.6pt] (1) -- (2) -- (4) -- (3) -- (1);
	\draw[double, double distance between line centers=1.6pt] (1) -- (0) -- (4);
	\draw[double, double distance between line centers=1.6pt] (2) -- (0) -- (3);
}

\newcommand{\cub}[2]{
	\draw[] ({#1-0.25},{#2-1}) -- ({#1+3.25},{#2+2});
	\draw[] ({#1-0.25},{#2+2.5}) -- ({#1+3.25},{#2-1.5});
	\draw[] ({#1+2.25},{#2+2.5}) -- ({#1+0.75},{#2-1.5});
	\draw[] ({#1+0.75},{#2+2}) -- ({#1+2.25},{#2-1});
	\centerarc[fill= white]({#1+1.5},{#2+0.5})(0:360:0.8)
	\centerarc[peps2]({#1+1.5},{#2+0.5})(0:360:0.8)
	\centerarc[]({#1+1.5},{#2+0.5})(0:360:0.8)
}

\newcommand{\tetraCor}[2]{
	\coordinate (0) at (#1,#2);
	\coordinate (1) at ({#1+5},#2);
	\coordinate (2) at ({#1+1.5},{#2+4});
	\coordinate (3) at ({#1+1.5},{#2-3});
	\coordinate (4) at ({#1-2},{#2+1});
	\coordinate (5) at ({#1+3},{#2+1});
	
	\draw[line width=1pt, shabMax] (1) -- (2) -- (5);
	\draw[line width=1pt, shbMax] (1) -- (3);
	\draw[line width=1pt, shabMax] (2) -- (0) -- (3);
	\draw[line width=1pt, shabMax] (2) -- (4)-- (3);
	\draw[line width=1pt, sha] (3) -- (intersection of 3--5 and 0--1);
	\draw[line width=1pt, shb, shaMax] (intersection of 3--5 and 0--1) -- (5);
	\draw[line width=1pt, shaMax] (0) -- (1);
	\draw[line width=1pt, sha] (4) -- (intersection of 4--5 and 0--2);
	\draw[line width=1pt, shb, shaMax] (intersection of 4--5 and 0--2) -- (5);
	\node[circle, fill = black, inner sep=1.2pt] at (2) {};
	\node[circle, fill = black, inner sep=1.2pt] at (3) {};
	\node[circle, fill = black, inner sep=1.2pt] at (0) {};
	\node[circle, fill = black, inner sep=1.2pt] at (4) {};
}

\def\centerarc[#1](#2)(#3:#4:#5)
{ \draw[#1] ($(#2)+({#5*cos(#3)},{#5*sin(#3)})$) arc (#3:#4:#5); }

\newcommand\marktopleft[1]{%
	\tikz[overlay,remember picture] 
	\node (marker-#1-a) at (0,1.5ex) {};%
}

\newcommand\markbottomright[2]{%
	\tikz[overlay,remember picture] 
	\node (marker-#1-b) at (0,0) {};%
	\tikz[overlay,remember picture,color=#2,inner sep=2pt]
	\node[draw,rectangle, rounded corners=2.5pt, fit=(marker-#1-a.center) (marker-#1-b.center)] {};%
}

\newcommand{\pulling}[2]{
	\begin{tikzpicture}[scale=#1,baseline={([yshift=-.5ex]current bounding box.center)}]
		\ifthenelse{#2=1}{
			\draw[] (0,0) -- (-0.2,0.2);
			\centerarc[fill=white,draw=none](0,0)(0:360:0.1)
			\centerarc[peps2b](0,0)(0:360:0.1)
			\centerarc[](0,0)(0:360:0.1)
			\draw[] (0,0.1) -- (0,0.9);
			\draw[] (0,-0.1) -- (0,-0.9);
			\draw[] (-0.1,0) -- (-0.9,0);
			\draw[] (0.1,0) -- (0.9,0);
			\draw[decorate,decoration=snake,segment amplitude=.2mm,segment length=1mm] (-0.5,0) to[out=90, in=180, looseness=1] (0,0.5);
			\draw[decorate,decoration=snake,segment amplitude=.2mm,segment length=1mm] (-0.5,0) -- (-0.5,-0.4);
			\draw[decorate,decoration=snake,segment amplitude=.2mm,segment length=1mm] (0,0.5) -- (0.4,0.5);
			\centerarc[fill=white](-0.5,0)(0:360:0.1)
			\centerarc[fill=white](0,0.5)(0:360:0.1)	
		}{}
		\ifthenelse{#2=2}{
			\draw[] (0,0) -- (-0.2,0.2);
			\centerarc[fill=white,draw=none](0,0)(0:360:0.1)
			\centerarc[peps2b](0,0)(0:360:0.1)
			\centerarc[](0,0)(0:360:0.1)
			\draw[] (0,0.1) -- (0,0.9);
			\draw[] (0,-0.1) -- (0,-0.9);
			\draw[] (-0.1,0) -- (-0.9,0);
			\draw[] (0.1,0) -- (0.9,0);
			\draw[decorate,decoration=snake,segment amplitude=.2mm,segment length=1mm] (0,-0.5) to[out=0, in=-90, looseness=1] (0.5,0);
			\draw[decorate,decoration=snake,segment amplitude=.2mm,segment length=1mm] (0,-0.5) -- (-0.4,-0.5);
			\draw[decorate,decoration=snake,segment amplitude=.2mm,segment length=1mm] (0.5,0) -- (0.5,0.4);
			\centerarc[fill=white](0.5,0)(0:360:0.1)
			\centerarc[fill=white](0,-0.5)(0:360:0.1)	
		}{}
	\end{tikzpicture}
}

\newcommand{\PEPS}[1]{
	\begin{tikzpicture}[scale=#1,baseline=0.6em]
	\draw[] (0,0) rectangle (1,1);
	\draw[] (0,0.5) -- (-0.5,0.5);
	\draw[] (1,0.5) -- (1.5,0.5);
	\draw[] (0.5,0) -- (0.5,-0.5);
	\draw[] (0.5,1) -- (0.5,1.5);
	
	\node[] at (0.5,0.5) {${\sss B}$};
	\node[left] at (-0.5,0.5) {${\sss a_1}$};
	\node[right] at (1.5,0.5) {${\sss a_4}$};
	\node[above] at (0.5,1.5) {${\sss a_2}$};
	\node[below] at (0.5,-0.5) {${\sss a_3}$};
	\node[circle, fill = black, inner sep=1.2pt] at (0.2,0.8) {};
	\node[above left, xshift=-0.1em, yshift=0.1em] at (0.2,0.8) {${\sss i}$};
	\end{tikzpicture}
}

\newcommand{\PEPO}[1]{
	\begin{tikzpicture}[scale=#1,baseline=0.6em]
	\draw[] (0,0) rectangle (1,1);
	\draw[] (0,0.5) -- (-0.5,0.5);
	\draw[] (1,0.5) -- (1.5,0.5);
	\draw[] (0.5,0) -- (0.5,-0.5);
	\draw[] (0.5,1) -- (0.5,1.5);
	
	\node[] at (0.5,0.5) {${\sss C}$};
	\node[left] at (-0.5,0.5) {${\sss a_1}$};
	\node[right] at (1.5,0.5) {${\sss a_4}$};
	\node[above] at (0.5,1.5) {${\sss a_2}$};
	\node[below] at (0.5,-0.5) {${\sss a_3}$};
	\ddott{0.2,0.8}
	\node[above left, xshift=-0.1em, yshift=0.1em] at (0.2,0.8) {${\sss i,j}$};
	\end{tikzpicture}
}

\newcommand{\PEPSNet}[1]{
	\begin{tikzpicture}[scale=#1,baseline=3.8em]
	\coordinate (1) at (0.5,-0.5);
	\coordinate (2) at (2,-0.5);
	\coordinate (3) at (4.5,-0.5);
	\coordinate (4) at (0.5,5.5);
	\coordinate (5) at (2,5.5);
	\coordinate (6) at (4.5,5.5);
	\coordinate (1b) at (-0.25,-0.5);
	\coordinate (2b) at (1.25,-0.5);
	\coordinate (3b) at (3.75,-0.5);
	\coordinate (4b) at (-0.25,5.5);
	\coordinate (4c) at (-0.25,3.5);
	\coordinate (4d) at (-0.25,3);
	\coordinate (5b) at (1.25,5.5);
	\coordinate (5c) at (1.25,3.5);
	\coordinate (5d) at (1.25,3);
	\coordinate (6b) at (3.75,5.5);
	\coordinate (6c) at (3.75,3.5);
	\coordinate (6d) at (3.75,3);
	\coordinate (7) at (-0.5,0.5);
	\coordinate (8) at (-0.5,2);
	\coordinate (9) at (-0.5,4.5);
	\coordinate (7b) at (-0.5,1.25);
	\coordinate (8b) at (-0.5,2.75);
	\coordinate (9b) at (-0.5,5.25);
	\coordinate (7c) at (1,1.25);
	\coordinate (8c) at (1,2.75);
	\coordinate (9c) at (1,5.25);
	\coordinate (10) at (5.5,0.5);
	\coordinate (11) at (5.5,2);
	\coordinate (12) at (5.5,4.5);
	\coordinate (10b) at (5.5,1.25);
	\coordinate (11b) at (5.5,2.75);
	\coordinate (12b) at (5.5,5.25);
	\coordinate (10c) at (3.5,1.25);
	\coordinate (11c) at (3.5,2.75);
	\coordinate (12c) at (3.5,5.25);
	\coordinate (10d) at (3,1.25);
	\coordinate (11d) at (3,2.75);
	\coordinate (12d) at (3,5.25);

	\draw[] (0,0) rectangle (1,1);
	\draw[] (1.5,0) rectangle (2.5,1);
	\draw[] (0,1.5) rectangle (1,2.5);
	\draw[] (1.5,1.5) rectangle (2.5,2.5);
	\draw[] (4,0) rectangle (5,1);
	\draw[] (4,1.5) rectangle (5,2.5);
	\draw[] (0,4) rectangle (1,5);
	\draw[] (1.5,4) rectangle (2.5,5);
	\draw[] (4,4) rectangle (5,5);
	\draw[] (0,0.5) -- (-0.5,0.5);
	\draw[] (1,0.5) -- (1.5,0.5);
	\draw[] (0.5,0) -- (0.5,-0.5);
	\draw[] (0.5,1) -- (0.5,1.5);
	\draw[] (2,1) -- (2,1.5);
	\draw[] (1,2) -- (1.5,2);
	\draw[] (2,0) -- (2,-0.5);
	\draw[] (0,2) -- (-0.5,2);
	\draw[] (0.5,2.5) -- (0.5,3);
	\draw[] (2.5,0.5) -- (3,0.5);
	\draw[] (2.5,2) -- (3,2);
	\draw[] (2,2.5) -- (2,3);
	\draw[dot] (3,0.5) -- (3.5,0.5);
	\draw[dot] (3,2) -- (3.5,2);
	\draw[dot] (0.5,3) -- (0.5,3.5);
	\draw[dot] (2,3) -- (2,3.5); 
	\draw[] (3.5,0.5) -- (4,0.5);
	\draw[] (3.5,2) -- (4,2);
	\draw[] (0.5,3.5) -- (0.5,4);
	\draw[] (2,3.5) -- (2,4); 
	\draw[] (4.5,1) -- (4.5,1.5);
	\draw[] (1,4.5) -- (1.5,4.5);
	\draw[] (4.5,2.5) -- (4.5,3);
	\draw[dot] (4.5,3) -- (4.5,3.5);
	\draw[] (4.5,3.5) -- (4.5,4);
	\draw[] (2.5,4.5) -- (3,4.5);
	\draw[dot] (3,4.5) -- (3.5,4.5);
	\draw[] (3.5,4.5) -- (4,4.5);
	\draw[] (-0.5,4.5) -- (0,4.5);
	\draw[] (4.5,0) -- (4.5,-0.5);
	\draw[] (0.5,5) -- (0.5,5.5);
	\draw[] (2,5) -- (2,5.5);
	\draw[] (4.5,5) -- (4.5,5.5);
	\draw[] (5,0.5) -- (5.5,0.5);
	\draw[] (5,2) -- (5.5,2);
	\draw[] (5,4.5) -- (5.5,4.5);
	
	\interLL{1b}{4d}{7}{10}{8}{11}{};
	\draw[dot] (4d) -- (4c);
	\interL{4c}{4b}{9}{12}{};
	\interLL{2b}{5d}{7}{10}{8}{11}{};
	\draw[dot] (5d) -- (5c);
	\interL{5c}{5b}{9}{12}{};
	\interLL{3b}{6d}{7}{10}{8}{11}{};
	\draw[dot] (6d) -- (6c);
	\interL{6c}{6b}{9}{12}{};
	
	\interLL{7b}{7c}{1b}{4b}{1}{4}{};
	\interLL{7c}{10d}{2b}{5b}{2}{5}{};
	\draw[dot] (10d) -- (10c);
	\interLL{10c}{10b}{3b}{6b}{3}{6}{};
	\interLL{8b}{8c}{1b}{4b}{1}{4}{};
	\interLL{8c}{11d}{2b}{5b}{2}{5}{};
	\draw[dot] (11d) -- (11c);
	\interLL{11c}{11b}{3b}{6b}{3}{6}{};
	\interLL{9b}{9c}{1b}{4b}{1}{4}{};
	\interLL{9c}{12d}{2b}{5b}{2}{5}{};
	\draw[dot] (12d) -- (12c);
	\interLL{12c}{12b}{3b}{6b}{3}{6}{};
	
	\draw[] (1) to[out=-90, in =-90, looseness=1.5] (1b);
	\draw[] (4) to[out=90, in =90, looseness=1.5] (4b);
	\draw[] (2) to[out=-90, in =-90, looseness=1.5] (2b);
	\draw[] (5) to[out=90, in =90, looseness=1.5] (5b);
	\draw[] (3) to[out=-90, in =-90, looseness=1.5] (3b);
	\draw[] (6) to[out=90, in =90, looseness=1.5] (6b);
	\draw[] (7) to[out=-180, in =-180, looseness=1.5] (7b);
	\draw[] (10) to[out=0, in =0, looseness=1.5] (10b);
	\draw[] (8) to[out=-180, in =-180, looseness=1.5] (8b);
	\draw[] (11) to[out=0, in =0, looseness=1.5] (11b);
	\draw[] (9) to[out=-180, in =-180, looseness=1.5] (9b);
	\draw[] (12) to[out=0, in =0, looseness=1.5] (12b);
	
	\node[] at (0.5,0.5) {${\sss B}$};
	\node[circle, fill = black, inner sep=1.2pt] at (0.2,0.8) {};
	\node[] at (2,0.5) {${\sss B}$};
	\node[circle, fill = black, inner sep=1.2pt] at (1.7,0.8) {};
	\node[] at (0.5,2) {${\sss B}$};
	\node[circle, fill = black, inner sep=1.2pt] at (1.7,2.3) {};
	\node[] at (2,2) {${\sss B}$};
	\node[circle, fill = black, inner sep=1.2pt] at (0.2,2.3) {};
	\node[] at (4.5,0.5) {${\sss B}$};
	\node[circle, fill = black, inner sep=1.2pt] at (4.2,0.8) {};
	\node[] at (4.5,2) {${\sss B}$};
	\node[circle, fill = black, inner sep=1.2pt] at (4.2,2.3) {};
	\node[] at (4.5,4.5) {${\sss B}$};
	\node[circle, fill = black, inner sep=1.2pt] at (4.2,4.8) {};
	\node[] at (0.5,4.5) {${\sss B}$};
	\node[circle, fill = black, inner sep=1.2pt] at (0.2,4.8) {};
	\node[] at (2,4.5) {${\sss B}$};
	\node[circle, fill = black, inner sep=1.2pt] at (1.7,4.8) {};
	
	\end{tikzpicture}
}

\newcommand{\PEPOB}[1]{
	\begin{tikzpicture}[scale=#1,baseline={([yshift=-.5ex]current bounding box.center)}]
	\draw[dot] (0,0) rectangle ({0+2},{0+2});
	\plaqOp{0}{0}{1}{peps1};
	\end{tikzpicture}
}

\newcommand{\PEPOBHalf}[1]{
	\begin{tikzpicture}[scale=#1,baseline=0.1em]
	\univPEPO{{sqrt(0.08)},{sqrt(0.08)}}{{sqrt(3.92)-sqrt(0.08)},{sqrt(0.08)}}{peps1}
	\draw[] ({0.5*sqrt(3.92)},0.55) -- ({0.5*sqrt(3.92)},1);
	\ddott{{sqrt(0.08)},{sqrt(0.08)}}
	\ddott{{sqrt(3.92)-sqrt(0.08)},{sqrt(0.08)}}
	\node[above] at ({0.5*sqrt(3.92)},1) {${\sss a}$};
	\end{tikzpicture}
}

\newcommand{\PEPOBHalfProp}[2]{
	\begin{tikzpicture}[scale=#1,baseline=0.1em]
	\ifthenelse{#2=1}{
		\univPEPO{{sqrt(0.08)},{sqrt(0.08)}}{{sqrt(3.92)-sqrt(0.08)},{sqrt(0.08)}}{peps1}
		\draw[] ({0.5*sqrt(3.92)},0.55) -- ({0.5*sqrt(3.92)},1.5);
		\ddott{{sqrt(0.08)},{sqrt(0.08)}}
		\ddott{{sqrt(3.92)-sqrt(0.08)},{sqrt(0.08)}}
		\node[pauliX] at ({0.5*sqrt(3.92)},1.02) {${\sss X}$}; 
	}{}
	\ifthenelse{#2=2}{
		\univPEPO{{sqrt(0.08)},{sqrt(0.08)}}{{sqrt(3.92)-sqrt(0.08)},{sqrt(0.08)}}{peps1}
		\draw[] ({0.5*sqrt(3.92)},0.55) -- ({0.5*sqrt(3.92)},1.5);
		\node[pauliX] at ({sqrt(0.08)},{sqrt(0.08)}) {${\sss X}$}; 
		\node[pauliX] at ({sqrt(3.92)-sqrt(0.08)},{sqrt(0.08)}) {${\sss X}$}; 
	}{}
	\ifthenelse{#2=3}{
		\univPEPS{{sqrt(0.08)},{sqrt(0.08)}}{{sqrt(3.92)-sqrt(0.08)},{sqrt(0.08)}}{peps1}
		\draw[] ({0.5*sqrt(3.92)},0.55) -- ({0.5*sqrt(3.92)},1.5);
		\node[pauliZ] at ({0.5*sqrt(3.92)},1.02) {${\sss Z}$}; 
	}{}
	\ifthenelse{#2=4}{
		\univPEPS{{sqrt(0.08)},{sqrt(0.08)}}{{sqrt(3.92)-sqrt(0.08)},{sqrt(0.08)}}{peps1}
		\draw[] ({0.5*sqrt(3.92)},0.55) -- ({0.5*sqrt(3.92)},1.5);
		\node[pauliZ] at ({sqrt(0.08)},{sqrt(0.08)}) {${\sss Z}$}; 
	}{}
	\ifthenelse{#2=5}{
		\univPEPS{{sqrt(0.08)},{sqrt(0.08)}}{{sqrt(3.92)-sqrt(0.08)},{sqrt(0.08)}}{peps1}
		\draw[] ({0.5*sqrt(3.92)},0.55) -- ({0.5*sqrt(3.92)},1.5);
		\node[pauliZ] at ({sqrt(3.92)-sqrt(0.08)},{sqrt(0.08)}) {${\sss Z}$}; 
	}{}
	\end{tikzpicture}
}

\newcommand{\PEPOAHalfProp}[2]{
	\begin{tikzpicture}[scale=#1,baseline=0.1em]
	\ifthenelse{#2=1}{
		\univPEPO{{sqrt(0.08)},{sqrt(0.08)}}{{sqrt(3.92)-sqrt(0.08)},{sqrt(0.08)}}{peps2}
		\draw[] ({0.5*sqrt(3.92)},0.55) -- ({0.5*sqrt(3.92)},1.5);
		\ddott{{sqrt(0.08)},{sqrt(0.08)}}
		\ddott{{sqrt(3.92)-sqrt(0.08)},{sqrt(0.08)}}
		\node[pauliZ] at ({0.5*sqrt(3.92)},1.02) {${\sss Z}$}; 
	}{}
	\ifthenelse{#2=2}{
		\univPEPO{{sqrt(0.08)},{sqrt(0.08)}}{{sqrt(3.92)-sqrt(0.08)},{sqrt(0.08)}}{peps2}
		\draw[] ({0.5*sqrt(3.92)},0.55) -- ({0.5*sqrt(3.92)},1.5);
		\node[pauliZ] at ({sqrt(0.08)},{sqrt(0.08)}) {${\sss Z}$}; 
		\node[pauliZ] at ({sqrt(3.92)-sqrt(0.08)},{sqrt(0.08)}) {${\sss Z}$}; 
	}{}
	\ifthenelse{#2=3}{
		\univPEPS{{sqrt(0.08)},{sqrt(0.08)}}{{sqrt(3.92)-sqrt(0.08)},{sqrt(0.08)}}{peps2}
		\draw[] ({0.5*sqrt(3.92)},0.55) -- ({0.5*sqrt(3.92)},1.5);
		\node[pauliX] at ({0.5*sqrt(3.92)},1.02) {${\sss X}$}; 
	}{}
	\ifthenelse{#2=4}{
		\univPEPS{{sqrt(0.08)},{sqrt(0.08)}}{{sqrt(3.92)-sqrt(0.08)},{sqrt(0.08)}}{peps2}
		\draw[] ({0.5*sqrt(3.92)},0.55) -- ({0.5*sqrt(3.92)},1.5);
		\node[pauliX] at ({sqrt(0.08)},{sqrt(0.08)}) {${\sss X}$}; 
	}{}
	\ifthenelse{#2=5}{
		\univPEPS{{sqrt(0.08)},{sqrt(0.08)}}{{sqrt(3.92)-sqrt(0.08)},{sqrt(0.08)}}{peps2}
		\draw[] ({0.5*sqrt(3.92)},0.55) -- ({0.5*sqrt(3.92)},1.5);
		\node[pauliX] at ({sqrt(3.92)-sqrt(0.08)},{sqrt(0.08)}) {${\sss X}$}; 
	}{}
	\end{tikzpicture}
}

\newcommand{\PEPOA}[1]{
	\begin{tikzpicture}[scale=#1,baseline={([yshift=-.5ex]current bounding box.center)}]
	\draw[dot] (1,-1) -- (1,3);
	\draw[dot] (-1,1) -- (3,1);
	\plaqOp{0}{0}{1}{peps2};
	\end{tikzpicture}
}

\newcommand{\PEPOAHalf}[1]{
	\begin{tikzpicture}[scale=#1,baseline=0.1em]
	\univPEPO{{sqrt(0.08)},{sqrt(0.08)}}{{sqrt(3.92)-sqrt(0.08)},{sqrt(0.08)}}{peps2}
	\draw[] ({0.5*sqrt(3.92)},0.55) -- ({0.5*sqrt(3.92)},1);
	\ddott{{sqrt(0.08)},{sqrt(0.08)}}
	\ddott{{sqrt(3.92)-sqrt(0.08)},{sqrt(0.08)}}
	\node[above] at ({0.5*sqrt(3.92)},1) {${\sss \alpha}$};
	\end{tikzpicture}
}

\newcommand{\toricProj}[1]{
	\begin{tikzpicture}[scale=#1,baseline={([yshift=-.5ex]current bounding box.center)}]
	\draw[dot] (0,-1) -- (0,7);
	\draw[dot] (2,-1) -- (2,7);
	\draw[dot] (4,-1) -- (4,7);
	\draw[dot] (6,-1) -- (6,7);
	\draw[dot] (-1,0) -- (7,0);
	\draw[dot] (-1,2) -- (7,2);
	\draw[dot] (-1,4) -- (7,4);
	\draw[dot] (-1,6) -- (7,6);
	
	\begin{scope}
		\clip (-1,-1) rectangle (7,7);
		\foreach \x in {-2,0,2,4,6}
			\plaqOp{\x}{\x}{1}{peps1};
		\plaqOp{4}{0}{1}{peps1}
		\plaqOp{0}{4}{1}{peps1}
		\plaqOp{-2}{2}{1}{peps1}
		\plaqOp{2}{-2}{1}{peps1}
		\plaqOp{6}{2}{1}{peps1}
		\plaqOp{2}{6}{1}{peps1}
		\plaqOp{-2}{6}{1}{peps1}
		\plaqOp{6}{-2}{1}{peps1}
		\plaqOp{0}{2}{2}{peps1}
		\plaqOp{2}{0}{2}{peps1}
		\plaqOp{4}{2}{2}{peps1}
		\plaqOp{2}{4}{2}{peps1}
		\plaqOp{0}{-2}{2}{peps1}
		\plaqOp{-2}{0}{2}{peps1}
		\plaqOp{4}{-2}{2}{peps1}
		\plaqOp{-2}{4}{2}{peps1}
		\plaqOp{6}{0}{2}{peps1}
		\plaqOp{0}{6}{2}{peps1}
		\plaqOp{6}{4}{2}{peps1}
		\plaqOp{4}{6}{2}{peps1}
	\end{scope}								
	\end{tikzpicture}
}

\newcommand{\toricProjLocB}[2]{
	\begin{tikzpicture}[scale=#1,baseline=-0.25em]
	\ifthenelse{#2=1}{
		\univPEPO{0,1}{1,0}{peps1}
		\univPEPO{-1,0}{0,-1}{peps1}
		\univPEPO{-1,0}{0,1}{peps1}
		\univPEPO{0,-1}{1,0}{peps1}
		
		\draw[] (0.68,0.68) -- (1,1) node[anchor=center, pos=1.7] {${\sss a_2}$};
		\draw[] (-0.68,0.68) -- (-1,1) node[anchor=center, pos=1.7] {${\sss a_1}$};
		\draw[] (0.68,-0.68) -- (1,-1) node[anchor=center, pos=1.7] {${\sss a_4}$};
		\draw[] (-0.68,-0.68) -- (-1,-1) node[anchor=center, pos=1.7] {${\sss a_3}$};
		
		\ddott{1,0};
		\ddott{0,1};
		\ddott{-1,0};
		\ddott{0,-1};		
	}{}	
	\ifthenelse{#2=2}{
		\draw[dot] (-2,0) -- (2,0);
		\draw[dot] (0,-2) -- (0,2);
		\node[ket] at (1,0) {${\sss |0\ra}$};
		\node[ket] at (-1,0) {${\sss |0\ra}$};
		\node[ket] at (0,1) {${\sss |0\ra}$};
		\node[ket] at (0,-1) {${\sss |0\ra}$};
	}{}
	\ifthenelse{#2=3}{
		\univTensor{-1,0}{0,1}{1,0}{0,-1}{peps1}		
		\draw[] (0.68,0.68) -- (1,1) node[anchor=center, pos=1.7] {${\sss a_2}$};
		\draw[] (-0.68,0.68) -- (-1,1) node[anchor=center, pos=1.7] {${\sss a_1}$};
		\draw[] (0.68,-0.68) -- (1,-1) node[anchor=center, pos=1.7] {${\sss a_4}$};
		\draw[] (-0.68,-0.68) -- (-1,-1) node[anchor=center, pos=1.7] {${\sss a_3}$};	
		\node[] at (0,0) {${\sss T^X_{\rm 2d}}$};	
	}{}	
	\ifthenelse{#2=4}{
		\univTensor{-1,0}{0,1}{1,0}{0,-1}{peps1}		
		\draw[] (0.68,0.68) -- (1,1) node[anchor=center, pos=1.7] {${\sss a_2}$};
		\draw[] (-0.68,0.68) -- (-1,1) node[anchor=center, pos=1.7] {${\sss a_1}$};
		\draw[] (0.68,-0.68) -- (1,-1) node[anchor=center, pos=1.7] {${\sss a_4}$};
		\draw[] (-0.68,-0.68) -- (-1,-1) node[anchor=center, pos=1.7] {${\sss a_3}$};	
		\node[left] at (-1.2,0) {${\sss a_1+a_3}$};
		\node[above] at (0,1.2) {${\sss a_1+a_2}$};
		\node[right] at (1.2,0) {${\sss a_2+a_4}$};
		\node[below] at (0,-1.2) {${\sss a_3+a_4}$};	
	}{}	
	\ifthenelse{#2=5}{
		\fill[peps1b] (-2,0) -- (0,2) -- (2,0) -- (0,-2) -- (-2,0);
		\draw[] (-2,0) -- (0,2) -- (2,0) -- (0,-2) -- (-2,0);
		\path[] (0.68,0.68) -- (1,1) node[anchor=center, pos=1.7] {${\sss a_2}$};
		\path[] (-0.68,0.68) -- (-1,1) node[anchor=center, pos=1.7] {${\sss a_1}$};
		\path[] (0.68,-0.68) -- (1,-1) node[anchor=center, pos=1.7] {${\sss a_4}$};
		\path[] (-0.68,-0.68) -- (-1,-1) node[anchor=center, pos=1.7] {${\sss a_3}$};	
	}{}	
	\end{tikzpicture}
}

\newcommand{\toricProjLocA}[2]{
	\begin{tikzpicture}[scale=#1,baseline=-0.25em]
	\ifthenelse{#2=1}{
		\univPEPO{0,1}{1,0}{peps2}
		\univPEPO{-1,0}{0,-1}{peps2}
		\univPEPO{-1,0}{0,1}{peps2}
		\univPEPO{0,-1}{1,0}{peps2}
		
		\draw[] (0.68,0.68) -- (1,1) node[anchor=center, pos=1.7] {${\sss \alpha_2}$};
		\draw[] (-0.68,0.68) -- (-1,1) node[anchor=center, pos=1.7] {${\sss \alpha_1}$};
		\draw[] (0.68,-0.68) -- (1,-1) node[anchor=center, pos=1.7] {${\sss \alpha_4}$};
		\draw[] (-0.68,-0.68) -- (-1,-1) node[anchor=center, pos=1.7] {${\sss \alpha_3}$};
		
		\ddott{1,0};
		\ddott{0,1};
		\ddott{-1,0};
		\ddott{0,-1};		
	}{}			
	\ifthenelse{#2=2}{
		\draw[dot] (-1,-1) rectangle (1,1);
		\node[ket] at (1,0) {${\sss |+\ra}$};
		\node[ket] at (-1,0) {${\sss |+\ra}$};
		\node[ket] at (0,1) {${\sss |+\ra}$};
		\node[ket] at (0,-1) {${\sss |+\ra}$};
	}{}
	\ifthenelse{#2=3}{
		\univTensor{-1,0}{0,1}{1,0}{0,-1}{peps2}		
		\draw[] (0.68,0.68) -- (1,1) node[anchor=center, pos=1.7] {${\sss \alpha_2}$};
		\draw[] (-0.68,0.68) -- (-1,1) node[anchor=center, pos=1.7] {${\sss \alpha_1}$};
		\draw[] (0.68,-0.68) -- (1,-1) node[anchor=center, pos=1.7] {${\sss \alpha_4}$};
		\draw[] (-0.68,-0.68) -- (-1,-1) node[anchor=center, pos=1.7] {${\sss \alpha_3}$};	
		\node[] at (0,0) {${\sss T^Z_{\rm 2d}}$};	
	}{}	
	\ifthenelse{#2=5}{
		\draw[] (-1,-1) -- (1,1);
		\draw[] (-1,1) -- (1,-1);
		\centerarc[fill= white](0,0)(0:360:0.7)
		\centerarc[peps2b](0,0)(0:360:0.7)
		\centerarc[](0,0)(0:360:0.7)
		\path[] (0.68,0.68) -- (1,1) node[anchor=center, pos=1.7] {${\sss \alpha_2}$};
		\path[] (-0.68,0.68) -- (-1,1) node[anchor=center, pos=1.7] {${\sss \alpha_1}$};
		\path[] (0.68,-0.68) -- (1,-1) node[anchor=center, pos=1.7] {${\sss \alpha_4}$};
		\path[] (-0.68,-0.68) -- (-1,-1) node[anchor=center, pos=1.7] {${\sss \alpha_3}$};		
	}{}	
	\end{tikzpicture}
}

\newcommand{\SymTwoDPepsOne}[2]{
	\begin{tikzpicture}[scale=#1,baseline=-0.23em]
	\ifthenelse{#2=1}{
		\fill[peps1b] (-1,-1) rectangle (1,1);
		\draw[] (-1,-1) rectangle (1,1);	
	}{}	
	\ifthenelse{#2=2}{
		\fill[peps1b] (-1,-1) rectangle (1,1);
		\draw[] (-1,-1) rectangle (1,1);
		\node[pauliX] at (-1,0) {${\sss X}$}; 
		\node[pauliX] at (0,1) {${\sss X}$}; 
		\node[pauliX] at (0,-1) {${\sss X}$}; 	
		\node[pauliX] at (1,0) {${\sss X}$}; 	
	}{}	
	\ifthenelse{#2=3}{
		\fill[peps1b] (-1,-1) rectangle (1,1);
		\draw[] (-1,-1) rectangle (1,1);
		\node[pauliX] at (-1,0) {${\sss X}$}; 
		\node[pauliX] at (0,1) {${\sss X}$}; 	
	}{}
	\ifthenelse{#2=4}{
		\fill[peps1b] (-1,-1) rectangle (1,1);
		\draw[] (-1,-1) rectangle (1,1);
		\node[pauliX] at (0,-1) {${\sss X}$}; 	
		\node[pauliX] at (1,0) {${\sss X}$}; 	
	}{}
	\ifthenelse{#2=5}{
		\fill[peps1b] (-1,-1) rectangle (3,1);
		\draw[] (-1,-1) rectangle (3,1);
		\draw[] (1,-1) -- (1,1);
		\node[pauliX] at (-1,0) {${\sss X}$}; 
		\node[pauliX] at (0,1) {${\sss X}$};
		\node[pauliX] at (2,1) {${\sss X}$}; 
		\node[pauliX] at (0,-1) {${\sss X}$};
		\node[pauliX] at (2,-1) {${\sss X}$}; 	
		\node[pauliX] at (3,0) {${\sss X}$}; 	
	}{}	
	\ifthenelse{#2=6}{
		\fill[peps1b] (-1,-1) rectangle (3,1);
		\draw[] (-1,-1) rectangle (3,1);
		\draw[] (1,-1) -- (1,1);
		\node[pauliX] at (-1,0) {${\sss X}$}; 
		\node[pauliX] at (0,1) {${\sss X}$};
		\node[pauliX] at (2,1) {${\sss X}$}; 
		\node[pauliX] at (0,-1) {${\sss X}$};
		\node[pauliX] at (2,-1) {${\sss X}$}; 	
		\node[pauliX] at (3,0) {${\sss X}$};
		\node[pauliX] at (1,-0.4) {${\sss X}$}; 
		\node[pauliX] at (1,0.4) {${\sss X}$};  	
	}{}	
	\ifthenelse{#2=7}{
		\fill[peps1b] (-1,-1) rectangle (3,1);
		\draw[] (-1,-1) rectangle (3,1);
		\draw[] (1,-1) -- (1,1);	
	}{}	
	\end{tikzpicture}
}

\newcommand{\SymTwoDPepsTwo}[2]{
	\begin{tikzpicture}[scale=#1,baseline=-0.23em]
	\ifthenelse{#2=1}{
		\centerarc[peps2b](0,0)(0:360:0.6)
		\centerarc[](0,0)(0:360:0.6)
		\draw[] (0,0.6) -- (0,1.6);
		\draw[] (0,-0.6) -- (0,-1.6);
		\draw[] (-0.6,0) -- (-1.6,0);
		\draw[] (0.6,0) -- (1.6,0);
		\node[pauliZ] at (0,1.1) {${\sss Z}$};
		\node[pauliZ] at (0,-1.1) {${\sss Z}$};
		\node[pauliZ] at (1.1,0) {${\sss Z}$};
		\node[pauliZ] at (-1.1,0) {${\sss Z}$};
	}{}	
	\ifthenelse{#2=2}{
		\centerarc[peps2b](0,0)(0:360:0.6)
		\centerarc[](0,0)(0:360:0.6)
		\draw[] (0,0.6) -- (0,1.6);
		\draw[] (0,-0.6) -- (0,-1.6);
		\draw[] (-0.6,0) -- (-1.6,0);
		\draw[] (0.6,0) -- (1.6,0);
	}{}	
	\ifthenelse{#2=3}{
		\centerarc[peps2b](0,0)(0:360:0.6)
		\centerarc[](0,0)(0:360:0.6)
		\draw[] (0,0.6) -- (0,1.6);
		\draw[] (0,-0.6) -- (0,-1.6);
		\draw[] (-0.6,0) -- (-1.6,0);
		\draw[] (0.6,0) -- (1.6,0);
		\node[pauliZ] at (0,1.1) {${\sss Z}$};
		\node[pauliZ] at (-1.1,0) {${\sss Z}$};	
	}{}
	\ifthenelse{#2=4}{
		\centerarc[peps2b](0,0)(0:360:0.6)
		\centerarc[](0,0)(0:360:0.6)
		\draw[] (0,0.6) -- (0,1.6);
		\draw[] (0,-0.6) -- (0,-1.6);
		\draw[] (-0.6,0) -- (-1.6,0);
		\draw[] (0.6,0) -- (1.6,0);
		\node[pauliZ] at (0,-1.1) {${\sss Z}$};
		\node[pauliZ] at (1.1,0) {${\sss Z}$};
	}{}
	\end{tikzpicture}
}

\newcommand{\TwoDAnyons}[1]{
	\begin{tikzpicture}[scale=#1,baseline=2.7em]
	\fill[peps1b] (-2,0) rectangle (8,4);
	\draw[] (-2,0) rectangle (8,4);
	\draw[] (-2,2) -- (8,2);
	\draw[] (0,4) -- (0,0);
	\draw[] (2,4) -- (2,0);
	\draw[] (4,4) -- (4,0);
	\draw[] (6,4) -- (6,0);
	\node[circle, fill = black, inner sep=1.2pt] at (0,2) {};
	\node[left, yshift=0.5em] at (0,2) {${\sss e}$};
	\node[right, yshift=0.5em] at (4,2) {${\sss e}$};
	\node[circle, fill = black, inner sep=1.2pt] at (4,2) {};
	\draw[snake=snake,
	segment amplitude=.2mm,
	segment length=1mm] (0,2) -- (0,0) -- (2,0) -- (2,2) -- (4,2);
	\node[pauliX] at (1,0) {${\sss X}$}; 
	\node[pauliX] at (0,1) {${\sss X}$}; 
	\node[pauliX] at (2,1) {${\sss X}$}; 
	\node[pauliX] at (3,2) {${\sss X}$}; 
	\node[pauliX] at (4,1) {${\sss Z}$}; 
	\node[right, xshift=0.1em, yshift=-0.5em] at (4,1) {${\sss m}$}; 
	\node[pauliX] at (6,3) {${\sss Z}$}; 
	\node[right, xshift=0.1em, yshift=-0.5em] at (6,3) {${\sss m}$};
	\end{tikzpicture}
}

\newcommand{\TwoDFlow}[2]{
	\begin{tikzpicture}[scale=#1,baseline=-0.15em]
	\ifthenelse{#2=0}{
		\fill[peps1b] (-2,-1) rectangle (2,1);
		\draw[] (-2,-1) rectangle (2,1);
		\draw[] (0,1) -- (0,-1);
		\node[above] at (-1,1) {${\sss 2}$};
		\node[above] at (1,1) {${\sss 3}$};
		\node[right] at (2,0) {${\sss 4}$};
		\node[below] at (1,-1) {${\sss 6}$};
		\node[below] at (-1,-1) {${\sss 5}$};
		\node[left] at (-2,0) {${\sss 1}$};	
	}{}	
	\ifthenelse{#2=1}{
		\fill[peps1b] (-2,-2) rectangle (2,2);
		\draw[] (-2,-2) rectangle (2,2);
		\draw[] (-2,0) -- (2,0);
		\draw[] (0,2) -- (0,-2);	
	}{}	
	\ifthenelse{#2=2}{
		\fill[peps1b] (-1,-1) rectangle (1,1);
		\draw[] (-1,-1) rectangle (1,1);
	}{}	
	\ifthenelse{#2=3}{
		\draw[shab, line width=1pt] (-1,-1) -- (1,-1);
		\draw[shab, line width=1pt] (-1,-1) -- (-1,1);
		\draw[shab, line width=1pt] (1,-1) -- (1,1);
		\draw[shab, line width=1pt] (-1,1) -- ( 1,1);
	}{}	
	\ifthenelse{#2=4}{
		\draw[shab, line width=1pt] (-1,-1) -- (1,-1);
		\draw[shab, line width=1pt] (-1,1) -- ( 1,1);
	}{}
	\ifthenelse{#2=5}{
		\draw[shab, line width=1pt] (-1,0) -- (1,0);
	}{}
	\end{tikzpicture}
}

\newcommand{\transferTwoD}[1]{
	\begin{tikzpicture}[scale=#1,baseline={([yshift=-.5ex]current bounding box.center)}]
		\path[name path = P1] (-2,2.5) -- (8,2.5);
		\path[name path = P1b] (-2,2.1) -- (8,2.1);
		\path[name path = P2] (-2,1.1) -- (8,1.1);
		\path[name path = P2b] (-2,0.7) -- (8,0.7);
		\path[name path = P3] (-2,-0.7) -- (8,-0.7);
		\path[name path = P3b] (-2,-1.1) -- (8,-1.1);	
		\path[name path = P4] (-2,-2.1) -- (8,-2.1);
		\path[name path = P4b] (-2,-2.5) -- (8,-2.5);
		\path[name path = C2r] (6,-2.7) to[out=180, in=180, looseness=0.55] (6,2.7);
		\path[name path = C2l] (6,-2.7) to[out=0, in=0, looseness=0.55] (6,2.7);
		\path[name path = C1l] (0,-2.7) to[out=0, in=0, looseness=0.55] (0,2.7);
		\path[name path = C1r] (0,-2.7) to[out=180, in=180, looseness=0.55] (0,2.7);
		\path [name intersections={of=P1 and C1l,by={sp1a}}];
		\path [name intersections={of=P1 and C2l,by={sp1b}}];
		\path [name intersections={of=P2 and C1l,by={sp2a}}];
		\path [name intersections={of=P2 and C2l,by={sp2b}}];
		\path [name intersections={of=P3 and C1l,by={sp3a}}];
		\path [name intersections={of=P3 and C2l,by={sp3b}}];
		\path [name intersections={of=P4 and C1l,by={sp4a}}];
		\path [name intersections={of=P4 and C2l,by={sp4b}}];
		\path [name intersections={of=P1b and C1r,by={sp1c}}];
		\path [name intersections={of=P1b and C2r,by={sp1d}}];
		\path [name intersections={of=P2b and C1r,by={sp2c}}];
		\path [name intersections={of=P2b and C2r,by={sp2d}}];
		\path [name intersections={of=P3b and C1r,by={sp3c}}];
		\path [name intersections={of=P3b and C2r,by={sp3d}}];
		\path [name intersections={of=P4b and C1r,by={sp4c}}];
		\path [name intersections={of=P4b and C2r,by={sp4d}}];

		\fill[peps1b] (0,-2.7) to[out=180, in=180, looseness=0.55] (0,2.7) -- (6,2.7) to[out=180, in=180, looseness=0.55] (6,-2.7) -- (0,-2.7);
		\draw[] (sp1c) -- (sp1d);
		\draw[] (sp2c) -- (sp2d);
		\draw[] (sp3c) -- (sp3d);
		\draw[] (sp4c) -- (sp4d);
		\draw[] (2,-2.7) to[out=180, in=180, looseness=0.55] (2,2.7);
		\draw[] (4,-2.7) to[out=180, in=180, looseness=0.55] (4,2.7);
		\draw[] (6,-2.7) to[out=180, in=180, looseness=0.55] (6,2.7);
		\fill[peps0] (0,-2.7) to[out=0, in=0, looseness=0.55] (0,2.7) -- (6,2.7) to[out=0, in=0, looseness=0.55] (6,-2.7) -- (0,-2.7);
		\fill[peps1b] (0,-2.7) to[out=0, in=0, looseness=0.55] (0,2.7) -- (6,2.7) to[out=0, in=0, looseness=0.55] (6,-2.7) -- (0,-2.7);	
		\draw[] (2,-2.7) to[out=0, in=0, looseness=0.55] (2,2.7);
		\draw[] (4,-2.7) to[out=0, in=0, looseness=0.55] (4,2.7);		
		\draw[] (6,-2.7) to[out=0, in=0, looseness=0.55] (6,2.7);
		\draw[] (0,-2.7) to[out=0, in=0, looseness=0.55] (0,2.7);
		\draw[] (0,-2.7) to[out=180, in=180, looseness=0.55] (0,2.7);
		\draw[] (sp1a) -- (sp1b);
		\draw[] (sp2a) -- (sp2b);
		\draw[] (sp3a) -- (sp3b);
		\draw[] (sp4a) -- (sp4b);
		\draw[snake=snake,
		segment amplitude=.2mm,
		segment length=1mm] (0.85,-0.7) -- (6.85,-0.7);
		\node[circle, fill = black, inner sep=1.2pt] at (0.85,-0.7) {};
		\node[circle, fill = black, inner sep=1.2pt] at (6.85,-0.7) {};
		\node[pauliX] at (1.8,-0.7) {${\sss X^g}$}; 
		\node[pauliX] at (3.8,-0.7) {${\sss X^g}$}; 
		\node[pauliX] at (5.8,-0.7) {${\sss X^g}$}; 
	\end{tikzpicture}
}

\newcommand{\transferTwoDBis}[2]{
	\begin{tikzpicture}[scale=#1,baseline={([yshift=-.5ex]current bounding box.center)}]
	\ifthenelse{#2 = 1}{
		\path[name path = P1] (-2,2.5) -- (8,2.5);
		\path[name path = P1b] (-2,2.1) -- (8,2.1);
		\path[name path = P2] (-2,1.1) -- (8,1.1);
		\path[name path = P2b] (-2,0.7) -- (8,0.7);
		\path[name path = P3] (-2,-0.7) -- (8,-0.7);
		\path[name path = P3b] (-2,-1.1) -- (8,-1.1);	
		\path[name path = P4] (-2,-2.1) -- (8,-2.1);
		\path[name path = P4b] (-2,-2.5) -- (8,-2.5);
		\path[name path = C2r] (3,-2.7) to[out=180, in=180, looseness=0.55] (3,2.7);
		\path[name path = C2l] (3,-2.7) to[out=0, in=0, looseness=0.55] (3,2.7);
		\path[name path = C1l] (0,-2.7) to[out=0, in=0, looseness=0.55] (0,2.7);
		\path[name path = C1r] (0,-2.7) to[out=180, in=180, looseness=0.55] (0,2.7);
		\path [name intersections={of=P1 and C1l,by={sp1a}}];
		\path [name intersections={of=P1 and C2l,by={sp1b}}];
		\path [name intersections={of=P2 and C1l,by={sp2a}}];
		\path [name intersections={of=P2 and C2l,by={sp2b}}];
		\path [name intersections={of=P3 and C1l,by={sp3a}}];
		\path [name intersections={of=P3 and C2l,by={sp3b}}];
		\path [name intersections={of=P4 and C1l,by={sp4a}}];
		\path [name intersections={of=P4 and C2l,by={sp4b}}];
		\path [name intersections={of=P1b and C1r,by={sp1c}}];
		\path [name intersections={of=P1b and C2r,by={sp1d}}];
		\path [name intersections={of=P2b and C1r,by={sp2c}}];
		\path [name intersections={of=P2b and C2r,by={sp2d}}];
		\path [name intersections={of=P3b and C1r,by={sp3c}}];
		\path [name intersections={of=P3b and C2r,by={sp3d}}];
		\path [name intersections={of=P4b and C1r,by={sp4c}}];
		\path [name intersections={of=P4b and C2r,by={sp4d}}];
		
		\fill[peps1b] (0,-2.7) to[out=180, in=180, looseness=0.55] (0,2.7) -- (3,2.7) to[out=180, in=180, looseness=0.55] (3,-2.7) -- (0,-2.7);
		\draw[] (sp1c) -- (sp1d);
		\draw[] (sp2c) -- (sp2d);
		\draw[] (sp3c) -- (sp3d);
		\draw[] (sp4c) -- (sp4d);
		\draw[] (2,-2.7) to[out=180, in=180, looseness=0.55] (2,2.7);
		\fill[peps0] (0,-2.7) to[out=0, in=0, looseness=0.55] (0,2.7) -- (3,2.7) to[out=0, in=0, looseness=0.55] (3,-2.7) -- (0,-2.7);
		\fill[peps1b] (0,-2.7) to[out=0, in=0, looseness=0.55] (0,2.7) -- (3,2.7) to[out=0, in=0, looseness=0.55] (3,-2.7) -- (0,-2.7);	
		\draw[] (2,-2.7) to[out=0, in=0, looseness=0.55] (2,2.7);		
		\draw[] (0,-2.7) to[out=0, in=0, looseness=0.55] (0,2.7);
		\draw[] (0,-2.7) to[out=180, in=180, looseness=0.55] (0,2.7);
		\draw[] (sp1a) -- (sp1b);
		\draw[] (sp2a) -- (sp2b);
		\draw[] (sp3a) -- (sp3b);
		\draw[] (sp4a) -- (sp4b);
		\draw[snake=snake,
		segment amplitude=.2mm,
		segment length=1mm] (0.85,-0.7) -- (3.8,-0.7);
		\node[circle, fill = black, inner sep=1.2pt] at (0.85,-0.7) {};
		\node[pauliX] at (1.8,-0.7) {${\sss X^g}$}; 
	}{}
	\ifthenelse{#2 = 2}{
		\fill[peps1b] (-2,-4) rectangle (1,4);
		\draw[] (-2,-4) -- (-2,4);
		\draw[] (0,-4) -- (0,4);
		\draw[] (-2,-1) -- (1,-1);
		\draw[] (-2,1) -- (1,1);
		\draw[] (-2,3) -- (1,3);
		\draw[] (-2,-3) -- (1,-3);
		\draw[snake=snake,
		segment amplitude=.2mm,
		segment length=1mm] (-2,-1) -- (1,-1);
		\node[circle, fill = black, inner sep=1.2pt] at (-2,-1) {};
		\node[pauliX] at (-1,-1) {${\sss X^g}$}; 
	}{}
	\ifthenelse{#2 = 3}{
		\fill[peps1b] (-2,-4) rectangle (1,4);
		\draw[] (-2,-4) -- (-2,4);
		\draw[] (0,-4) -- (0,4);
		\draw[] (-2,-1) -- (1,-1);
		\draw[] (-2,1) -- (1,1);
		\draw[] (-2,3) -- (1,3);
		\draw[] (-2,-3) -- (1,-3);
		\draw[snake=snake,
		segment amplitude=.2mm,
		segment length=1mm] (-2,-1) -- (-2,4);
		\node[circle, fill = black, inner sep=1.2pt] at (-2,-1) {};
		\node[pauliX] at (-2,0) {${\sss X^g}$}; 
		\node[pauliX] at (-2,2) {${\sss X^g}$}; 
	}{}
	\end{tikzpicture}
}

\newcommand{\mixedTwoD}[2]{
	\begin{tikzpicture}[scale=#1,baseline={([yshift=-.5ex]current bounding box.center)}]
	\begin{scope}
		\clip (-1,-1) rectangle (11,7);
		\draw[dot] (0,-1) -- (0,7);
		\draw[dot] (2,-1) -- (2,7);
		\draw[dot] (4,-1) -- (4,7);
		\draw[dot] (6,-1) -- (6,7);
		\draw[dot] (8,-1) -- (8,7);
		\draw[dot] (10,-1) -- (10,7);
		\draw[dot] (-1,0) -- (11,0);
		\draw[dot] (-1,2) -- (11,2);
		\draw[dot] (-1,4) -- (11,4);
		\draw[dot] (-1,6) -- (11,6);
	\end{scope}
	
	\ifthenelse{#2 = 1}{
		\foreach \x in {1,3,5,7}{
			\node[ket] at (\x,0) {${\sss |+\ra}$};
		}
		\foreach \x in {1,3,5}{
			\node[ket] at (\x,2) {${\sss |+\ra}$};	
			\node[ket] at (0,\x) {${\sss |+\ra}$};
			\node[ket] at (2,\x) {${\sss |+\ra}$};
			\node[ket] at (10,\x) {${\sss |0\ra}$};
			\node[ket] at (8,\x) {${\sss |0\ra}$};
		}
		\foreach \x in {1,3}{
			\node[ket] at (\x,4) {${\sss |+\ra}$};
			\node[ket] at (4,\x) {${\sss |+\ra}$};
		}
		\foreach \x in {1}{
			\node[ket] at (\x,6) {${\sss |+\ra}$};
			\node[ket] at (6,\x) {${\sss |+\ra}$};		
		}
		\foreach \x in {3,5,7,9}{
			\node[ket] at (\x,6) {${\sss |0\ra}$};
		}
		\foreach \x in {5,7,9}{
			\node[ket] at (\x,4) {${\sss |0\ra}$};	
		}
		\foreach \x in {7,9}{
			\node[ket] at (\x,2) {${\sss |0\ra}$};
		}
		\foreach \x in {9}{
			\node[ket] at (\x,0) {${\sss |0\ra}$};	
		}
		\foreach \x in {3,5}{
			\node[ket] at (6,\x) {${\sss |0\ra}$};	
		}
		\foreach \x in {5}{
			\node[ket] at (4,\x) {${\sss |0\ra}$};	
		}
	}{}
	\ifthenelse{#2 = 2}{
		\begin{scope}
			\clip (-1,-1) rectangle (11,7);
			\plaqOp{-3}{5}{1}{peps2}
			\plaqOp{-3}{1}{1}{peps2}
			\plaqOp{-1}{7}{1}{peps2}
			\plaqOp{-1}{3}{1}{peps2}
			\plaqOp{-1}{-1}{1}{peps2}
			\plaqOp{3}{-1}{1}{peps2}
			\plaqOp{1}{1}{1}{peps2}
			\plaqOp{1}{-3}{1}{peps2}
			\plaqOp{5}{-3}{1}{peps2}
			\univPEPO{-1,8}{0,7}{peps2}
			\univPEPO{1,6}{2,5}{peps2}
			\univPEPO{3,4}{4,3}{peps2}
			\univPEPO{5,2}{6,1}{peps2}
			\univPEPO{7,0}{8,-1}{peps2}
			
			\plaqOp{-1}{5}{2}{peps2}
			\plaqOp{1}{3}{2}{peps2}
			\plaqOp{3}{1}{2}{peps2}
			\plaqOp{5}{-1}{2}{peps2}
			\plaqOp{-3}{3}{2}{peps2}
			\plaqOp{-1}{1}{2}{peps2}
			\plaqOp{1}{-1}{2}{peps2}
			\plaqOp{3}{-3}{2}{peps2}
			\plaqOp{-3}{-1}{2}{peps2}
			\plaqOp{-1}{-3}{2}{peps2}

			\plaqOp{4}{6}{1}{peps1}
			\plaqOp{8}{6}{1}{peps1}
			\plaqOp{6}{4}{1}{peps1}
			\plaqOp{10}{4}{1}{peps1}
			\plaqOp{8}{2}{1}{peps1}
			\plaqOp{10}{0}{1}{peps1}
			\univPEPO{1,8}{2,7}{peps1}
			\univPEPO{3,6}{4,5}{peps1}
			\univPEPO{5,4}{6,3}{peps1}
			\univPEPO{7,2}{8,1}{peps1}
			\univPEPO{9,0}{10,-1}{peps1}
			
			\plaqOp{6}{6}{2}{peps1}
			\plaqOp{2}{6}{2}{peps1}
			\plaqOp{8}{4}{2}{peps1}
			\plaqOp{4}{4}{2}{peps1}
			\plaqOp{10}{2}{2}{peps1}
			\plaqOp{6}{2}{2}{peps1}
			\plaqOp{8}{0}{2}{peps1}
			\plaqOp{10}{-2}{2}{peps1}
			\plaqOp{10}{6}{2}{peps1}
			
			\centerarc[peps1](1,7)(0:360:0.2)
			\centerarc[](1,7)(0:360:0.2)
			\centerarc[peps1](3,5)(0:360:0.2)
			\centerarc[](3,5)(0:360:0.2)
			\centerarc[peps1](5,3)(0:360:0.2)
			\centerarc[](5,3)(0:360:0.2)
			\centerarc[peps1](7,1)(0:360:0.2)
			\centerarc[](7,1)(0:360:0.2)
			\centerarc[peps1](9,-1)(0:360:0.2)
			\centerarc[](9,-1)(0:360:0.2)

			\centerarc[peps0](2,6)(0:360:0.2)
			\centerarc[peps2](2,6)(0:360:0.2)
			\centerarc[](2,6)(0:360:0.2)
			\centerarc[peps0](4,4)(0:360:0.2)
			\centerarc[peps2](4,4)(0:360:0.2)
			\centerarc[](4,4)(0:360:0.2)
			\centerarc[peps0](6,2)(0:360:0.2)
			\centerarc[peps2](6,2)(0:360:0.2)
			\centerarc[](6,2)(0:360:0.2)
			\centerarc[peps0](8,0)(0:360:0.2)
			\centerarc[peps2](8,0)(0:360:0.2)
			\centerarc[](8,0)(0:360:0.2)
			
			\draw[shorten <= 3pt, shorten >= 3pt] (1,7) -- (2,6);
			\draw[shorten <= 3pt, shorten >= 3pt] (2,6) -- (3,5);
			\draw[shorten <= 3pt, shorten >= 3pt] (3,5) -- (4,4);
			\draw[shorten <= 3pt, shorten >= 3pt] (4,4) -- (5,3);
			\draw[shorten <= 3pt, shorten >= 3pt] (5,3) -- (6,2);
			\draw[shorten <= 3pt, shorten >= 3pt] (6,2) -- (7,1);
			\draw[shorten <= 3pt, shorten >= 3pt] (7,1) -- (8,0);
			\draw[shorten <= 3pt, shorten >= 3pt] (8,0) -- (9,-1);
			\draw[shorten >= 3pt] (1.68,5.68) -- (2,6);
			\draw[shorten >= 3pt] (3.68,3.68) -- (4,4);
			\draw[shorten >= 3pt] (5.68,1.68) -- (6,2);
			\draw[shorten >= 3pt] (7.68,-0.32) -- (8,0);
			\draw[shorten >= 3pt] (3.32,5.32) -- (3,5);
			\draw[shorten >= 3pt] (5.32,3.32) -- (5,3);
			\draw[shorten >= 3pt] (7.32,1.32) -- (7,1);
			\draw[shorten >= 3pt] (9.32,-0.68) -- (9,-1);
		\end{scope}	
	}{}	
	\ifthenelse{#2 = 3}{	
		\begin{scope}
			\clip (-1,-1) rectangle (11,7);
			\fill[peps0] (1,7) -- (9,-1) -- (11,7) -- (1,7);
			\fill[peps1b] (1,7) -- (9,-1) -- (11,-1) -- (11,7) -- (1,7);
			\draw[] (5,7) -- (12,0);
			\draw[] (8,8) -- (12,4);
			\draw[] (2,6) -- (4,8);
			\draw[] (4,4) -- (8,8);
			\draw[] (6,2) -- (12,8);
			\draw[] (8,0) -- (12,4);
			\draw[] (-1,7) -- (7,-1);
			\draw[] (-1,3) -- (3,-1);
			\draw[] (-1,-1) -- (3.75,3.75);
			\draw[] (-1,3) -- (1.75,5.75);
			\draw[] (3,-1) -- (5.75,1.75);
			\draw[] (1,7) -- (9,-1);
			\draw[] (7,-1) -- (7.75,-0.25);
			\draw[] (1,7) -- (1.75,5.75) -- (3,5) -- (3.75,3.75) -- (5,3) -- (5.75,1.75) -- (7,1) -- (7.75,-0.25) -- (9,-1);
			
			\centerarc[fill=white](1,7)(0:360:0.2)
			\centerarc[peps1](1,7)(0:360:0.2)
			\centerarc[](1,7)(0:360:0.2)
			\centerarc[fill=white](3,5)(0:360:0.2)
			\centerarc[peps1](3,5)(0:360:0.2)
			\centerarc[](3,5)(0:360:0.2)
			\centerarc[fill=white](5,3)(0:360:0.2)
			\centerarc[peps1](5,3)(0:360:0.2)
			\centerarc[](5,3)(0:360:0.2)
			\centerarc[fill=white](7,1)(0:360:0.2)
			\centerarc[peps1](7,1)(0:360:0.2)
			\centerarc[](7,1)(0:360:0.2)
			\centerarc[fill=white](9,-1)(0:360:0.2)
			\centerarc[peps1](9,-1)(0:360:0.2)
			\centerarc[](9,-1)(0:360:0.2)
			
			\centerarc[fill=white](1.75,5.75)(0:360:0.2)
			\centerarc[peps2](1.75,5.75)(0:360:0.2)
			\centerarc[](1.75,5.75)(0:360:0.2)
			\centerarc[fill=white](3.75,3.75)(0:360:0.2)
			\centerarc[peps2](3.75,3.75)(0:360:0.2)
			\centerarc[](3.75,3.75)(0:360:0.2)
			\centerarc[fill=white](5.75,1.75)(0:360:0.2)
			\centerarc[peps2](5.75,1.75)(0:360:0.2)
			\centerarc[](5.75,1.75)(0:360:0.2)
			\centerarc[fill=white](7.75,-0.25)(0:360:0.2)
			\centerarc[peps2](7.75,-0.25)(0:360:0.2)
			\centerarc[](7.75,-0.25)(0:360:0.2)
						
			\centerarc[fill=white](-1,7)(0:360:0.7)
			\centerarc[peps2b](-1,7)(0:360:0.7)
			\centerarc[](-1,7)(0:360:0.7)
			\centerarc[fill=white](-1,3)(0:360:0.7)
			\centerarc[peps2b](-1,3)(0:360:0.7)
			\centerarc[](-1,3)(0:360:0.7)
			\centerarc[fill=white](-1,-1)(0:360:0.7)
			\centerarc[peps2b](-1,-1)(0:360:0.7)
			\centerarc[](-1,-1)(0:360:0.7)
			\centerarc[fill=white](1,5)(0:360:0.7)
			\centerarc[peps2b](1,5)(0:360:0.7)
			\centerarc[](1,5)(0:360:0.7)
			\centerarc[fill=white](1,1)(0:360:0.7)
			\centerarc[peps2b](1,1)(0:360:0.7)
			\centerarc[](1,1)(0:360:0.7)
			\centerarc[fill=white](3,3)(0:360:0.7)
			\centerarc[peps2b](3,3)(0:360:0.7)
			\centerarc[](3,3)(0:360:0.7)
			\centerarc[fill=white](3,-1)(0:360:0.7)
			\centerarc[peps2b](3,-1)(0:360:0.7)
			\centerarc[](3,-1)(0:360:0.7)
			\centerarc[fill=white](5,1)(0:360:0.7)
			\centerarc[peps2b](5,1)(0:360:0.7)
			\centerarc[](5,1)(0:360:0.7)
			\centerarc[fill=white](7,-1)(0:360:0.7)
			\centerarc[peps2b](7,-1)(0:360:0.7)
			\centerarc[](7,-1)(0:360:0.7)
		\end{scope}		
	}{}							
	\end{tikzpicture}
}

\newcommand{\deltaT}[2]{
	\begin{tikzpicture}[scale=#1,baseline=-3pt]
	\draw[] (-1.2,0) -- (1.2,0);
	\ifthenelse{#2 = 0}{
		\draw[] (0,1.2) -- (0,0);
		\centerarc[fill=white, draw=none](0,0)(0:360:0.2)	
		\centerarc[peps1](0,0)(0:360:0.2)
		\centerarc[](0,0)(0:360:0.2)	
	}{}		
	\ifthenelse{#2 = 1}{
		\draw[] (0,1.2) -- (0,0);	
		\centerarc[fill=white, draw=none](0,0)(0:360:0.2)	
		\centerarc[peps1](0,0)(0:360:0.2)
		\centerarc[](0,0)(0:360:0.2)
		\node[left] at (-1.2,0) {${\sss a_1}$};
		\node[above] at (0,1.2) {${\sss a_2}$};	
		\node[right] at (1.2,0) {${\sss a_3}$};		
	}{}	
	\ifthenelse{#2 = 2}{
		\draw[] (0,1.2) -- (0,0);	
		\centerarc[fill=white, draw=none](0,0)(0:360:0.2)	
		\centerarc[peps1](0,0)(0:360:0.2)
		\centerarc[](0,0)(0:360:0.2)
		\node[pauliX] at (-0.7,0) {${\sss X}$}; 
		\node[pauliX] at (0.7,0) {${\sss X}$}; 
		\node[pauliX] at (0,0.7) {${\sss X}$}; 
	}{}	
	\ifthenelse{#2 = 3}{
		\draw[] (0,1.2) -- (0,0);	
		\centerarc[fill=white, draw=none](0,0)(0:360:0.2)	
		\centerarc[peps1](0,0)(0:360:0.2)
		\centerarc[](0,0)(0:360:0.2)
		\node[pauliZ] at (0.7,0) {${\sss Z}$}; 
		\node[pauliZ] at (0,0.7) {${\sss Z}$}; 
	}{}
	\ifthenelse{#2 = 4}{
		\draw[] (0,1.2) -- (0,0);
		\centerarc[fill=white, draw=none](0,0)(0:360:0.2)	
		\centerarc[peps1](0,0)(0:360:0.2)
		\centerarc[](0,0)(0:360:0.2)
		\node[pauliZ] at (-0.7,0) {${\sss Z}$}; 
		\node[pauliZ] at (0,0.7) {${\sss Z}$}; 
	}{}
	\ifthenelse{#2 = 5}{
		\draw[] (0,1.2) -- (0,0);	
		\centerarc[fill=white, draw=none](0,0)(0:360:0.2)	
		\centerarc[peps1](0,0)(0:360:0.2)
		\centerarc[](0,0)(0:360:0.2)
		\node[pauliZ] at (-0.7,0) {${\sss Z}$}; 
		\node[pauliZ] at (0.7,0) {${\sss Z}$}; 
	}{}
	\ifthenelse{#2 = 10}{	
		\draw[] (0,-1.2) -- (0,0);	
		\centerarc[fill=white, draw=none](0,0)(0:360:0.2)
		\centerarc[peps2](0,0)(0:360:0.2)
		\centerarc[](0,0)(0:360:0.2)		
	}{}		
	\ifthenelse{#2 = 11}{
		\draw[] (0,-1.2) -- (0,0);		
		\centerarc[fill=white, draw=none](0,0)(0:360:0.2)
		\centerarc[peps2](0,0)(0:360:0.2)
		\centerarc[](0,0)(0:360:0.2)
		\node[left] at (-1.2,0) {${\sss \alpha_1}$};
		\node[below] at (0,-1.2) {${\sss \alpha_2}$};	
		\node[right] at (1.2,0) {${\sss \alpha_3}$};		
	}{}	
	\ifthenelse{#2 = 12}{	
		\draw[] (0,-1.2) -- (0,0);
		\centerarc[fill=white, draw=none](0,0)(0:360:0.2)
		\centerarc[peps2](0,0)(0:360:0.2)
		\centerarc[](0,0)(0:360:0.2)	
		\node[pauliZ] at (-0.7,0) {${\sss Z}$}; 
		\node[pauliZ] at (0.7,0) {${\sss Z}$}; 
		\node[pauliZ] at (0,-0.7) {${\sss Z}$}; 	
	}{}	
	\ifthenelse{#2 = 13}{	
		\draw[] (0,-1.2) -- (0,0);
		\centerarc[fill=white, draw=none](0,0)(0:360:0.2)
		\centerarc[peps2](0,0)(0:360:0.2)
		\centerarc[](0,0)(0:360:0.2)
		\node[pauliX] at (0.7,0) {${\sss X}$}; 
		\node[pauliX] at (0,-0.7) {${\sss X}$}; 	
	}{}	
	\ifthenelse{#2 = 14}{	
		\draw[] (0,-1.2) -- (0,0);
		\centerarc[fill=white, draw=none](0,0)(0:360:0.2)
		\centerarc[peps2](0,0)(0:360:0.2)
		\centerarc[](0,0)(0:360:0.2)
		\node[pauliX] at (-0.7,0) {${\sss X}$}; 
		\node[pauliX] at (0,-0.7) {${\sss X}$}; 	
	}{}	
	\ifthenelse{#2 = 15}{	
		\draw[] (0,-1.2) -- (0,0);
		\centerarc[fill=white, draw=none](0,0)(0:360:0.2)
		\centerarc[peps2](0,0)(0:360:0.2)
		\centerarc[](0,0)(0:360:0.2)	
		\node[pauliX] at (-0.7,0) {${\sss X}$}; 
		\node[pauliX] at (0.7,0) {${\sss X}$}; 	
	}{}						
	\end{tikzpicture}
}

\newcommand{\mixedTwoDMPO}[2]{
	\begin{tikzpicture}[scale=#1,baseline={([yshift=-.5ex]current bounding box.center)}]
	\draw[] ({-0.5*sqrt(2)},0) -- ({9.5*sqrt(2)},0);
	\foreach \x in {0,2,4,6,8}{
		\draw[] ({\x*sqrt(2)},1.2) -- ({\x*sqrt(2)},0);
		\centerarc[fill=white, draw=none]({\x*sqrt(2)},0)(0:360:0.2)	
		\centerarc[peps1]({\x*sqrt(2)},0)(0:360:0.2)
		\centerarc[]({\x*sqrt(2)},0)(0:360:0.2)
	}
	\foreach \x in {1,3,5,7,9}{
		\draw[] ({\x*sqrt(2)},-1.2) -- ({\x*sqrt(2)},0);
		\centerarc[fill=white, draw=none]({\x*sqrt(2)},0)(0:360:0.2)	
		\centerarc[peps2]({\x*sqrt(2)},0)(0:360:0.2)
		\centerarc[]({\x*sqrt(2)},0)(0:360:0.2)
	}	
	\ifthenelse{#2 = 0}{	
		\foreach \x in {2,4,6,8}
			\node[pauliX] at ({\x*sqrt(2)},0.7) {${\sss X}$}; 
	}{}		
	\ifthenelse{#2 = 1}{	
		\foreach \x in {1.5,2.5,3.5,4.5,5.5,6.5,7.5,8.5}
			\node[pauliX] at ({\x*sqrt(2)},0) {${\sss X}$}; 
	}{}	
	\ifthenelse{#2 = 2}{	
		\foreach \x in {1,9}
		\node[pauliX] at ({\x*sqrt(2)},-0.7) {${\sss X}$}; 
	}{}				
	\end{tikzpicture}
}

\newcommand{\interMPOSym}[2]{
	\begin{tikzpicture}[scale=#1,baseline={([yshift=-.5ex]current bounding box.center)}]
	\draw[dot] (-1,0) -- (3,0);
	\draw[dot] (-1,2) -- (3,2);
	\draw[dot] (0,-1) -- (0,3);
	\draw[dot] (2,-1) -- (2,3); 
	\ifthenelse{#2 = 0}{
		\node[pauliX] at (0,1) {${\sss X}$};
		\node[pauliX] at (1,0) {${\sss X}$};
		\node[pauliX] at (1,2) {${\sss X}$};
		\node[pauliX] at (2,1) {${\sss X}$};
	}{}
	\ifthenelse{#2 = -1}{
		\node[ket] at (0,1) {${\sss |+\ra}$};
		\node[ket] at (1,0) {${\sss |+\ra}$};
		\node[ket] at (1,2) {${\sss |0\ra}$};
		\node[ket] at (2,1) {${\sss |0\ra}$};
	}{}
	\ifthenelse{#2 = 1}{
		\begin{scope}
			\clip (-1,-1) rectangle (3,3);
			\univPEPO{1,2}{2,1}{peps1}
			\univPEPO{-1,2}{0,1}{peps2}
			\univPEPO{1,0}{2,-1}{peps2}
			\centerarc[peps0](0,2)(0:360:0.2)
			\centerarc[peps2](0,2)(0:360:0.2)
			\centerarc[](0,2)(0:360:0.2)
			\centerarc[peps0](2,0)(0:360:0.2)
			\centerarc[peps2](2,0)(0:360:0.2)
			\centerarc[](2,0)(0:360:0.2)
			\centerarc[peps1](1,1)(0:360:0.2)
			\centerarc[](1,1)(0:360:0.2)
			\draw[shorten <= 3pt, shorten >= 3pt] (-1,3) -- (0,2);
			\draw[shorten <= 3pt, shorten >= 3pt] (0,2) -- (1,1);
			\draw[shorten <= 3pt, shorten >= 3pt] (1,1) -- (2,0);
			\draw[shorten <= 3pt, shorten >= 3pt] (2,0) -- (3,-1);
			\draw[shorten >= 3pt] (-0.32,1.68) -- (0,2);
			\draw[shorten >= 3pt] (1.68,-0.32) -- (2,0);
			\draw[shorten >= 3pt] (1.32,1.32) -- (1,1);
		\end{scope}
	}{}
	\ifthenelse{#2 = 11}{
		\begin{scope}
			\clip (-1,-1) rectangle (3,3);
			\univPEPS{1,2}{2,1}{peps1}
			\univPEPS{-1,2}{0,1}{peps2}
			\univPEPS{1,0}{2,-1}{peps2}
			\centerarc[peps0](0,2)(0:360:0.2)
			\centerarc[peps2](0,2)(0:360:0.2)
			\centerarc[](0,2)(0:360:0.2)
			\centerarc[peps0](2,0)(0:360:0.2)
			\centerarc[peps2](2,0)(0:360:0.2)
			\centerarc[](2,0)(0:360:0.2)
			\centerarc[peps1](1,1)(0:360:0.2)
			\centerarc[](1,1)(0:360:0.2)
			\draw[shorten <= 3pt, shorten >= 3pt] (-1,3) -- (0,2);
			\draw[shorten <= 3pt, shorten >= 3pt] (0,2) -- (1,1);
			\draw[shorten <= 3pt, shorten >= 3pt] (1,1) -- (2,0);
			\draw[shorten <= 3pt, shorten >= 3pt] (2,0) -- (3,-1);
			\draw[shorten >= 3pt] (-0.32,1.68) -- (0,2);
			\draw[shorten >= 3pt] (1.68,-0.32) -- (2,0);
			\draw[shorten >= 3pt] (1.32,1.32) -- (1,1);
		\end{scope}
	}{}
	\ifthenelse{#2 = 2}{
		\begin{scope}
			\clip (-1,-1) rectangle (3,3);
			\univPEPS{1,2}{2,1}{peps1}
			\univPEPS{-1,2}{0,1}{peps2}
			\univPEPS{1,0}{2,-1}{peps2}
			\centerarc[peps0](0,2)(0:360:0.2)
			\centerarc[peps2](0,2)(0:360:0.2)
			\centerarc[](0,2)(0:360:0.2)
			\centerarc[peps0](2,0)(0:360:0.2)
			\centerarc[peps2](2,0)(0:360:0.2)
			\centerarc[](2,0)(0:360:0.2)
			\centerarc[peps1](1,1)(0:360:0.2)
			\centerarc[](1,1)(0:360:0.2)
			\draw[shorten <= 3pt, shorten >= 3pt] (-1,3) -- (0,2);
			\draw[shorten <= 3pt, shorten >= 3pt] (0,2) -- (1,1);
			\draw[shorten <= 3pt, shorten >= 3pt] (1,1) -- (2,0);
			\draw[shorten <= 3pt, shorten >= 3pt] (2,0) -- (3,-1);
			\draw[shorten >= 3pt] (-0.32,1.68) -- (0,2);
			\draw[shorten >= 3pt] (1.68,-0.32) -- (2,0);
			\draw[shorten >= 3pt] (1.32,1.32) -- (1,1);
			\node[pauliX] at (0,1) {${\sss X}$};
			\node[pauliX] at (1,0) {${\sss X}$};
			\node[pauliX] at (1,2) {${\sss X}$};
			\node[pauliX] at (2,1) {${\sss X}$};
		\end{scope}
	}{}
	\ifthenelse{#2 = 3}{
		\begin{scope}
			\clip (-1,-1) rectangle (3,3);
			\univPEPS{1,2}{2,1}{peps1}
			\univPEPS{-1,2}{0,1}{peps2}
			\univPEPS{1,0}{2,-1}{peps2}
			\centerarc[peps0](0.47,2.47)(0:360:0.2)
			\centerarc[peps2](0.47,2.47)(0:360:0.2)
			\centerarc[](0.47,2.47)(0:360:0.2)
			\centerarc[peps0](2.47,0.47)(0:360:0.2)
			\centerarc[peps2](2.47,0.47)(0:360:0.2)
			\centerarc[](2.47,0.47)(0:360:0.2)
			\centerarc[peps1](0.53,0.53)(0:360:0.2)
			\centerarc[](0.53,0.53)(0:360:0.2)
			\draw[shorten <= 3pt, shorten >= 3pt] (-1.47,2.53) to[out=-35, in=145, looseness=0.8] (0.47,2.47);
			\draw[shorten <= 3pt, shorten >= 3pt] (0.47,2.47) to[out=-55, in=125, looseness=0.8] (0.53,0.53);
			\draw[shorten <= 3pt, shorten >= 3pt] (0.53,0.53) to[out=-35, in=145, looseness=0.8] (2.47,0.47);
			\draw[shorten <= 3pt, shorten >= 3pt] (2.47,0.47) to[out=-55, in=125, looseness=0.8] (2.53,-1.47);
			\draw[shorten >= 3pt] (-0.32,1.68) -- (0.47,2.47);
			\draw[shorten >= 3pt] (1.68,-0.32) -- (2.47,0.47);
			\draw[shorten >= 3pt] (1.32,1.32) -- (0.53,0.53);
			\node[pauliX] at (0,2) {${\sss X}$};
			\node[pauliX] at (2,0) {${\sss X}$};
			\node[pauliX] at (1,1) {${\sss X}$};
		\end{scope}
	}{}
	\ifthenelse{#2 = 4}{
		\begin{scope}
			\clip (-1,-1) rectangle (3,3);
			\univPEPS{1,2}{2,1}{peps1}
			\univPEPS{-1,2}{0,1}{peps2}
			\univPEPS{1,0}{2,-1}{peps2}
			\centerarc[peps0](0.47,2.47)(0:360:0.2)
			\centerarc[peps2](0.47,2.47)(0:360:0.2)
			\centerarc[](0.47,2.47)(0:360:0.2)
			\centerarc[peps0](2.47,0.47)(0:360:0.2)
			\centerarc[peps2](2.47,0.47)(0:360:0.2)
			\centerarc[](2.47,0.47)(0:360:0.2)
			\centerarc[peps1](0.53,0.53)(0:360:0.2)
			\centerarc[](0.53,0.53)(0:360:0.2)
			\draw[shorten <= 3pt, shorten >= 3pt] (-1.47,2.53) to[out=-35, in=145, looseness=0.8] (0.47,2.47);
			\draw[shorten <= 3pt, shorten >= 3pt] (0.47,2.47) to[out=-55, in=125, looseness=0.8, edge node={coordinate[pos=0.5] (sp1)}] (0.53,0.53);
			\draw[shorten <= 3pt, shorten >= 3pt] (0.53,0.53) to[out=-35, in=145, looseness=0.8, edge node={coordinate[pos=0.5] (sp2)}] (2.47,0.47);
			\draw[shorten <= 3pt, shorten >= 3pt] (2.47,0.47) to[out=-55, in=125, looseness=0.8] (2.53,-1.47);
			\draw[shorten >= 3pt] (-0.32,1.68) -- (0.47,2.47);
			\draw[shorten >= 3pt] (1.68,-0.32) -- (2.47,0.47);
			\draw[shorten >= 3pt] (1.32,1.32) -- (0.53,0.53);
			\node[pauliX] at (sp1) {${\sss X}$};
			\node[pauliX] at (sp2) {${\sss X}$};
			\node[pauliX] at (1,1) {${\sss X}$};
		\end{scope}
	}{}
	\end{tikzpicture}
}

\newcommand{\defectTwoD}[2]{
	\begin{tikzpicture}[scale=#1,baseline=5.5em]
	\draw[dot] (0,4) -- (0,9);
	\draw[dot] (2,4) -- (2,9);
	\draw[dot] (4,4) -- (4,9);
	\draw[dot] (6,4) -- (6,9);
	\draw[dot] (8,4) -- (8,9);
	\draw[dot] (10,4) -- (10,9);
	\draw[dot] (12,4) -- (12,9);
	\draw[dot] (-1,-1) -- (-1,4);
	\draw[dot] (1,-1) -- (1,4);
	\draw[dot] (3,-1) -- (3,4);
	\draw[dot] (5,-1) -- (5,4);
	\draw[dot] (7,-1) -- (7,4);
	\draw[dot] (9,-1) -- (9,4);
	\draw[dot] (11,-1) -- (11,4);
	\draw[dot] (13,-1) -- (13,4);
	\draw[dot] (-1,0) -- (13,0);
	\draw[dot] (-1,2) -- (13,2);
	\draw[dot] (-1,6) -- (13,6);
	\draw[dot] (-1,8) -- (13,8);
	\draw[snake=zigzag,
	segment amplitude=.5mm,
	segment length=2mm,
	line width=1pt] (-1,4) -- (13,4);
	
	\ifthenelse{#2 = 1}{
		\foreach \x in {0,2,4,6,8,12}
			\node[ket] at (\x,0) {${\sss |+\ra}$};
		\foreach \x in {-1,1,3,5,7,13}
			\node[ket] at (\x,1) {${\sss |+\ra}$};
		\foreach \x in {3,7,9,11,13}
			\node[ket] at (\x,3) {${\sss |+\ra}$};
		\foreach \x in {2,4,6,8,12}
			\node[ket] at (\x,2) {${\sss |+\ra}$};	

		\foreach \x in {0,2,4,6,8,12}
			\node[ket] at (\x,7) {${\sss |0\ra}$};
		\foreach \x in {2,8,12}
			\node[ket] at (\x,5) {${\sss |0\ra}$};	
		\foreach \x in {1,3,7}
			\node[ket] at (\x,6) {${\sss |0\ra}$};
		\foreach \x in {1,3,5,7,9,11}
			\node[ket] at (\x,8) {${\sss |0\ra}$};
									
		\node[pauliX] at (-1,3) {${\sss X}$}; 
		\node[pauliX] at (1,3) {${\sss X}$}; 
		\node[pauliX] at (0,2) {${\sss X}$}; 
		\node[pauliZ] at (0,5) {${\sss Z}$}; 
		\node[pauliX] at (4,5) {${\sss X}$}; 
		\node[pauliX] at (6,5) {${\sss X}$}; 
		\node[pauliX] at (5,6) {${\sss X}$}; 
		\node[pauliZ] at (5,3) {${\sss Z}$}; 
		\node[pauliZ] at (9,6) {${\sss Z}$};
		\node[pauliZ] at (11,6) {${\sss Z}$};
		\node[pauliZ] at (10,7) {${\sss Z}$};
		\node[pauliZ] at (10,5) {${\sss Z}$};
		\node[pauliX] at (9,1) {${\sss X}$};
		\node[pauliX] at (11,1) {${\sss X}$};
		\node[pauliX] at (10,2) {${\sss X}$};
		\node[pauliX] at (10,0) {${\sss X}$};	
	}{}
	\ifthenelse{#2 = 2}{
		\begin{scope}
			\clip (-1,-1) rectangle (13,9);
			\foreach \x in {-2,2,6,10}
				\plaqOp{\x}{6}{1}{peps1};
			\foreach \x in {-2,2,6,10}
				\plaqOp{\x}{1}{1}{peps2};
			\foreach \x in {0,4,8,12}
				\plaqOp{\x}{8}{1}{peps1};
			\foreach \x in {0,4,8,12}
				\plaqOp{\x}{-1}{1}{peps2};	
				
			\foreach \x in {-2,2,6,10}
				\plaqOp{\x}{8}{2}{peps1};
			\foreach \x in {0,4,8,12}
				\plaqOp{\x}{6}{2}{peps1};
			\foreach \x in {0,4,8,12}
				\plaqOp{\x}{1}{2}{peps2};
			\foreach \x in {-2,2,6,10}
				\plaqOp{\x}{-1}{2}{peps2};	
			
			\univPEPO{0,5}{1,3}{peps3};
			\univPEPO{-1,3}{0,5}{peps3};
			\univPEPO{4,5}{5,3}{peps3};
			\univPEPO{3,3}{4,5}{peps3};
			\univPEPO{8,5}{9,3}{peps3};
			\univPEPO{7,3}{8,5}{peps3};
			\univPEPO{12,5}{13,3}{peps3};
			\univPEPO{11,3}{12,5}{peps3};
			
			\univPEPO{5,6}{6,5}{peps1};
			\univPEPO{1,6}{2,5}{peps1};
			\univPEPO{9,6}{10,5}{peps1};
			\univPEPO{13,6}{14,5}{peps1};
			\univPEPO{6,5}{7,6}{peps1};
			\univPEPO{2,5}{3,6}{peps1};
			\univPEPO{10,5}{11,6}{peps1};
			\univPEPO{-2,5}{-1,6}{peps1};
			
			\draw[shorten <= 4.5pt, shorten >= 14.2pt] (0.5,4) -- (2,6);
			\draw[shorten <= 4.5pt, shorten >= 14.2pt] (4.5,4) -- (6,6);
			\draw[shorten <= 4.5pt, shorten >= 14.2pt] (8.5,4) -- (10,6);
			\draw[shorten <= 4.5pt, shorten >= 14.2pt] (12.5,4) -- (14,6);
			\draw[shorten <= 4.5pt, shorten >= 14.2pt] (-0.5,4) -- (-2,6);
			\draw[shorten <= 4.5pt, shorten >= 14.2pt] (3.5,4) -- (2,6);
			\draw[shorten <= 4.5pt, shorten >= 14.2pt] (7.5,4) -- (6,6);
			\draw[shorten <= 4.5pt, shorten >= 14.2pt] (11.5,4) -- (10,6);		
		\end{scope}
	}{}	
	\ifthenelse{#2 = 3}{
		\begin{scope}
			\clip (-1,-1) rectangle (13,9);
			\fill[peps0] (-2,4) -- (0,6) -- (2,4) -- (4,6) -- (6,4) -- (8,6) -- (10,4) -- (12,6) -- (14,4) -- (13,9) -- (-1,9) -- (-2,4);
			\fill[peps1] (-2,4) -- (0,6) -- (2,4) -- (4,6) -- (6,4) -- (8,6) -- (10,4) -- (12,6) -- (14,4) -- (13,9) -- (-1,9) -- (-2,4);
			\fill[peps0] (-2,4) -- (0,6) -- (2,4) -- (4,6) -- (6,4) -- (8,6) -- (10,4) -- (12,6) -- (14,4) arc(0:-180:2) (10,4) arc(0:-180:2) (6,4) arc(0:-180:2) (2,4) arc(0:-180:2) (-2,4);
			\fill[peps] (-2,4) -- (0,6) -- (2,4) -- (4,6) -- (6,4) -- (8,6) -- (10,4) -- (12,6) -- (14,4) arc(0:-180:2) (10,4) arc(0:-180:2) (6,4) arc(0:-180:2) (2,4) arc(0:-180:2) (-2,4);
			\draw[] (14,4) arc(0:-180:2) (10,4) arc(0:-180:2) (6,4) arc(0:-180:2) (2,4) arc(0:-180:2) (-2,4);
			
			\draw[] (-2,1) -- (0,-1) -- (2,1) -- (4,-1) -- (6,1) -- (8,-1) -- (10,1) -- (12,-1) -- (14,1);
			\draw[shorten >= 17pt] (-2,1) -- (0,3);
			\draw[shorten >= 17pt] (2,1) -- (0,3);
			\draw[shorten >= 17pt] (2,1) -- (4,3);
			\draw[shorten >= 17pt] (6,1) -- (4,3);
			\draw[shorten >= 17pt] (6,1) -- (8,3);
			\draw[shorten >= 17pt] (10,1) -- (8,3);
			\draw[shorten >= 17pt] (10,1) -- (12,3);
			\draw[shorten >= 17pt] (14,1) -- (12,3);

			\centerarc[fill=white](0,-1)(0:360:0.7)
			\centerarc[fill=white](4,-1)(0:360:0.7)
			\centerarc[fill=white](8,-1)(0:360:0.7)
			\centerarc[fill=white](12,-1)(0:360:0.7)
			\centerarc[fill=white](2,1)(0:360:0.7)
			\centerarc[fill=white](6,1)(0:360:0.7)
			\centerarc[fill=white](10,1)(0:360:0.7)
			\centerarc[peps2](0,-1)(0:360:0.7)
			\centerarc[peps2](4,-1)(0:360:0.7)
			\centerarc[peps2](8,-1)(0:360:0.7)
			\centerarc[peps2](12,-1)(0:360:0.7)
			\centerarc[peps2](2,1)(0:360:0.7)
			\centerarc[peps2](6,1)(0:360:0.7)
			\centerarc[peps2](10,1)(0:360:0.7)
			\centerarc[](0,-1)(0:360:0.7)
			\centerarc[](4,-1)(0:360:0.7)
			\centerarc[](8,-1)(0:360:0.7)
			\centerarc[](12,-1)(0:360:0.7)
			\centerarc[](2,1)(0:360:0.7)
			\centerarc[](6,1)(0:360:0.7)
			\centerarc[](10,1)(0:360:0.7)
			
			\draw[] (-2,4) -- (4,10);
			\draw[] (2,4) -- (8,10);
			\draw[] (6,4) -- (12,10);
			\draw[] (10,4) -- (16,10);
			\draw[] (2,4) -- (-4,10);
			\draw[] (6,4) -- (0,10);
			\draw[] (10,4) -- (4,10);
			\draw[] (14,4) -- (8,10);
		\end{scope}
	}{}							
	\end{tikzpicture}
}

\newcommand{\toricProjLocC}[2]{
	\begin{tikzpicture}[scale=#1,baseline=1.1em]
	\ifthenelse{#2=1}{
		\univPEPO{-1,0}{0,-1}{peps2}
		\univPEPO{0,-1}{1,0}{peps2}
		\univPEPO{0,2}{1,0}{peps3}
		\univPEPO{-1,0}{0,2}{peps3}
		
		\draw[shorten <= 4.5pt, shorten >= 21pt] (0.5,1) -- (2,3) node[anchor=center, pos=0.53] {${\sss a_2}$};
		\draw[shorten <= 4.5pt, shorten >= 21pt] (-0.5,1) -- (-2,3) node[anchor=center, pos=0.53] {${\sss a_1}$};		

		\draw[] (0.68,-0.68) -- (1.2,-1.2) node[anchor=center, pos=1.5] {${\sss \alpha_2}$};
		\draw[] (-0.68,-0.68) -- (-1.2,-1.2) node[anchor=center, pos=1.5] {${\sss \alpha_1}$};		
	}{}	
	\ifthenelse{#2=2}{
		\draw[dot] (-1,1) -- (-1,-1) -- (1,-1) -- (1,1);
		\draw[dot] (0,1) -- (0,3);
		\draw[snake=zigzag,
		segment amplitude=.5mm,
		segment length=2mm,
		line width=1pt] (-1.5,1) -- (1.5,1);
		\node[ket] at (1,0) {${\sss |+\ra}$};
		\node[ket] at (-1,0) {${\sss |+\ra}$};
		\node[ket] at (0,-1) {${\sss |+\ra}$};
		\node[ket] at (0,2) {${\sss |0\ra}$};
	}{}
	\ifthenelse{#2=3}{
		\draw[] (0.68,-0.68) -- (1.2,-1.2) node[anchor=center, pos=1.5] {${\sss \alpha_2}$};
		\draw[] (-0.68,-0.68) -- (-1.2,-1.2) node[anchor=center, pos=1.5] {${\sss \alpha_1}$};	
		\fill[white] (-2,1) -- (0,3) -- (2,1) arc(0:-180:2) (-2,1);	
		\fill[peps] (-2,1) -- (0,3) -- (2,1) arc(0:-180:2) (-2,1);	
		\draw[] (-2,1) -- (0,3) -- (2,1) arc(0:-180:2) (-2,1);
			
		\path[shorten <= 4.5pt, shorten >= 21pt] (0.5,1) -- (2,3) node[anchor=center, pos=0.53, xshift=2pt] {${\sss a_2}$};
		\path[shorten <= 4.5pt, shorten >= 21pt] (-0.5,1) -- (-2,3) node[anchor=center, pos=0.53, xshift=-2pt] {${\sss a_1}$};	
		\node[] at (0,1) {${\sss \mathcal{D}_{\rm 2d}}$};
	}{}		
	\end{tikzpicture}
}

\newcommand{\symTwoDPepsThree}[2]{
	\begin{tikzpicture}[scale=#1,baseline={([yshift=-.5ex]current bounding box.center)}]	
	\ifthenelse{#2=1}{
		\draw[shorten <= 2pt, shorten >= 2pt]  (0,0) -- (-2,-2);
		\draw[shorten <= 2pt, shorten >= 2pt]  (0,0) -- (2,-2);
		\fill[white] (-2,1) -- (0,3) -- (2,1) arc(0:-180:2) (-2,1);	
		\fill[peps] (-2,1) -- (0,3) -- (2,1) arc(0:-180:2) (-2,1);	
		\draw[] (-2,1) -- (0,3) -- (2,1) arc(0:-180:2) (-2,1);
		\node[pauliX] at (-1,2) {${\sss X}$};
		\node[pauliX] at (1,2) {${\sss X}$};
		\node[pauliZ] at (-1.35,-1.35) {${\sss Z}$};
		\node[pauliZ] at (1.35,-1.35) {${\sss Z}$};	
	}{}	
	\ifthenelse{#2=2}{
		\draw[shorten <= 2pt, shorten >= 2pt]  (0,0) -- (-2,-2);
		\draw[shorten <= 2pt, shorten >= 2pt]  (0,0) -- (2,-2);
		\fill[white] (-2,1) -- (0,3) -- (2,1) arc(0:-180:2) (-2,1);	
		\fill[peps] (-2,1) -- (0,3) -- (2,1) arc(0:-180:2) (-2,1);	
		\draw[] (-2,1) -- (0,3) -- (2,1) arc(0:-180:2) (-2,1);	
	}{}	
	\ifthenelse{#2=3}{
		\draw[shorten <= 2pt, shorten >= 2pt]  (0,0) -- (-2,-2);
		\draw[shorten <= 2pt, shorten >= 2pt]  (0,0) -- (2,-2);
		\fill[white] (-2,1) -- (0,3) -- (2,1) arc(0:-180:2) (-2,1);	
		\fill[peps] (-2,1) -- (0,3) -- (2,1) arc(0:-180:2) (-2,1);	
		\draw[] (-2,1) -- (0,3) -- (2,1) arc(0:-180:2) (-2,1);
		\node[pauliX] at (-1,2) {${\sss X}$};
		\node[pauliX] at (1,2) {${\sss X}$};	
	}{}	
	\ifthenelse{#2=4}{
		\draw[shorten <= 2pt, shorten >= 2pt]  (0,0) -- (-2,-2);
		\draw[shorten <= 2pt, shorten >= 2pt]  (0,0) -- (2,-2);
		\fill[white] (-2,1) -- (0,3) -- (2,1) arc(0:-180:2) (-2,1);	
		\fill[peps] (-2,1) -- (0,3) -- (2,1) arc(0:-180:2) (-2,1);	
		\draw[] (-2,1) -- (0,3) -- (2,1) arc(0:-180:2) (-2,1);
		\node[pauliZ] at (-1.35,-1.35) {${\sss Z}$};
		\node[pauliZ] at (1.35,-1.35) {${\sss Z}$};	
	}{}	
	\end{tikzpicture}
}

\newcommand{\PEPOCHalf}[2]{
	\begin{tikzpicture}[scale=#1,baseline=1.5em]
	\ifthenelse{#2=1}{
		\univPEPO{0,2}{1,0}{peps3}
		\draw[shorten <= 4.5pt, shorten >= 21pt] (0.5,1) -- (2,3) node[anchor=center, pos=0.53] {${\sss a}$};
	}{}	
	\ifthenelse{#2=2}{
		\univPEPO{-1,0}{0,2}{peps3}
		\draw[shorten <= 4.5pt, shorten >= 21pt] (-0.5,1) -- (-2,3) node[anchor=center, pos=0.53] {${\sss a}$};		
	}{}	
	\end{tikzpicture}
}

\newcommand{\tetNetwork}[2]{
	\begin{tikzpicture}[scale=#1,baseline={([yshift=-.5ex]current bounding box.center)}]
	\ifthenelse{#2 = 1}{
		\tetra{6.5}{-3};
		\tetra{3}{1};
		\tetra{-2}{1};
		\tetra{1.5}{-3};
		\tetra{5}{0};
		\tetra{0}{0};
		\tetra{-3.5}{-3};
	}{}
	\ifthenelse{#2 = 2}{
		\coordinate (0) at (6.5,-3);
		\coordinate (1) at ({6.5+5},-3);
		\coordinate (2) at ({6.5+1.5},{-3+4});
		\coordinate (3) at ({6.5+1.5},{-3-3});
		\coordinate (4) at ({6.5-2},{-3+1});
		\coordinate (5) at ({6.5+3},{-3+1});
		
		\draw[] (1) -- (5) -- (4);
		\draw[] (2) -- (5) -- (3);
		\node[pauliZ] at (10.5,-2.5) {${\sss Z}$};
		\fill[peps0] (1) -- (3) -- (4) -- (2) -- (1);
		\fill[peps1] (1) -- (3) -- (4) -- (2) -- (1);
		\draw[] (1) -- (2) -- (4) -- (3) -- (1);
		\draw[] (1) -- (0) -- (4);
		\draw[] (2) -- (0) -- (3);
		
		\tetra{3}{1};
		\tetra{-2}{1};
		\tetra{1.5}{-3};
		\tetra{-3.5}{-3};
		\draw[snake=snake,
		segment amplitude=.2mm,
		segment length=1mm] (-3.5,-3) -- (1.5,-3) -- (-2,1) -- (3,1) -- (4.5,5);
		\node[circle, fill = black, inner sep=1.2pt] at (-3.5,-3) {};
		\node[circle, fill = black, inner sep=1.2pt] at (4.5,5) {};
		\node[pauliX] at (-1,-3) {${\sss X}$};
		\node[pauliX] at (3.75,3) {${\sss X}$};
		\node[pauliX] at (-0.25,-1) {${\sss X}$};
		\node[pauliX] at (0.5,1) {${\sss X}$};
		\node[pauliZ] at (4,-3) {${\sss Z}$};
		\node[pauliZ] at (9.75,-1) {${\sss Z}$};
		\node[pauliZ] at (9.75,-4.5) {${\sss Z}$};
		\node[pauliZ] at (4.75,-1) {${\sss Z}$};
		\node[pauliZ] at (7.25,-1) {${\sss Z}$};
		\node[pauliZ] at (4.75,-4.5) {${\sss Z}$};
		\node[pauliZ] at (7.25,-4.5) {${\sss Z}$};
		\node[left, yshift=-0.1em] at (-3.5,-3) {${\sss e}$};
		\node[right] at (4.5,5) {${\sss e}$};
		\node[] at (6,-4.5) {${\sss m}$};
	}{}
	\end{tikzpicture}
}

\newcommand{\cubNetwork}[2]{
	\begin{tikzpicture}[scale=#1,baseline={([yshift=-.5ex]current bounding box.center)}]
	\ifthenelse{#2 = 1}{
		\node at (-3.75,3) (a) {};
		\node at (-2.75,2.5) (d) {};
		\node at (4.75,2.5) (c) {};
		\node at (3.75,3) (b) {};
	
		\cub{1.5}{0.5};
		\cub{5}{-3.5};
		\cub{0}{-3.5};
		\cub{-3.5}{0.5};
		\cub{6.5}{0.5};
		
		\node at (-5.75,4) (a) {};
		\node at (-2.75,2.5) (d) {};
		\node at (4.75,2.5) (c) {};
		\node at (1.75,4) (b) {};
		\node at ({-3.5-0.25},{0.5-1.2}) (e) {};
		\node at ({1.5+2.35},{0.5-1.2}) (f) {};
		\fill[peps0] \convexpath{a,b,c,d}{11pt};
		\fill[black, opacity=0.15] \convexpath{a,b,c,d}{11pt};

		\node[pauliX] at (2.25,-1) {${\sss X}$};
		\node[below] at (2.25,-1.25) {${\sss e}$};
		\node[pauliX] at (7.25,-1) {${\sss X}$};
		\node[below] at (7.25,-1.25) {${\sss e}$};
		\node[] at (1,1.6) {${\sss m}$};
		\node[pauliZ] at (-5.75,4) {${\sss Z}$};
		\node[pauliZ] at (-0.75,4) {${\sss Z}$};
		\node[pauliZ] at (-3.25,4) {${\sss Z}$};
		\node[pauliZ] at (1.75,4) {${\sss Z}$};
		\node[pauliZ] at (-4.75,3.5) {${\sss Z}$};
		\node[pauliZ] at (0.25,3.5) {${\sss Z}$};
		\node[pauliZ] at (-2.25,3.5) {${\sss Z}$};
		\node[pauliZ] at (2.75,3.5) {${\sss Z}$};
		\node[pauliZ] at (-3.75,3) {${\sss Z}$};
		\node[pauliZ] at (1.25,3) {${\sss Z}$};
		\node[pauliZ] at (-1.25,3) {${\sss Z}$};
		\node[pauliZ] at (3.75,3) {${\sss Z}$};
		\node[pauliZ] at (-2.75,2.5) {${\sss Z}$};
		\node[pauliZ] at (2.25,2.5) {${\sss Z}$};
		\node[pauliZ] at (-0.25,2.5) {${\sss Z}$};
		\node[pauliZ] at (4.75,2.5) {${\sss Z}$};
	}{}
	\end{tikzpicture}
}

\newcommand{\symThreeDPepsOne}[2]{
	\begin{tikzpicture}[scale=#1,baseline={([yshift=-.5ex]current bounding box.center)}]
	\ifthenelse{#2 = 1}{
		\tetra{0}{0};
		\node[pauliX] at (2.5,0) {${\sss X}$}; 
		\node[pauliX] at (0.75,2) {${\sss X}$};
		\node[pauliX] at (3.25,2) {${\sss X}$};
	}{}	
	\ifthenelse{#2 = 2}{
		\tetra{0}{0};
		\node[pauliX] at (-1,0.5) {${\sss X}$}; 
		\node[pauliX] at (0.75,2) {${\sss X}$};
		\node[pauliX] at (-0.25,2.5) {${\sss X}$};
	}{}
	\ifthenelse{#2 = 3}{
		\coordinate (0) at (0,0);
		\coordinate (1) at (5,0);
		\coordinate (2) at (1.5,4);
		\coordinate (3) at (1.5,-3);
		\coordinate (4) at (-2,1);
		\coordinate (5) at (3,1);
		
		\draw[] (1) -- (5) -- (4);
		\draw[] (2) -- (5) -- (3);
		\node[pauliX] at (0.5,1) {${\sss X}$}; 
		\node[pauliX] at (4,0.5) {${\sss X}$};
		\fill[peps0] (1) -- (3) -- (4) -- (2) -- (1);
		\fill[peps1] (1) -- (3) -- (4) -- (2) -- (1);
		\draw[] (1) -- (2) -- (4) -- (3) -- (1);
		\draw[] (1) -- (0) -- (4);
		\draw[] (2) -- (0) -- (3);
		\node[pauliX] at (2.5,0) {${\sss X}$};
		\node[pauliX] at (-1,0.5) {${\sss X}$};
	}{}\
	\ifthenelse{#2 = 4}{
		\tetra{0}{0};
		\node[pauliX] at (-0.25,2.5) {${\sss X}$};
		\node[pauliX] at (3.25,2) {${\sss X}$};
		\node[pauliX] at (3.25,-1.5) {${\sss X}$};
		\node[pauliX] at (-0.25,-1) {${\sss X}$};
	}{}
	\ifthenelse{#2 = 5}{
		\tetra{0}{0};
	}{}
	\ifthenelse{#2 = 6}{
		\tetra{0}{0};
		\node[pauliX] at (-0.25,2.5) {${\sss X}$};
		\node[pauliX] at (-0.25,-1) {${\sss X}$};
	}{}
	\ifthenelse{#2 = 7}{
		\tetra{0}{0};
		\node[pauliX] at (3.25,2) {${\sss X}$};
		\node[pauliX] at (3.25,-1.5) {${\sss X}$};
	}{}
	\ifthenelse{#2 = 8}{
		\tetra{0}{0};
		\node[pauliX] at (0.75,2) {${\sss X}$};
		\node[pauliX] at (2.5,0) {${\sss X}$};
		\node[pauliX] at (3.25,-1.5) {${\sss X}$};
	}{}
	\end{tikzpicture}
}

\newcommand{\symThreeDPepsTwo}[2]{
	\begin{tikzpicture}[scale=#1,baseline={([yshift=-.5ex]current bounding box.center)}]
	\ifthenelse{#2 = 1}{
		\draw[shorten <= 8pt, shorten >= 8pt] (-2,-2.5) -- (5,3.5) node[pauliZ, anchor=center, pos=0.25] {${\sss Z}$} node[pauliZ, anchor=center, pos=0.75] {${\sss Z}$};
		\draw[shorten <= 8pt, shorten >= 8pt] (-2,4.5) -- (5,-3.5) node[pauliZ, anchor=center, pos=0.25] {${\sss Z}$} node[pauliZ, anchor=center, pos=0.75] {${\sss Z}$};
		\draw[shorten <= 8pt, shorten >= 8pt] (3,4.5) -- (0,-3.5) node[pauliZ, anchor=center, pos=0.25] {${\sss Z}$} node[pauliZ, anchor=center, pos=0.75] {${\sss Z}$};
		\draw[shorten <= 8pt, shorten >= 8pt] (0,3.5) -- (3,-2.5) node[pauliZ, anchor=center, pos=0.25] {${\sss Z}$} node[pauliZ, anchor=center, pos=0.75] {${\sss Z}$};
		\centerarc[fill= white](1.5,0.5)(0:360:0.8)
		\centerarc[peps2](1.5,0.5)(0:360:0.8)
		\centerarc[](1.5,0.5)(0:360:0.8)
	}{}	
	\ifthenelse{#2 = 2}{
		\draw[shorten <= 8pt, shorten >= 8pt] (-2,-2.5) -- (5,3.5);
		\draw[shorten <= 8pt, shorten >= 8pt] (-2,4.5) -- (5,-3.5);
		\draw[shorten <= 8pt, shorten >= 8pt] (3,4.5) -- (0,-3.5);
		\draw[shorten <= 8pt, shorten >= 8pt] (0,3.5) -- (3,-2.5);
		\centerarc[fill= white](1.5,0.5)(0:360:0.8)
		\centerarc[peps2](1.5,0.5)(0:360:0.8)
		\centerarc[](1.5,0.5)(0:360:0.8)
	}{}	
	\ifthenelse{#2 = 3}{
		\draw[shorten <= 8pt, shorten >= 8pt] (-2,-2.5) -- (5,3.5) node[pauliZ, anchor=center, pos=0.75] {${\sss Z}$};
		\draw[shorten <= 8pt, shorten >= 8pt] (-2,4.5) -- (5,-3.5) node[pauliZ, anchor=center, pos=0.25] {${\sss Z}$};
		\draw[shorten <= 8pt, shorten >= 8pt] (3,4.5) -- (0,-3.5) node[pauliZ, anchor=center, pos=0.25] {${\sss Z}$};
		\draw[shorten <= 8pt, shorten >= 8pt] (3,-2.5) -- (0,3.5) node[pauliZ, anchor=center, pos=0.75] {${\sss Z}$};
		\centerarc[fill= white](1.5,0.5)(0:360:0.8)
		\centerarc[peps2](1.5,0.5)(0:360:0.8)
		\centerarc[](1.5,0.5)(0:360:0.8)
	}{}	
	\ifthenelse{#2 = 4}{
		\draw[shorten <= 8pt, shorten >= 8pt] (-2,-2.5) -- (5,3.5) node[pauliZ, anchor=center, pos=0.25] {${\sss Z}$};
		\draw[shorten <= 8pt, shorten >= 8pt] (-2,4.5) -- (5,-3.5) node[pauliZ, anchor=center, pos=0.75] {${\sss Z}$};
		\draw[shorten <= 8pt, shorten >= 8pt] (3,4.5) -- (0,-3.5) node[pauliZ, anchor=center, pos=0.75] {${\sss Z}$};
		\draw[shorten <= 8pt, shorten >= 8pt] (3,-2.5) -- (0,3.5) node[pauliZ, anchor=center, pos=0.25] {${\sss Z}$};
		\centerarc[fill= white](1.5,0.5)(0:360:0.8)
		\centerarc[peps2](1.5,0.5)(0:360:0.8)
		\centerarc[](1.5,0.5)(0:360:0.8)
	}{}	
	\end{tikzpicture}
}

\newcommand{\ThreeDFlow}[2]{
	\begin{tikzpicture}[scale=#1,baseline={([yshift=-.5ex]current bounding box.center)}]
	\ifthenelse{#2 = 0}{
		\begin{scope}
			\clip (-4.5,-4.5) rectangle (11,6);
			\begin{scope}[opacity=0.2]
				\tetra{6.5}{-3};
				\tetra{1.5}{-3};
				\tetra{6.5}{4};
				\tetra{1.5}{4};	
			\end{scope}
			\node[above] at (-0.25,2.5) {${\sss 1}$};
			\node[left, xshift=1pt] at (2.25,2.5) {${\sss 2}$};
			\node[above] at (4.75,2.5) {${\sss 3}$};
			\node[left, xshift=1pt] at (7.25,2.5) {${\sss 4}$};
			\node[above, yshift=-2pt, xshift=10pt] at (0.5,1) {${\sss 10}$};
			\node[above, yshift=-2pt, xshift=10pt] at (5.5,1) {${\sss 11}$};
			\node[below, yshift=4.5pt, xshift=-1pt] at (8.5,0.5) {${\sss 12}$};
			\node[right, xshift=-2pt] at (2.25,-1) {${\sss 16}$};
			\node[right, xshift=-2pt] at (7.25,-1) {${\sss 18}$};
			\draw[color=LimeGreen, line width=2.5pt] (-2,1) -- (8,1);
			\draw[color=BurntOrange, line width=2.5pt] (8,1) -- (6.5,4) -- (3,1) -- (1.5,4) -- (5,0) -- (6.5,4) -- (10,0);
			\draw[color=BurntOrange, line width=2.5pt] (8,1) -- (6.5,-3) -- (3,1) -- (1.5,-3) -- (5,0) -- (6.5,-3) -- (10,0);
			\draw[color = LimeGreen, line width=2.5pt] (0,0) -- (10,0);
			\tetra{5}{0};
			\node[above, xshift=-0.25em, yshift=-0.12em] at (4,0.5) {${\sss 0}$};
			\tetra{0}{0};
			\node[left,xshift=1pt] at (0.75,2) {${\sss 5}$};
			\node[above] at (3.25,2) {${\sss 6}$};
			\node[left,xshift=1pt] at (5.75,2) {${\sss 7}$};
			\node[above] at (8.25,2) {${\sss 8}$};
			\node[below, xshift=9pt] at (-1.5,0.5) {${\sss 9}$};
			\node[above, yshift=-2pt, xshift=-10pt] at (2.5,0) {${\sss 13}$};
			\node[above, yshift=-2pt, xshift=-10pt] at (7.5,0) {${\sss 14}$};
			\node[below, xshift=-2pt] at (-0.25,-1) {${\sss 15}$};
			\node[below, xshift=-2pt] at (4.75,-1) {${\sss 17}$};
			\node[right, xshift=-3pt] at (0.75,-1) {${\sss 19}$};
			\node[right, xshift=-3pt] at (5.75,-1) {${\sss 21}$};
			\node[below, xshift=0.5pt] at (3.25,-1.5) {${\sss 20}$};
			\node[below, xshift=0.5pt] at (8.25,-1.5) {${\sss 22}$};
		\end{scope}	
	}{}	
	\ifthenelse{#2 = 1}{
		\tetra{5}{0};
		\tetra{0}{0};
	}{}	
	\ifthenelse{#2 = 2}{
		\tetra{0}{0}
	}{}
	\ifthenelse{#2 = 3}{
		\tetraCor{0}{0}
	}{}
	\ifthenelse{#2 = 4}{
		\draw[line width=1pt] (0,0) -- (3.6,0);
		\draw[line width=1pt] (2.5,2.5) -- (0,0) -- (2.5,-2.5);
		\node[circle, fill = black, inner sep=1.2pt] at (0,0) {};
	}{}
	\ifthenelse{#2 = 5}{
		\draw[line width=1pt] (-1.4,2) -- (0,-1) -- (1.4,2);
		\node[circle, fill = black, inner sep=1.2pt] at (0,-1) {};
	}{}
	\ifthenelse{#2 = 6}{
		\draw[line width=1pt] (-1.4,2) -- (0,-1) -- (1.4,2);
		\node[circle, fill = black, inner sep=1.2pt] at (0,-1) {};
	}{}
	\ifthenelse{#2 = 7}{
		\draw[line width=1pt] (0,0) -- (3.6,0) node[pos=0.7, above] {${\sss 2}$};
		\draw[line width=1pt] (0,0) -- (2.5,-2.5) node[pos=0.6, left, yshift=-1pt] {${\sss 3}$};
		\draw[line width=1pt] (0,0) -- (2.5,2.5) node[pos=0.6, left, yshift=1pt] {${\sss 1}$};
		\node[circle, fill = black, inner sep=1.2pt] at (0,0) {};
	}{}
	\end{tikzpicture}
}

\newcommand{\exampleSymFlow}[1]{
	\begin{tikzpicture}[scale=#1,baseline={([yshift=-.5ex]current bounding box.center)}]	
		\tetra{5}{0};
		\tetra{0}{0};
		\node[pauliX] at (2.5,0) {${\sss X}$}; 
		\node[pauliX] at (0.75,2) {${\sss X}$};
		\node[pauliX] at (3.25,2) {${\sss X}$};
		\node[pauliX] at (7.5,0) {${\sss X}$}; 
		\node[pauliX] at (5.75,2) {${\sss X}$};
		\node[pauliX] at (8.25,2) {${\sss X}$};
	\end{tikzpicture}
}

\newcommand{\ThreeToricProjLocB}[2]{
	\begin{tikzpicture}[scale=#1,baseline={([yshift=-.5ex]current bounding box.center)}]	
	\ifthenelse{#2=1}{
		\draw[shorten <= 14pt] (0,0) -- (1.4,0.9) node[anchor=center, pos=1.15] {${\sss a_8}$};
		\draw[shorten <= 22pt] (0,0) -- (1.8,-1.8) node[anchor=center, pos=1.15] {${\sss a_{12}}$};
		\draw[shorten <= 14pt] (0,0) -- (-0.9,-1.4) node[anchor=center, pos=1.15] {${\sss a_{10}}$};
		\draw[shorten <= 23pt] (0,0) -- (-2.5,0.7) node[anchor=center, pos=1.15] {${\sss a_6}$};
		\draw[shorten <= 23.5pt] (0,0) -- (-0.7,2.5) node[anchor=center, pos=1.1] {${\sss a_2}$};
		
		\univPEPO{-0.7,0.7}{0,1.8}{peps1}
		\univPEPO{-1.8,0}{-0.7,0.7}{peps1}
		\univPEPO{-0.7,0.7}{0,-1.8}{peps1}
		\univPEPO{-0.7,0.7}{1.8,0}{peps1}

		\univPEPO{-1.8,0}{0,-1.8}{peps1}
		\univPEPO{0,-1.8}{1.8,0}{peps1}
		\univPEPO{0,-1.8}{0.7,-0.7}{peps1}
		\univPEPO{-1.8,0}{0.7,-0.7}{peps1}
		\univPEPO{0,1.8}{1.8,0}{peps1}
		\univPEPO{-1.8,0}{0,1.8}{peps1}
		\univPEPO{0,1.8}{0.7,-0.7}{peps1}
		\univPEPO{0.7,-0.7}{1.8,0}{peps1}
		
		\draw[shorten <= 22pt] (0,0) -- (1.8,1.8) node[anchor=center, pos=1.1] {${\sss a_4}$};
		\draw[shorten <= 22pt] (0,0) -- (-1.8,-1.8) node[anchor=center, pos=1.1] {${\sss a_9}$};
		\draw[shorten <= 22pt] (0,0) -- (-1.8,1.8) node[anchor=center, pos=1.15] {${\sss a_1}$};
		\draw[shorten <= 14pt] (0,0) -- (-1.4,-0.9) node[anchor=center, pos=1.2] {${\sss a_5}$};
		\draw[shorten <= 23pt] (0,0) -- (2.5,-0.7) node[anchor=center, pos=1.15] {${\sss a_7}$};
		\draw[shorten <= 23.5pt] (0,0) -- (0.7,-2.5) node[anchor=center, pos=1.1] {${\sss a_{11}}$};
		\draw[shorten <= 14pt] (0,0) -- (0.9,1.4) node[anchor=center, pos=1.15] {${\sss a_3}$}; 
	}{}	
	\ifthenelse{#2=2}{
		\draw[dot] (-3.6,0) -- (3.6,0);
		\draw[dot] (0,3.6) -- (0,-3.6);
		\draw[dot] (1.4,-1.4) -- (-1.4,1.4);
		\node[ket] at (1.8,0) {${\sss |0\ra}$};
		\node[ket] at (-1.8,0) {${\sss |0\ra}$};
		\node[ket] at (0,1.8) {${\sss |0\ra}$};
		\node[ket] at (0,-1.8) {${\sss |0\ra}$};
		\node[ket] at (0.7,-0.7) {${\sss |0\ra}$};
		\node[ket] at (-0.7,0.7) {${\sss |0\ra}$};
	}{}
	\ifthenelse{#2=3}{
		\node at (-1.8,0) (a) {};
		\node at (0,1.8) (b) {};
		\node at (1.8,0) (c) {};
		\node at (0,-1.8) (d) {};
		\node at (-0.7,0.7) (e) {};
		\node at (0.7,-0.7) (f) {};
		
		\draw[shorten <= 14pt] (0,0) -- (1.4,0.9) node[anchor=center, pos=1.15] {${\sss a_8}$};
		\draw[shorten <= 22pt] (0,0) -- (1.8,-1.8) node[anchor=center, pos=1.15] {${\sss a_{12}}$};
		\draw[shorten <= 14pt] (0,0) -- (-0.9,-1.4) node[anchor=center, pos=1.15] {${\sss a_{10}}$};
		\draw[shorten <= 23pt] (0,0) -- (-2.5,0.7) node[anchor=center, pos=1.15] {${\sss a_6}$};
		\draw[shorten <= 23.5pt] (0,0) -- (-0.7,2.5) node[anchor=center, pos=1.1] {${\sss a_2}$};

		\node[circle, fill = black, inner sep=1.2pt] at (-0.7,0.7) {};
		\draw[] \convexpath{a,e,c,f}{7pt};
		\draw[] \convexpath{e,b,f,d}{7pt};
		\fill[peps0] \convexpath{a,b,c,d}{7pt};
		\fill[peps1] \convexpath{a,b,c,d}{7pt};
		\draw[] \convexpath{a,b,c,d}{7pt};
		\draw[] \convexpath{a,b,c,f}{7pt};
		\draw[] \convexpath{a,b,f,d}{7pt};
		
		\draw[shorten <= 22pt] (0,0) -- (1.8,1.8) node[anchor=center, pos=1.1] {${\sss a_4}$};
		\draw[shorten <= 22pt] (0,0) -- (-1.8,-1.8) node[anchor=center, pos=1.1] {${\sss a_9}$};
		\draw[shorten <= 22pt] (0,0) -- (-1.8,1.8) node[anchor=center, pos=1.15] {${\sss a_1}$};
		\draw[shorten <= 14pt] (0,0) -- (-1.4,-0.9) node[anchor=center, pos=1.2] {${\sss a_5}$};
		\draw[shorten <= 23pt] (0,0) -- (2.5,-0.7) node[anchor=center, pos=1.15] {${\sss a_7}$};
		\draw[shorten <= 23.5pt] (0,0) -- (0.7,-2.5) node[anchor=center, pos=1.1] {${\sss a_{11}}$};
		\draw[shorten <= 14pt] (0,0) -- (0.9,1.4) node[anchor=center, pos=1.15] {${\sss a_3}$}; 
		
		\node[circle, fill = black, inner sep=1.2pt] at (-1.8,0) {};
		\node[circle, fill = black, inner sep=1.2pt] at (1.8,0) {};	
		\node[circle, fill = black, inner sep=1.2pt] at (0,1.8) {};	
		\node[circle, fill = black, inner sep=1.2pt] at (0,-1.8) {};		
		\node[circle, fill = black, inner sep=1.2pt] at (0.7,-0.7) {};		
		\node[] at (0,0) {${\sss T^X_{\rm 3d}}$};	
	}{}
	\ifthenelse{#2=4}{	
		\draw[] (-3.6,0) --node[pos=0.6, below] {${\sss a_6}$} (-1.4,1.4) --node[pos=0.6, above, yshift=-1pt] {${\sss a_8}$} (3.6,0);
		\draw[] (0,3.6) --node[pos=0.6, right] {${\sss a_2}$} (-1.4,1.4) --node[pos=0.32, right,xshift=-2pt] {${\sss a_{10}}$} (0,-3.6);
		\fill[peps0] (-3.6,0) -- (0,3.6) -- (3.6,0) -- (0,-3.6) --cycle;
		\fill[peps1] (-3.6,0) -- (0,3.6) -- (3.6,0) -- (0,-3.6) --cycle;
		\draw[] (-3.6,0) --node[pos=0.5, above, xshift=-1pt] {${\sss a_1}$} (0,3.6) --node[pos=0.5, above] {${\sss a_4}$} (3.6,0) --node[pos=0.5, below,xshift=3pt] {${\sss a_{12}}$} (0,-3.6) --node[pos=0.5, below] {${\sss a_9}$} (-3.6,0);
		\draw[] (-3.6,0) --node[pos=0.4, below] {${\sss a_5}$} (1.4,-1.4) --node[pos=0.4, above] {${\sss a_7}$} (3.6,0);
		\draw[] (0,3.6) --node[pos=0.65, left, xshift=2pt] {${\sss a_3}$} (1.4,-1.4) --node[pos=0.4, left, xshift=2pt] {${\sss a_{11}}$} (0,-3.6);
	}{}
	\end{tikzpicture}
}

\newcommand{\PEPOAThree}[1]{
	\begin{tikzpicture}[scale=#1,baseline={([yshift=-.5ex]current bounding box.center)}]
	\draw[dot] (-3.6,0) -- (3.6,0);
	\draw[dot] (0,3.6) -- (0,-3.6);
	\draw[dot] (-1.4,1.4) -- (1.4,-1.4);
	
	\univTriPEPO{-1.8,0}{-0.7,0.7}{0,-1.8}{peps2}
	\draw[] (-0.7,-0.6) -- (0.7,0.6);
	\univTriPEPO{1.8,0}{0.7,-0.7}{0,1.8}{peps2}
	
	\ddott{-1.8,0};
	\ddott{1.8,0};
	\ddott{0,1.8};
	\ddott{0,-1.8};
	\ddott{0.7,-0.7};
	\ddott{-0.7,0.7};
	\end{tikzpicture}
}

\newcommand{\PEPOAThreeHalf}[1]{
	\begin{tikzpicture}[scale=#1,baseline=0em]	
		\node at ({-0.5*sqrt(2*(1.8)^2)},0) (a) {};
		\node at ({0.5*sqrt(2*(1.8)^2)},0) (b) {};
		\node at (0,-1) (c) {};
		\fill[peps0] \convexpath{a,b,c}{7pt};
		\fill[peps2] \convexpath{a,b,c}{7pt};
		\draw[] \convexpath{a,b,c}{7pt};
		\draw[] (0,-0.3) -- (0,0.7);	
		\ddott{a};
		\ddott{b};
		\ddott{c};
		\node[above] at (0,0.7) {${\sss \alpha}$};
	\end{tikzpicture}
}

\newcommand{\ThreeToricProjLocA}[2]{
	\begin{tikzpicture}[scale=#1,baseline={([yshift=-.5ex]current bounding box.center)}]	
	\ifthenelse{#2=1}{
		\draw[shorten <= 27pt] (1.1,2.5) -- (-1.4,1.4) node[anchor=center, pos=1.15] {${\sss \alpha_5}$};
		\draw[shorten <= 15pt] (1.1,2.5) -- (2.3,1.6) node[anchor=center, pos=1.2] {${\sss \alpha_6}$};
		\draw[shorten <= 32pt] (1.1,2.5) -- (-1.4,5) node[anchor=center, pos=1.1] {${\sss \alpha_1}$};
		\draw[shorten <= 28pt] (1.1,2.5) -- (2.2,5) node[anchor=center, pos=1.1] {${\sss \alpha_2}$};	
		\univTriPEPO{0.4,1.4}{2.2,3.2}{2.9,0.7}{peps2};
		\univTriPEPO{-0.7,0.7}{-1.4,3.2}{0.4,1.4}{peps2};
		\univTriPEPO{3.6,1.8}{2.9,0.7}{1.8,0}{peps2};
		\univTriPEPO{-0.7,0.7}{0,1.8}{1.8,0}{peps2};
		
		\univTriPEPO{0.4,5}{2.9,4.3}{2.2,3.2}{peps2};
		\univTriPEPO{1.8,3.6}{2.9,4.3}{3.6,1.8}{peps2};
		\univTriPEPO{-1.4,3.2}{-0.7,4.3}{0.4,5}{peps2};
		\univTriPEPO{-0.7,4.3}{1.8,3.6}{0,1.8}{peps2};
		\draw[shorten <= 27pt] (1.1,2.5) -- (3.6,3.6) node[anchor=center, pos=1.15] {${\sss \alpha_4}$};
		\draw[shorten <= 15pt] (1.1,2.5) -- (-0.1,3.4) node[anchor=center, pos=1.2] {${\sss \alpha_3}$};
		\draw[shorten <= 32pt] (1.1,2.5) -- (3.6,0) node[anchor=center, pos=1.1] {${\sss \alpha_8}$};
		\draw[shorten <= 28pt] (1.1,2.5) -- (0,0) node[anchor=center, pos=1.1] {${\sss \alpha_7}$};
	}{}	
	\ifthenelse{#2=2}{
		\draw[dot] (-1.4,1.4) -- (2.2,1.4) -- (2.2,5) -- (-1.4,5) --cycle;
		\draw[dot] (0,0) -- (3.6,0) -- (3.6,3.6) -- (0,3.6) --cycle;
		\draw[dot] (-1.4,1.4) -- (0,0);
		\draw[dot] (2.2,1.4) -- (3.6,0);
		\draw[dot] (2.2,5) -- (3.6,3.6);
		\draw[dot] (-1.4,5) -- (0,3.6);
		\node[ket] at (1.8,0) {${\sss |+\ra}$};
		\node[ket] at (1.5,3.6) {${\sss |+\ra}$};
		\node[ket] at (3.6,1.8) {${\sss |+\ra}$};
		\node[ket] at (0,2.1) {${\sss |+\ra}$};
		\node[ket] at (-0.7,0.7) {${\sss |+\ra}$};
		\node[ket] at (-0.7,4.3) {${\sss |+\ra}$};
		\node[ket] at (2.9,0.7) {${\sss |+\ra}$};
		\node[ket] at (0.7,1.4) {${\sss |+\ra}$};
		\node[ket] at (0.4,5) {${\sss |+\ra}$};
		\node[ket] at (-1.4,3.2) {${\sss |+\ra}$};
		\node[ket] at (2.2,2.9) {${\sss |+\ra}$};
		\node[ket] at (2.9,4.3) {${\sss |+\ra}$};
	}{}
	\ifthenelse{#2=3}{
		\node at (-0.7,0.7) (a) {};
		\node at (-1.4,3.2) (b) {};
		\node at (-0.7,4.3) (c) {};
		\node at (0.4,5) (d) {};
		\node at (2.9,4.3) (e) {};
		\node at (3.6,1.8) (f) {};
		\node at (2.9,0.7) (g) {};
		\node at (1.8,0) (h) {};
		\node at (0,1.8) (i) {};
		\node at (1.8,3.6) (j) {};
		\node at (0.4,1.4) (k) {};
		\node at (2.2,3.2) (l) {};
		
		\foreach \x in {k,l}
			\node[circle, fill = black, inner sep=1.2pt] at (\x) {};
		\draw[shorten <= 27pt] (1.1,2.5) -- (-1.4,1.4) node[anchor=center, pos=1.15] {${\sss \alpha_5}$};
		\draw[shorten <= 15pt] (1.1,2.5) -- (2.3,1.6) node[anchor=center, pos=1.2] {${\sss \alpha_6}$};
		\draw[shorten <= 32pt] (1.1,2.5) -- (-1.4,5) node[anchor=center, pos=1.1] {${\sss \alpha_1}$};
		\draw[shorten <= 28pt] (1.1,2.5) -- (2.2,5) node[anchor=center, pos=1.1] {${\sss \alpha_2}$};
		\draw[] \convexpath{b,k,a}{7pt};
		\draw[] \convexpath{b,c,d}{7pt};
		\draw[] \convexpath{d,e,l}{7pt};	
		\draw[] \convexpath{k,l,g}{7pt};
		\fill[peps0] \convexpath{a,b,c,d,e,f,g,h}{7pt};
		\fill[peps2] \convexpath{a,b,c,d,e,f,g,h}{7pt};
		\draw[] \convexpath{c,j,i}{7pt};
		\draw[] \convexpath{a,i,h}{7pt};
		\draw[] \convexpath{h,f,g}{7pt};
		\draw[] \convexpath{j,e,f}{7pt};
		\draw[] \convexpath{a,b,c,d,e,f,g,h}{7pt};
		\draw[shorten <= 27pt] (1.1,2.5) -- (3.6,3.6) node[anchor=center, pos=1.15] {${\sss \alpha_4}$};
		\draw[shorten <= 15pt] (1.1,2.5) -- (-0.1,3.4) node[anchor=center, pos=1.2] {${\sss \alpha_3}$};
		\draw[shorten <= 32pt] (1.1,2.5) -- (3.6,0) node[anchor=center, pos=1.1] {${\sss \alpha_8}$};
		\draw[shorten <= 28pt] (1.1,2.5) -- (0,0) node[anchor=center, pos=1.1] {${\sss \alpha_7}$};
		
		\foreach \x in {a,b,c,d,e,f,h,i,j,g}
			\node[circle, fill = black, inner sep=1.2pt] at (\x) {};	
		\node[] at (0.3,2.5) {${\sss T^Z_{\rm 3d}}$};
	}{}
	\ifthenelse{#2=4}{
		\draw[] (1.1,2.5) -- (-1.4,1.4) node[anchor=center, pos=1.15] {${\sss \alpha_5}$};
		\draw[] (1.1,2.5) -- (2.3,1.6) node[anchor=center, pos=1.15, xshift=2pt] {${\sss \alpha_6}$};
		\draw[] (1.1,2.5) -- (-1.4,5) node[anchor=center, pos=1.1] {${\sss \alpha_1}$};
		\draw[] (1.1,2.5) -- (2.2,5) node[anchor=center, pos=1.15] {${\sss \alpha_2}$};
		\draw[] (1.1,2.5) -- (3.6,3.6) node[anchor=center, pos=1.15] {${\sss \alpha_4}$};
		\draw[] (1.1,2.5) -- (-0.1,3.4) node[anchor=center, pos=1.15,xshift=-2pt] {${\sss \alpha_3}$};
		\draw[] (1.1,2.5) -- (3.6,0) node[anchor=center, pos=1.1] {${\sss \alpha_8}$};
		\draw[] (1.1,2.5) -- (0,0) node[anchor=center, pos=1.15] {${\sss \alpha_7}$};
		\centerarc[fill= white](1.1,2.5)(0:360:0.75)
		\centerarc[peps2](1.1,2.5)(0:360:0.75)
		\centerarc[](1.1,2.5)(0:360:0.75)
	}{}		
	\end{tikzpicture}
}

\newcommand{\correlationNetwork}[2]{
	\begin{tikzpicture}[scale=#1,baseline={([yshift=-.5ex]current bounding box.center)}]
	\ifthenelse{#2 = 1}{
		\tetra{3}{1};
		\tetra{-2}{1};
		\tetra{6.5}{-3};
		\tetra{1.5}{-3};
		\tetra{6.5}{4};
		\tetra{1.5}{4};
		\tetra{5}{0};
		\tetra{0}{0};
	}{}
	\ifthenelse{#2 = 2}{
		\tetra{-2}{1};
		\tetra{1.5}{-3};
		\tetra{1.5}{4};
		\tetra{0}{0};
	}{}
	\ifthenelse{#2 = 3}{
		\coordinate (0) at (5,0);
		\coordinate (1) at ({5+5},0);
		\coordinate (2) at ({5+1.5},{0+4});
		\coordinate (3) at ({5+1.5},{0-3});
		\coordinate (4) at ({5-2},{0+1});
		\coordinate (5) at ({5+3},{0+1});
		
		\coordinate (10) at (3,1);
		\coordinate (11) at ({3+5},1);
		\coordinate (12) at ({3+1.5},{1+4});
		\coordinate (13) at ({3+1.5},{1-3});
		\coordinate (14) at ({3-2},{1+1});
		\coordinate (15) at ({3+3},{1+1});
		
		\coordinate (20) at (1.5,4);
		\coordinate (21) at ({1.5+5},4);
		\coordinate (22) at ({1.5+1.5},{4+4});
		\coordinate (23) at ({1.5+1.5},{4-3});
		\coordinate (24) at ({1.5-2},{4+1});
		\coordinate (25) at ({1.5+3},{4+1});
		\coordinate (26) at ({6.5+1.5},{4+4});
		\coordinate (27) at ({6.5+1.5},{4-4});
		
		\coordinate (30) at (1.5,-3);
		\coordinate (31) at ({1.5+5},-3);
		\coordinate (32) at ({1.5+1.5},{-3+4});
		\coordinate (33) at ({1.5+1.5},{-3-3});
		\coordinate (34) at ({1.5-2},{-3+1});
		\coordinate (35) at ({1.5+3},{-3+1});
		\coordinate (36) at ({6.5+1.5},{-3+4});
		\coordinate (37) at ({6.5+1.5},{-3-4});
			
		\begin{scope}[line width =1pt]
			\draw[shbMax] (1) -- (2);
			\draw[line cap=round, dash pattern=on 0pt off 2pt, line width=1.1pt, shaMax] (2) -- (5);
			\draw[shbMax] (1) -- (3);
			\draw[shabMax] (2) -- (0) -- (3);
			\draw[line cap=round, dash pattern=on 0pt off 2pt, line width=1.1pt, shabMax] (2) -- (4)-- (3);
			\draw[line cap=round, dash pattern=on 0pt off 2pt, line width=1.1pt, sha] (3) -- (intersection of 3--5 and 0--1);
			\draw[line cap=round, dash pattern=on 0pt off 2pt, line width=1.1pt, shb, shaMax] (intersection of 3--5 and 0--1) -- (5);
			\draw[line width=1pt, shaMax] (0) -- (1);
			\draw[line cap=round, dash pattern=on 0pt off 2pt, line width=1.1pt, sha] (4) -- (intersection of 4--5 and 0--2);
			\draw[line cap=round, dash pattern=on 0pt off 2pt, line width=1.1pt, shb, shaMax] (intersection of 4--5 and 0--2) -- (5);
		
			\draw[shbMax] (20) -- (21);
			\draw[shaMax] (21) -- (26); 
			\draw[shabMax] (21) -- (22) -- (25);
			\draw[sha] (25) -- (intersection of 25--26 and 21--22);
			\draw[shb, shaMax] (intersection of 25--26 and 21--22) -- (26);
			\draw[shaMax] (25) -- (24);
		
			\draw[line cap=round, dash pattern=on 0pt off 2pt, line width=1.1pt, shbMax, sha] (11) -- (intersection of 11--12 and 0--2);
			\draw[line cap=round, dash pattern=on 0pt off 2pt, line width=1.1pt, shab] (intersection of 11--12 and 2--4) -- (intersection of 11--12 and 20--21);
			\draw[line cap=round, dash pattern=on 0pt off 2pt, line width=1.1pt, shb] (intersection of 11--12 and 20--21) -- (12);
			\draw[sha] (12) -- (intersection of 12--15 and 20--21);
			\draw[shab] (intersection of 12--15 and 20--21) -- (intersection of 12--15 and 2--4);
			\draw[shb, shaMax] (intersection of 12--15 and 2--4) -- (15);
			\draw[line cap=round, dash pattern=on 0pt off 2pt, line width=1.1pt, sha] (13) -- (intersection of 13--11 and 3--4);
			\draw[line cap=round, dash pattern=on 0pt off 2pt, line width=1.1pt, shab] (intersection of 13--11 and 3--4) -- (intersection of 13--11 and 0--3);
			\draw[line cap=round, dash pattern=on 0pt off 2pt, line width=1.1pt, shab] (intersection of 13--11 and 0--3) -- (intersection of 13--11 and 0--1);
			\draw[line cap=round, dash pattern=on 0pt off 2pt, line width=1.1pt, shb, shaMax] (intersection of 13--11 and 0--1) -- (11);
			\draw[line cap=round, dash pattern=on 0pt off 2pt, line width=1.1pt, shaMax] (10) -- (13);
			\draw[sha] (14) -- (intersection of 12--14 and 20--21);
			\draw[shb, shaMax] (intersection of 12--14 and 20--21) -- (12);
			\draw[line cap=round, dash pattern=on 0pt off 2pt, line width=1.1pt, sha] (10) -- (intersection of 10--12 and 20--21);
			\draw[line cap=round, dash pattern=on 0pt off 2pt, line width=1.1pt, shb, shaMax] (intersection of 10--12 and 20--21) -- (12);
			\draw[shbMax] (13) -- (14);
			\draw[sha] (13) -- (intersection of 13--15 and 3--4);
			\draw[shab] (intersection of 13--15 and 3--4) -- (intersection of 13--15 and 0--3); 
			\draw[shab] (intersection of 13--15 and 0--1) -- (intersection of 13--15 and 10--11);
			\draw[shb, shaMax] (intersection of 13--15 and 10--11) -- (15);
			\draw[sha] (14) -- (intersection of 14--15 and 10--12);
			\draw[shab] (intersection of 14--15 and 10--12) -- (intersection of 14--15 and 2--4);
			\draw[shb, shaMax] (intersection of 14--15 and 2--4) -- (15);
			
			\draw[shbMax] (34) -- (35);
			\draw[shbMax] (30) -- (31);
			\draw[shaMax] (33) -- (31);
			\draw[sha] (33) -- (intersection of 33--35 and 30--31);
			\draw[shb, shaMax] (intersection of 33--35 and 30--31) -- (35);
			\draw[shaMax] (31) -- (37);
			\draw[sha] (35) -- (intersection of 35--37 and 30--31);
			\draw[shab] (intersection of 35--37 and 30--31) -- (intersection of 35--37 and 31--33);
			\draw[shb, shaMax] (intersection of 35--37 and 31--33) -- (37);
			
			\node[circle, fill = black, inner sep=1.2pt] at (12) {};
			\node[circle, fill = black, inner sep=1.2pt] at (13) {};
			\node[circle, fill = black, inner sep=1.2pt] at (14) {};
			\node[circle, fill = black, inner sep=1.2pt] at (22) {};
			\node[circle, fill = black, inner sep=1.2pt] at (33) {};
			\node[circle, fill = black, inner sep=1.2pt] at (2) {};
			\node[circle, fill = black, inner sep=1.2pt] at (3) {};
			\node[circle, fill = black, inner sep=1.2pt] at (0) {};
			\node[circle, fill = black, inner sep=1.2pt] at (4) {};
		\end{scope}	
	}{}
	\end{tikzpicture}
}

\newcommand{\symCorrFlow}[2]{
	\begin{tikzpicture}[scale=#1,baseline={([yshift=-.5ex]current bounding box.center)}]
	\ifthenelse{#2 = 1}{
		\draw[line width=1pt] (0,0) -- (3.6,0);
		\draw[line width=1pt] (2.5,2.5) -- (0,0) -- (2.5,-2.5);
		\node[circle, fill = black, inner sep=1.2pt] at (0,0) {};
		\node[pauliX] at (1.8,0) {${\sss X}$};
		\node[pauliX] at (1.25,1.25) {${\sss X}$};
	}{}
	\ifthenelse{#2 = 2}{
		\draw[line width=1pt] (0,0) -- (3.6,0);
		\draw[line width=1pt] (2.5,2.5) -- (0,0) -- (2.5,-2.5);
		\node[circle, fill = black, inner sep=1.2pt] at (0,0) {};
		\node[pauliX] at (1.8,0) {${\sss X}$};
		\node[pauliX] at (1.25,-1.25) {${\sss X}$};
	}{}
	\ifthenelse{#2 = 3}{
		\draw[line width=1pt] (-1.4,2) -- (0,-1) -- (1.4,2);
		\node[circle, fill = black, inner sep=1.2pt] at (0,-1) {};
		\node[pauliX] at (-0.7,0.5) {${\sss X}$};
		\node[pauliX] at (0.7,0.5) {${\sss X}$};
	}{}
	\end{tikzpicture}
}

\newcommand{\onSiteTransfer}[2]{
	\begin{tikzpicture}[scale=#1,baseline={([yshift=-.5ex]current bounding box.center)}]
	\coordinate (0) at (0,0);
	\coordinate (1) at ({5},0);
	\coordinate (2) at ({1.5},{4});
	\coordinate (3) at ({1.5},{-3});
	\coordinate (4) at ({-2},{1});
	\coordinate (5) at ({3},{1});
	
	\ifthenelse{#2 = 0}{
		\draw[line width=1pt] (1) -- (5) -- (4);
		\draw[line width = 1pt] (2) -- (5) -- (3);
		\fill[peps0] (1) -- (3) -- (4) -- (2) -- (1);
		\fill[peps1spe] (1) -- (3) -- (4) -- (2) -- (1);
		\draw[line width=1pt] (1) -- (2) -- (4) -- (3) -- (1);
		\draw[line width=1pt] (1) -- (0) -- (4);
		\draw[line width=1pt] (2) -- (0) -- (3);	
	}{}	

	\ifthenelse{#2 = 5}{
		\draw[line cap=round, dash pattern=on 0pt off 2pt, line width=1.3pt] (1) -- (5) -- (4);
		\draw[line width = 1pt] (2) -- (5) -- (3);
		\fill[peps0] (1) -- (3) -- (4) -- (2) -- (1);
		\fill[peps1spe] (1) -- (3) -- (4) -- (2) -- (1);
		\draw[line width = 1pt] (1) -- (2) -- (4) -- (3) -- (1);
		\draw[line cap=round, dash pattern=on 0pt off 2pt, line width=1.3pt] (1) -- (0) -- (4);
		\draw[line width = 1pt] (2) -- (0) -- (3);
	}{}	

	\ifthenelse{#2 = 1}{
		\doubleTetra{0}{0};
	}{}	

	\ifthenelse{#2 = 2}{
		\coordinate (0) at (0,0);
		\coordinate (1) at ({5},0);
		\coordinate (2) at ({1.5},{4});
		\coordinate (3) at ({1.5},{-3});
		\coordinate (4) at ({-2},{1});
		\coordinate (5) at ({3},{1});
		
		\coordinate (10) at (2.5,0);
		\coordinate (11) at (0.75,2);
		\coordinate (12) at (0.75,-1.5);
		\coordinate (13) at (3.25,2);
		\coordinate (14) at (3.25,-1.5);
		\coordinate (15) at (0.5,1);
		\coordinate (16) at (-1,0.5);
		\coordinate (17) at (4,0.5);
		\coordinate (18) at (-0.25,2.5);
		\coordinate (19) at (-0.25,-1);
		\coordinate (20) at (2.25,2.5);
		\coordinate (21) at (2.25,-1);
		
		\draw[double, double distance between line centers=1.6pt] (1) -- (5) -- (4);
		\draw[double, double distance between line centers=1.6pt] (2) -- (5) -- (3);
		\foreach \x in {15,17,20,21}{
			\centerarc[fill=white, draw=none](\x)(0:360:0.3125)	
			\begin{scope}
				\clip ($(\x)+({0.3125*cos(0)},{0.3125*sin(0)})$) arc (0:360:0.3125);
				\ifthenelse{\x=15}{
					\draw[] (-2,1) -- (3,1);}{}
				\ifthenelse{\x=17}{
					\draw[] (3,1) -- (5,0);}{}
				\ifthenelse{\x=20}{
					\draw[] (3,1) -- (1.5,4);}{}
				\ifthenelse{\x=21}{
					\draw[] (3,1) -- (1.5,-3);}{}
			\end{scope}
			\centerarc[peps2](\x)(0:360:0.3125)
			\centerarc[](\x)(0:360:0.3125)
		}
		\fill[peps0] (1) -- (3) -- (4) -- (2) -- (1);
		\fill[peps1] (1) -- (3) -- (4) -- (2) -- (1);
		\draw[double, double distance between line centers=1.6pt] (1) -- (2) -- (4) -- (3) -- (1);
		\draw[double, double distance between line centers=1.6pt] (1) -- (0) -- (4);
		\draw[double, double distance between line centers=1.6pt] (2) -- (0) -- (3);
		
		\foreach \x in {10,11,12,13,14,16,18,19}{
			\centerarc[fill=white, draw=none](\x)(0:360:0.3125)	
			\centerarc[peps2](\x)(0:360:0.3125)
			\centerarc[](\x)(0:360:0.3125)
			\begin{scope}
				\clip ($(\x)+({0.3125*cos(0)},{0.3125*sin(0)})$) arc (0:360:0.3125);
				\ifthenelse{\x=10}{
					\draw[] (0,0) -- (5,0);}{}
				\ifthenelse{\x=11}{
					\draw[] (0,0) -- (1.5,4);}{}
				\ifthenelse{\x=12}{
					\draw[] (0,0) -- (1.5,-3);}{}
				\ifthenelse{\x=13}{
					\draw[] (1.5,4) -- (5,0);}{}
				\ifthenelse{\x=14}{
					\draw[] (1.5,-3) -- (5,0);}{}
				\ifthenelse{\x=16}{
					\draw[] (0,0) -- (-2,1);}{}
				\ifthenelse{\x=18}{
					\draw[] (-2,1) -- (1.5,4);}{}
				\ifthenelse{\x=19}{
					\draw[] (-2,1) -- (1.5,-3);}{}
			\end{scope}
		}
	
	}{}
	
	\ifthenelse{#2 = 3}{
		\fill[peps1] (0,0) rectangle (5,5);
		\draw[] (0,0) -- (5,5) node[pos=0.3, left] {${\sss \alpha_4}$} node[pos=0.7, right] {${\sss \alpha_2}$};
		\draw[] (0,5) -- (5,0) node[pos=0.3, above, xshift=2pt] {${\sss \alpha_1}$} node[pos=0.7, below, xshift=-2pt] {${\sss \alpha_3}$};
	}{}
	\ifthenelse{#2 = 4}{
		\fill[peps1] (0,0) rectangle (5,5);
		\draw[] (0,0) -- (5,5) node[pos=0.3, left] {${\sss  \kappa \alpha_4}$} node[pos=0.7, right] {${\sss \kappa \alpha_2}$};
		\draw[] (0,5) -- (5,0) node[pos=0.3, above, xshift=4pt] {${\sss \kappa \alpha_1}$} node[pos=0.7, below, xshift=-4pt] {${\sss  \kappa \alpha_3}$};
	}{}
	\end{tikzpicture}
}

\newcommand{\loopFlux}[2]{
	\begin{tikzpicture}[scale=#1,baseline={([yshift=-.5ex]current bounding box.center)}]	
	\ifthenelse{#2=1}{
		\coordinate (1) at (-2.2,-5);
		\coordinate (2) at (1.4,-5);
		\coordinate (3) at (5,-5);
		\coordinate (4) at (-2.2,-1.4);
		\coordinate (5) at (1.4,-1.4);
		\coordinate (6) at (5,-1.4);
		\coordinate (7) at (-2.2,2.2);
		\coordinate (8) at (1.4,2.2);
		\coordinate (9) at (5,2.2);
		\coordinate (11) at (-3.6,-3.6);
		\coordinate (12) at (0,-3.6);
		\coordinate (13) at (3.6,-3.6);
		\coordinate (14) at (-3.6,0);
		\coordinate (15) at (0,0);
		\coordinate (16) at (3.6,0);
		\coordinate (17) at (-3.6,3.6);
		\coordinate (18) at (0,3.6);
		\coordinate (19) at (3.6,3.6);
		
		\draw[dot] (11) -- (17) -- (19);
		\interLL{14}{16}{1}{7}{2}{8}{dot};
		\interLL{11}{13}{1}{7}{2}{8}{dot};
		\interLL{12}{18}{4}{5}{7}{8}{dot};
		\interLL{13}{19}{5}{6}{8}{9}{dot};

		\node at (-1.8,0) (a) {};
		\node at (0,1.8) (b) {};
		\node at (1.8,0) (c) {};
		\node at (0,-1.8) (d) {};
		\node at (-0.7,0.7) (e) {};
		\node at (0.7,-0.7) (f) {};
		\draw[shorten <= 14pt] (0,0) -- (1.4,0.9);
		\draw[shorten <= 22pt] (0,0) -- (1.8,-1.8);
		\draw[shorten <= 14pt] (0,0) -- (-0.9,-1.4);
		\draw[shorten <= 23pt] (0,0) -- (-2.5,0.7);
		\draw[shorten <= 23.5pt] (0,0) -- (-0.7,2.5);
		\node[circle, fill = black, inner sep=1.2pt] at (-0.7,0.7) {};
		\draw[] \convexpath{a,e,c,f}{7pt};
		\draw[] \convexpath{e,b,f,d}{7pt};
		\fill[peps0] \convexpath{a,b,c,d}{7pt};
		\fill[peps1] \convexpath{a,b,c,d}{7pt};
		\draw[] \convexpath{a,b,c,d}{7pt};
		\draw[] \convexpath{a,b,c,f}{7pt};
		\draw[] \convexpath{a,b,f,d}{7pt};
		\draw[shorten <= 22pt] (0,0) -- (1.8,1.8);
		\draw[shorten <= 22pt] (0,0) -- (-1.8,-1.8);
		\draw[shorten <= 22pt] (0,0) -- (-1.8,1.8);
		\draw[shorten <= 14pt] (0,0) -- (-1.4,-0.9);
		\draw[shorten <= 23pt] (0,0) -- (2.5,-0.7);
		\draw[shorten <= 23.5pt] (0,0) -- (0.7,-2.5);
		\draw[shorten <= 14pt] (0,0) -- (0.9,1.4); 
		\node[circle, fill = black, inner sep=1.2pt] at (-1.8,0) {};
		\node[circle, fill = black, inner sep=1.2pt] at (1.8,0) {};	
		\node[circle, fill = black, inner sep=1.2pt] at (0,1.8) {};	
		\node[circle, fill = black, inner sep=1.2pt] at (0,-1.8) {};		
		\node[circle, fill = black, inner sep=1.2pt] at (0.7,-0.7) {};

		\draw[dot] (1) -- (7) -- (9) -- (3) --cycle;
		\draw[dot] (2) -- (8);
		\draw[dot] (4) -- (6);
		\foreach \x in {1,2,3,4,5,6,7,8,9}
			\draw[dot] (\x) -- ({1\x});	
	}{}
	\ifthenelse{#2=2}{
		\coordinate (1) at (-2.2,-5);
		\coordinate (2) at (1.4,-5);
		\coordinate (3) at (5,-5);
		\coordinate (4) at (-2.2,-1.4);
		\coordinate (5) at (1.4,-1.4);
		\coordinate (6) at (5,-1.4);
		\coordinate (7) at (-2.2,2.2);
		\coordinate (8) at (1.4,2.2);
		\coordinate (9) at (5,2.2);
		\coordinate (11) at (-3.6,-3.6);
		\coordinate (12) at (0,-3.6);
		\coordinate (13) at (3.6,-3.6);
		\coordinate (14) at (-3.6,0);
		\coordinate (15) at (0,0);
		\coordinate (16) at (3.6,0);
		\coordinate (17) at (-3.6,3.6);
		\coordinate (18) at (0,3.6);
		\coordinate (19) at (3.6,3.6);
		
		\draw[dot] (11) -- (17) -- (19);
		\interLL{14}{16}{1}{7}{2}{8}{dot};
		\interLL{11}{13}{1}{7}{2}{8}{dot};
		\interLL{12}{18}{4}{5}{7}{8}{dot};
		\interLL{13}{19}{5}{6}{8}{9}{dot};

		\draw[dot] (1) -- (7) -- (9) -- (3) --cycle;
		\draw[dot] (2) -- (8);
		\draw[dot] (4) -- (6);
		\foreach \x in {1,2,3,4,5,6,7,8,9}
			\draw[dot] (\x) -- ({1\x});	
	}{}
	\ifthenelse{#2=3}{
		\node at (-1.8,0) (a) {};
		\node at (0,1.8) (b) {};
		\node at (1.8,0) (c) {};
		\node at (0,-1.8) (d) {};
		\node at (-0.7,0.7) (e) {};
		\node at (0.7,-0.7) (f) {};
		\draw[shorten <= 14pt] (0,0) -- (1.4,0.9);
		\draw[shorten <= 22pt] (0,0) -- (1.8,-1.8);
		\draw[shorten <= 14pt] (0,0) -- (-0.9,-1.4);
		\draw[shorten <= 23pt] (0,0) -- (-2.5,0.7);
		\draw[shorten <= 23.5pt] (0,0) -- (-0.7,2.5);
		\node[circle, fill = black, inner sep=1.2pt] at (-0.7,0.7) {};
		\draw[] \convexpath{a,e,c,f}{7pt};
		\draw[] \convexpath{e,b,f,d}{7pt};
		\fill[peps0] \convexpath{a,b,c,d}{7pt};
		\fill[peps1] \convexpath{a,b,c,d}{7pt};
		\draw[] \convexpath{a,b,c,d}{7pt};
		\draw[] \convexpath{a,b,c,f}{7pt};
		\draw[] \convexpath{a,b,f,d}{7pt};
		\draw[shorten <= 22pt] (0,0) -- (1.8,1.8);
		\draw[shorten <= 22pt] (0,0) -- (-1.8,-1.8);
		\draw[shorten <= 22pt] (0,0) -- (-1.8,1.8);
		\draw[shorten <= 14pt] (0,0) -- (-1.4,-0.9);
		\draw[shorten <= 23pt] (0,0) -- (2.5,-0.7);
		\draw[shorten <= 23.5pt] (0,0) -- (0.7,-2.5);
		\draw[shorten <= 14pt] (0,0) -- (0.9,1.4); 
		\node[circle, fill = black, inner sep=1.2pt] at (-1.8,0) {};
		\node[circle, fill = black, inner sep=1.2pt] at (1.8,0) {};	
		\node[circle, fill = black, inner sep=1.2pt] at (0,1.8) {};	
		\node[circle, fill = black, inner sep=1.2pt] at (0,-1.8) {};		
		\node[circle, fill = black, inner sep=1.2pt] at (0.7,-0.7) {};
		\node[pauliZ] at (0.7,-0.7) {${\sss Z}$};	
	}{}
	\ifthenelse{#2=4}{
		\node at (-1.8,0) (a) {};
		\node at (0,1.8) (b) {};
		\node at (1.8,0) (c) {};
		\node at (0,-1.8) (d) {};
		\node at (-0.7,0.7) (e) {};
		\node at (0.7,-0.7) (f) {};
		\draw[shorten <= 14pt] (0,0) -- (1.4,0.9);
		\draw[shorten <= 22pt] (0,0) -- (1.8,-1.8);
		\draw[shorten <= 14pt] (0,0) -- (-0.9,-1.4);
		\draw[shorten <= 23pt] (0,0) -- (-2.5,0.7);
		\draw[shorten <= 23.5pt] (0,0) -- (-0.7,2.5);
		\node[circle, fill = black, inner sep=1.2pt] at (-0.7,0.7) {};
		\draw[] \convexpath{a,e,c,f}{7pt};
		\draw[] \convexpath{e,b,f,d}{7pt};
		\fill[peps0] \convexpath{a,b,c,d}{7pt};
		\fill[peps1] \convexpath{a,b,c,d}{7pt};
		\draw[] \convexpath{a,b,c,d}{7pt};
		\draw[] \convexpath{a,b,c,f}{7pt};
		\draw[] \convexpath{a,b,f,d}{7pt};
		\draw[shorten <= 22pt] (0,0) -- (1.8,1.8);
		\draw[shorten <= 22pt] (0,0) -- (-1.8,-1.8);
		\draw[shorten <= 22pt] (0,0) -- (-1.8,1.8);
		\draw[shorten <= 14pt] (0,0) -- (-1.4,-0.9) node[anchor=center, pos=1, pauliZ] {${\sss Z}$};
		\draw[shorten <= 23pt] (0,0) -- (2.5,-0.7) node[anchor=center, pos=1, pauliZ] {${\sss Z}$};
		\draw[shorten <= 23.5pt] (0,0) -- (0.7,-2.5) node[anchor=center, pos=1, pauliZ] {${\sss Z}$};
		\draw[shorten <= 14pt] (0,0) -- (0.9,1.4) node[anchor=center, pos=1, pauliZ] {${\sss Z}$}; 
		\node[circle, fill = black, inner sep=1.2pt] at (-1.8,0) {};
		\node[circle, fill = black, inner sep=1.2pt] at (1.8,0) {};	
		\node[circle, fill = black, inner sep=1.2pt] at (0,1.8) {};	
		\node[circle, fill = black, inner sep=1.2pt] at (0,-1.8) {};		
		\node[circle, fill = black, inner sep=1.2pt] at (0.7,-0.7) {};	
	}{}
	\end{tikzpicture}
}

\newcommand{\transferThreeD}[2]{
	\begin{tikzpicture}[scale=#1,baseline={([yshift=-.5ex]current bounding box.center)}]
	
	\ifthenelse{#2 = 1}{
		\tetra{7.5}{-1};
		\tetra{2.5}{-1};
		\tetra{-2.5}{-1};
		\tetra{9.5}{-2};
		\tetra{4.5}{-2};
		\tetra{-0.5}{-2};
		\tetra{11.5}{-3};
		\tetra{6.5}{-3};
		\tetra{1.5}{-3};
		
		\tetra{6}{2};
		\tetra{1}{2};
		\tetra{-4}{2};
		\tetra{8}{1};
		\tetra{3}{1};
		\tetra{-2}{1};
		\tetra{10}{0};
		\tetra{5}{0};
		\tetra{0}{0};

	}{}
	\ifthenelse{#2 = 2}{
		\fill[peps1] (0,-0) rectangle (15,15);
		\draw[line cap=round, dash pattern=on 0pt off 2pt, line width=1.1pt] (0,0) rectangle (15,15);
		\draw[line cap=round, dash pattern=on 0pt off 2pt, line width=1.1pt] (5,0) -- (5,15);
		\draw[line cap=round, dash pattern=on 0pt off 2pt, line width=1.1pt] (10,0) -- (10,15);
		\draw[line cap=round, dash pattern=on 0pt off 2pt, line width=1.1pt] (0,5) -- (15,5);
		\draw[line cap=round, dash pattern=on 0pt off 2pt, line width=1.1pt] (0,10) -- (15,10);
		\draw[] (0,15) -- (15,0);
		\draw[] (0,0) -- (15,15);
		\draw[] (0,5) -- (10,15);
		\draw[] (5,0) -- (15,10);
		\draw[] (0,10) -- (10,0);
		\draw[] (0,10) -- (5,15);
		\draw[] (0,5) -- (5,0);
		\draw[] (10,0) -- (15,5);
		\draw[] (5,15) -- (15,5);
		\draw[] (10,15) -- (15,10);
		\node[pauliX] at (1.25,6.25) {${\sss X}$};
		\node[pauliX] at (3.75,6.25) {${\sss X}$};
		\node[pauliX] at (6.25,6.25) {${\sss X}$};
		\node[pauliX] at (8.75,6.25) {${\sss X}$};
		\node[pauliX] at (11.25,6.25) {${\sss X}$};
		\node[pauliX] at (13.75,6.25) {${\sss X}$};
		\node[pauliX] at (3.75,11.25) {${\sss X}$};
		\node[pauliX] at (3.75,13.75) {${\sss X}$};
		\node[pauliX] at (6.25,11.25) {${\sss X}$};
		\node[pauliX] at (6.25,13.75) {${\sss X}$};	
		
		\foreach \y in {0,15}{
			\node[] at (2.5,\y) {${\sss \wedge}$};
			\node[] at (7.4,\y) {${\sss \wedge}$};
			\node[] at (7.6,\y) {${\sss \wedge}$};
			\node[] at (12.3,\y) {${\sss \wedge}$};
			\node[] at (12.5,\y) {${\sss \wedge}$};
			\node[] at (12.7,\y) {${\sss \wedge}$};
		}
		\foreach \x in {0,15}{
			\node[] at (\x,2.7) {\rotatebox[origin=c]{-20}{${\sss /}$}};
			\node[] at (\x,2.5) {\rotatebox[origin=c]{-20}{${\sss /}$}};
			\node[] at (\x,2.3) {\rotatebox[origin=c]{-20}{${\sss /}$}};
			\node[] at (\x,7.6) {\rotatebox[origin=c]{-20}{${\sss /}$}};
			\node[] at (\x,7.4) {\rotatebox[origin=c]{-20}{${\sss /}$}};
			\node[] at (\x,12.5) {\rotatebox[origin=c]{-20}{${\sss /}$}};
		}

	}{}
	\end{tikzpicture}
}

\newcommand{\mixedThreeD}[2]{
	\begin{tikzpicture}[scale=#1,baseline={([yshift=-.5ex]current bounding box.center)}]
	\ifthenelse{#2 = 1}{
		\begin{scope}
			\clip ({11.75+0.8525},-3) -- ({4.25-0.3525},-3) -- (-2.25,4) -- ({1.25-0.8525},11) -- ({8.75+0.3525},11) -- (15.25,4) --cycle;
			\tetra{4.5}{12};
			\tetra{-0.5}{12};
			\tetra{8}{8};
			\tetra{3}{8};
			\tetra{-2}{8};
			\tetra{11.5}{4};
			\tetra{6.5}{4};
			\tetra{1.5}{4};
			\tetra{-3.5}{4};
			\tetra{15}{0};
			\tetra{10}{0};
			\tetra{5}{0};
			\tetra{0}{0};
			\tetra{13.5}{-4};
			\tetra{8.5}{-4};
			\tetra{3.5}{-4};
			\node[pauliZ] at (11.75,-2) {${\sss Z}$};
			\node[pauliZ] at (4.25,-2) {${\sss Z}$};
			\node[pauliZ] at (9.25,-2) {${\sss Z}$};
			\node[pauliZ] at (6.75,-2) {${\sss Z}$};
			\node[pauliZ] at (-1,4) {${\sss Z}$};
			\node[pauliZ] at (1.25,10) {${\sss Z}$};
			\node[pauliZ] at (3.75,10) {${\sss Z}$};
			\node[pauliZ] at (6.25,10) {${\sss Z}$};
			\node[pauliZ] at (8.75,10) {${\sss Z}$};
			\node[pauliZ] at (14,4) {${\sss Z}$};
			\node[pauliZ] at (-0.25,6) {${\sss Z}$};
			\node[pauliZ] at (0.5,8) {${\sss Z}$};
			\node[pauliZ] at (10.5,8) {${\sss Z}$};
			\node[pauliZ] at (12.25,6) {${\sss Z}$};
			\node[pauliZ] at (0.75,2) {${\sss Z}$};
			\node[pauliZ] at (2.5,0) {${\sss Z}$};
			\node[pauliZ] at (12.5,0) {${\sss Z}$};
			\node[pauliZ] at (13.25,2) {${\sss Z}$};
		\end{scope}
	}{}	
	\end{tikzpicture}
}

\newcommand{\mixedThreeDPEPO}[2]{
	\begin{tikzpicture}[scale=#1,baseline={([yshift=-.5ex]current bounding box.center)}]
		\coordinate (a) at (7.5,0);
		\coordinate (b) at (12.5,0);
		\coordinate (c) at (5,4.33);
		\coordinate (d) at (10,4.33);
		\coordinate (e) at (15,4.33);
		\coordinate (f) at (7.5,8.66);
		\coordinate (g) at (12.5,8.66);
		
		\begin{scope}
			\clip (1.25,4.3301) -- ({5+0.625},{8.6602+3.2476}) -- ({15-0.625},{8.6602+3.2476}) -- ({20-1.25},4.3301) -- ({15-0.625},{-3.2476}) -- ({5+0.625},{-3.2476}) --cycle;
			\clip (1.6,-10) rectangle (18.4,15);
			\draw[] (0,0) -- (20,0);
			\draw[] (0,4.3301) -- (20,4.3301);
			\draw[] (0,8.6602) -- (20,8.6602);
			\draw[ shift={(0,-4.3301)}, rotate=60] (0,0) -- (20,0);
			\draw[ shift={(5,-4.3301)}, rotate=60] (0,0) -- (20,0);
			\draw[ shift={(10,-4.3301)}, rotate=60] (0,0) -- (20,0);
			\draw[ shift={(0,12.9903)}, rotate=-60] (0,0) -- (20,0);
			\draw[ shift={(5,12.9903)}, rotate=-60] (0,0) -- (20,0);
			\draw[ shift={(10,12.9903)}, rotate=-60] (0,0) -- (20,0);
		
			\foreach \x in {2.5,7.5,12.5}{
				\foreach \y in {0,8.66}{
					\coordinate (u) at ({\x+2.5},\y);
					\coordinate (v) at ({\x+1.25},{\y+2.165});
					\coordinate (w) at ({\x+3.75},{\y+2.165});	
					\foreach \z in {u,v,w}{
						\centerarc[fill=white, draw=none](\z)(0:360:0.5)	
						\centerarc[peps1](\z)(0:360:0.5)
						\centerarc[](\z)(0:360:0.5)
						\node[circle, fill = black, inner sep=1.2pt] at (\z) {};
						\node[circle, fill = white, inner sep=0.7pt] at (\z) {};
					}
				}
			}
			\foreach \x in {0,5,10,15}{
				\foreach \y in {-4.33,4.33}{
					\coordinate (u) at ({\x+2.5},\y);
					\coordinate (v) at ({\x+1.25},{\y+2.165});
					\coordinate (w) at ({\x+3.75},{\y+2.165});	
					\foreach \z in {u,v,w}{
						\centerarc[fill=white, draw=none](\z)(0:360:0.5)	
						\centerarc[peps1](\z)(0:360:0.5)
						\centerarc[](\z)(0:360:0.5)
						\node[circle, fill = black, inner sep=1.2pt] at (\z) {};
						\node[circle, fill = white, inner sep=0.7pt] at (\z) {};
					}
				}
			}
			\foreach \z in {a,b,c,d,e,f,g}{
				\centerarc[fill=white, draw=none](\z)(0:360:0.5)	
				\centerarc[peps2](\z)(0:360:0.5)
				\centerarc[](\z)(0:360:0.5)
				\node[circle, fill = black, inner sep=1.2pt] at (\z) {};
			}
		\end{scope}
		
		\ifthenelse{#2 = 1}{	
		}{}	
		\ifthenelse{#2 = 2}{
			\foreach \z in {a,b,c,d,e,f,g}
				\node[pauliZ] at (\z) {${\sss Z}$}; 	
		}{}	
		\ifthenelse{#2 = 3}{
			\foreach \x in {7.5,12.5}{
				\foreach \y in {0,8.66}{
					\node[pauliZ] at ({\x+1.25},\y) {${\sss Z}$}; 
					\node[pauliZ] at ({\x-1.25},\y) {${\sss Z}$};
					\node[pauliZ] at ({\x+0.625},{\y+1.0825}) {${\sss Z}$};
					\node[pauliZ] at ({\x+0.625},{\y-1.0825}) {${\sss Z}$};
					\node[pauliZ] at ({\x-0.625},{\y+1.0825}) {${\sss Z}$};
					\node[pauliZ] at ({\x-0.625},{\y-1.0825}) {${\sss Z}$};
				}
			}
			\foreach \x in {5,10,15}{
				\node[pauliZ] at ({\x+1.25},4.33) {${\sss Z}$}; 
				\node[pauliZ] at ({\x-1.25},4.33) {${\sss Z}$};
				\node[pauliZ] at ({\x+0.625},{4.33+1.0825}) {${\sss Z}$};
				\node[pauliZ] at ({\x+0.625},{4.33-1.0825}) {${\sss Z}$};
				\node[pauliZ] at ({\x-0.625},{4.33+1.0825}) {${\sss Z}$};
				\node[pauliZ] at ({\x-0.625},{4.33-1.0825}) {${\sss Z}$};
			}	
		}{}	
		\ifthenelse{#2 = 4}{
			\foreach \x in {6.25, 8.75, 11.25, 13.75}
				\foreach \y in {-2.165, 10.825}
					\node[pauliZ] at (\x,\y) {${\sss Z}$}; 

			\foreach \x in {5, 15}
				\foreach \y in {0, 8.66}
					\node[pauliZ] at (\x,\y) {${\sss Z}$}; 
					
			\foreach \x in {3.75, 16.25}
					\foreach \y in {2.165, 6.495}
						\node[pauliZ] at (\x,\y) {${\sss Z}$}; 
											
			\foreach \x in {2.5,17.5}
				\node[pauliZ] at (\x,4.33) {${\sss Z}$}; 
	
		}{}	
	\end{tikzpicture}
}

\newcommand{\deltaTBis}[2]{
	\begin{tikzpicture}[scale=#1,baseline=-3pt]
	\coordinate (0) at (0,0);
	\coordinate (1) at (2.4,0);
	\ifthenelse{#2 = -1}{
		\draw[] (1.4,0) -- (0,0);	
		\draw[] (0,0) to[out=90, in=0, looseness=1] (-1.2,0.8);
		\draw[] (0,0) to[out=-90, in=0, looseness=1] (-1.2,-0.8);
		\draw[] (-1.5,0) to[out=90,in=-180,looseness=1] (-1.2,0.8);
		\draw[] (-1.5,0) to[out=-90,in=180,looseness=1] (-1.2,-0.8);
		\centerarc[fill=white, draw=none](0,0)(0:360:0.2)	
		\centerarc[peps2](0,0)(0:360:0.2)
		\centerarc[](0,0)(0:360:0.2)
		\node[pauliX] at (-1.5,0) {${\sss X^a}$}; 
	}{}
	\ifthenelse{#2 = -2}{
		\draw[] (1.4,0) -- (0,0);	
		\draw[] (0,0) to[out=90, in=0, looseness=1] (-1.2,0.8);
		\draw[] (0,0) to[out=-90, in=0, looseness=1] (-1.2,-0.8);
		\draw[] (-1.5,0) to[out=90,in=-180,looseness=1] (-1.2,0.8);
		\draw[] (-1.5,0) to[out=-90,in=180,looseness=1] (-1.2,-0.8);
		\centerarc[fill=white, draw=none](0,0)(0:360:0.2)	
		\centerarc[peps2](0,0)(0:360:0.2)
		\centerarc[](0,0)(0:360:0.2)
		\node[pauliX] at (0.8,0) {${\sss X^a}$}; 
	}{}
	\ifthenelse{#2 = 0}{
		\draw[] (1.2,0) -- (0,0);	
		\draw[] (0,0) to[out=90, in=0, looseness=1] (-1.2,0.8);
		\draw[] (0,0) to[out=-90, in=0, looseness=1] (-1.2,-0.8);
		\draw[] (-1.5,0) to[out=90,in=-180,looseness=1.1] (-1.2,0.8);
		\draw[] (-1.5,0) to[out=-90,in=-180,looseness=1.1] (-1.2,-0.8);
		\centerarc[fill=white, draw=none](0,0)(0:360:0.2)	
		\centerarc[peps2](0,0)(0:360:0.2)
		\centerarc[](0,0)(0:360:0.2)
		\node[pauliX] at (-1.5,0) {${\sss X_{\mathsf e}^{a_\mathsf{e}}}$}; 
	}{}
	\ifthenelse{#2 = 1}{
		\draw[] (-1.2,0) -- (0,0);	
		\draw[] (0,0) to[out=90, in=180, looseness=1] (1.2,0.8);
		\draw[] (0,0) to[out=-90, in=180, looseness=1] (1.2,-0.8);
		\centerarc[fill=white, draw=none](0,0)(0:360:0.2)	
		\centerarc[peps2](0,0)(0:360:0.2)
		\centerarc[](0,0)(0:360:0.2)
	}{}		
	\ifthenelse{#2 = 2}{
		\draw[] (1.2,0) -- (0,0);	
		\draw[] (0,0) to[out=90, in=0, looseness=1] (-1.2,0.8);
		\draw[] (0,0) to[out=-90, in=0, looseness=1] (-1.2,-0.8);
		\centerarc[fill=white, draw=none](0,0)(0:360:0.2)	
		\centerarc[peps2](0,0)(0:360:0.2)
		\centerarc[](0,0)(0:360:0.2)
	}{}	
	\ifthenelse{#2 = 3}{
		\draw[] (-1.2,0) -- (0,0);	
		\draw[] (0,0) to[out=90, in=180, looseness=1] (1.2,0.8);
		\draw[] (0,0) to[out=-90, in=180, looseness=1] (1.2,-0.8);
		\draw[] (2.4,0) -- (3.6,0);
		\draw[] (2.4,0) to[out=90, in=0, looseness=1] (1.2,0.8);
		\draw[] (2.4,0) to[out=-90, in=0, looseness=1] (1.2,-0.8);
		\foreach \x in {0,1}{	
			\centerarc[fill=white, draw=none](\x)(0:360:0.2)	
			\centerarc[peps2](\x)(0:360:0.2)
			\centerarc[](\x)(0:360:0.2)
		}
		\node[pauliZ] at (1.2,0.8) {${\sss X}$}; 
	}{}	
	\ifthenelse{#2 = 4}{
		\draw[] (-1.2,0) -- (0,0);	
		\draw[] (0,0) to[out=90, in=180, looseness=1] (1.2,0.8);
		\draw[] (0,0) to[out=-90, in=180, looseness=1] (1.2,-0.8);
		\draw[] (2.4,0) -- (3.6,0);
		\draw[] (2.4,0) to[out=90, in=0, looseness=1] (1.2,0.8);
		\draw[] (2.4,0) to[out=-90, in=0, looseness=1] (1.2,-0.8);
		\foreach \x in {0,1}{	
			\centerarc[fill=white, draw=none](\x)(0:360:0.2)	
			\centerarc[peps2](\x)(0:360:0.2)
			\centerarc[](\x)(0:360:0.2)
		}
		\node[pauliZ] at (1.2,0.8) {${\sss Z}$}; 
	}{}		
	\end{tikzpicture}
}

\newcommand{\threeCylinder}[2]{
	\begin{tikzpicture}[scale=#1,baseline={([yshift=-.5ex]current bounding box.center)}, rotate=-90]
	\def\ry{1};
	\def\rx{0.5};
	\def\ratio{0.4};
	\def\ratiob{0.675};

	\draw[] (3,0) arc(-90:-270:\rx cm and \ry cm);
	
	\ifthenelse{#2 = 2}{
		\draw[black, opacity=0.3, line width=0.5pt] (0,{(1-\ratiob)*\ry}) arc(-90:-270:\ratiob*\rx cm and \ratiob*\ry cm) arc(90:-90:\ratiob*\rx cm and \ratiob*\ry cm);
	}{}
	\ifthenelse{#2 = 3}{
		\fill[black, opacity=0.3] (0,{(1+\ratio)*\ry}) -- (0,2*\ry) -- (3,2*\ry) -- (3,{(1+\ratio)*\ry}) --cycle;
		\node[pauliX] at (1.5,1.7) {${\sss Z^g}$};
	}{}
	\ifthenelse{#2 = 4}{
		\draw[black, opacity=0.3, line width=0.5pt] (0,{(1-\ratiob)*\ry}) arc(-90:-270:\ratiob*\rx cm and \ratiob*\ry cm) arc(90:-90:\ratiob*\rx cm and \ratiob*\ry cm);
	}{}

	\fill[peps0] (0,{(1-\ratio)*\ry}) arc(-90:-270:\ratio*\rx cm and \ratio*\ry cm) -- (3,{(1+\ratio)*\ry}) arc(90:270:\ratio*\rx cm and \ratio*\ry cm) -- (0,{(1-\ratio)*\ry});
	\draw[] (3,{(1-\ratio)*\ry}) arc(-90:-270:\ratio*\rx cm and \ratio*\ry cm);
	\fill[peps0] (0,{(1-\ratio)*\ry}) arc(-90:90:\ratio*\rx cm and \ratio*\ry cm) -- (3,{(1+\ratio)*\ry}) arc(90:-90:\ratio*\rx cm and \ratio*\ry cm) -- (0,{(1-\ratio)*\ry});
	\draw[] (0,{(1-\ratio)*\ry}) -- (3,{(1-\ratio)*\ry});
	\draw[] (0,{(1+\ratio)*\ry}) -- (3,{(1+\ratio)*\ry});
	\draw[] (3,{(1-\ratio)*\ry}) arc(-90:90:\ratio*\rx cm and \ratio*\ry cm);
		
	\ifthenelse{#2 = 2}{
		\fill[black, opacity=0.3] (0,{(1-\ratiob)*\ry}) arc(-90:90:\ratiob*\rx cm and \ratiob*\ry cm) -- (3,{(1+\ratiob)*\ry}) arc(90:-90:\ratiob*\rx cm and \ratiob*\ry cm) -- (0,{(1-\ratiob)*\ry});
		\node[pauliX] at (1.5,\ry) {${\sss Z^h}$};
	}{}
	\ifthenelse{#2 = 4}{
		\fill[black, opacity=0.3] (0,{(1+\ratio)*\ry}) -- (0,2*\ry) -- (3,2*\ry) -- (3,{(1+\ratio)*\ry}) --cycle;
		\fill[black, opacity=0.3] (0,{(1-\ratiob)*\ry}) arc(-90:90:\ratiob*\rx cm and \ratiob*\ry cm) -- (3,{(1+\ratiob)*\ry}) arc(90:-90:\ratiob*\rx cm and \ratiob*\ry cm) -- (0,{(1-\ratiob)*\ry});
		\node[pauliX] at (1.5,1.7) {${\sss Z^g}$};
		\node[pauliX] at (1.5,\ry) {${\sss Z^h}$};
	}{}
	
	\fill[peps0] (0,0) arc(-90:-270:\rx cm and \ry cm) arc(90:-90:\rx cm and \ry cm);
	\fill[peps0] (0,0) -- (3,0) arc(-90:90:\rx cm and \ry cm) -- (0,2*\ry) arc(90:-90:\rx cm and \ry cm);
	\draw[] (3,0) arc(-90:90:\rx cm and \ry cm);
	\draw[] (0,0) arc(-90:90:\rx cm and \ry cm);
	\draw[] (0,0) arc(-90:-270:\rx cm and \ry cm);
	\draw[] (0,{(1-\ratio)*\ry}) arc(-90:90:\ratio*\rx cm and \ratio*\ry cm);
	\draw[] (0,{(1-\ratio)*\ry}) arc(-90:-270:\ratio*\rx cm and \ratio*\ry cm);
	\draw[] (0,0) -- (3,0);
	\draw[] (0,2*\ry) -- (3,2*\ry);
	\end{tikzpicture}
}

%% file: _Introduction.tex
\section{Introduction}

\noindent
Tensor networks have proven very powerful as a numerical as well as analytical framework for the study of strongly correlated quantum many-body systems. They provide a description of complex \emph{global} entanglement patterns in terms of \emph{local} tensors, which carry both \emph{physical} and \emph{virtual} entanglement degrees of freedom. The global quantum
correlations are then built up by contracting the entanglement degrees of freedom of the individual tensors following a pattern dictated by a lattice. This ability to describe complex quantum correlations makes tensor networks particularly suited for the study of systems that exhibit \emph{topological order}~\cite{SCHUCH20102153, PhysRevLett.111.090501, BUERSCHAPER2014447, csahinouglu2014characterizing, BULTINCK2017183, Bultinck_2017, Williamson:2017uzx}, i.e.\ where an interacting spin system does not order magnetically but rather in its long-range entanglement~\cite{Wen:1989iv, Chen:2010gda}. Models of topological order display a variety of unconventional phenomena, such as topology-dependent ground state degeneracy and excitations with exotic \emph{anyonic statistics} \cite{Kitaev:2006lla}, making them prime candidates
for \emph{quantum memories} and \emph{quantum computing} platforms~\cite{Kitaev1997,doi:10.1063/1.1499754, RevModPhys.80.1083}.  Nevertheless, topological order in two dimensions is not robust to finite temperature due to the resulting doping with point-like excitations that drives the system to a trivial phase~\cite{doi:10.1063/1.1499754,PhysRevB.76.184442, NUSSINOV2009977}.

In (2+1)d, topological order in a tensor network wave function manifests itself in an \emph{entanglement symmetry}, i.e. a symmetry acting solely on the virtual degrees of freedom. In the case of (untwisted) gauge models \cite{Kitaev1997}, whose input data are finite groups, this symmetry amounts to an invariance under the action of a group representation on all entanglement degrees of freedom. Equivalently, it can be phrased as a so-called \emph{pulling through} condition:
\begin{equation*}
	\includeTikz{0}{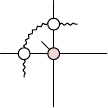}{\pulling{1}{1}} = \includeTikz{0}{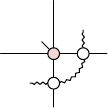}{\pulling{1}{2}}\, ,
\end{equation*} 
stating that a string of symmetry operators on the entanglement level can be pulled through  a tensor, making such a string freely deformable~\cite{SCHUCH20102153,BUERSCHAPER2014447,csahinouglu2014characterizing}.  Remarkably, this simple property can be exploited to encode the ground state subspace, the anyonic excitations and their statistics, as well as to understand topological corrections to the entanglement and the structure of renormalization group flows. It turns out that this description remains valid as one considers deformations, allowing to capture physical excitations away from the renormalization group fixed point. However, a sufficiently strong deformation might drive the system into a trivial phase---even though it respects the virtual symmetry---through the mechanism of \emph{anyon condensation}~\cite{doi:10.1146/annurev-conmatphys-033117-054154,PhysRevB.79.045316}. Once more, this can be understood solely from the point of view of the virtual symmetry. Indeed, given an open network, the boundary inherits the symmetry of the tensors, so that the boundary state (or entanglement spectrum) can order in different ways with respect to this symmetry. These `entanglement phases', which can be both \emph{symmetry-broken} and
\emph{symmetry-protected}, correspond to the different ways the topological order can be modified, as characterized by the behavior of the corresponding anyons~\cite{haegeman2015shadows, PhysRevB.95.235119, PhysRevB.97.195124, PhysRevB.100.245125}.  

Although it is not a \emph{sufficient} condition, the entanglement symmetry is a \emph{necessary} property for topological order in (2+1)d. Indeed, perturbations of the tensors that break this symmetry were shown to immediately destroy the topological order~\cite{PhysRevB.82.165119, PhysRevB.90.245116, PhysRevB.98.125112}. This can be understood from the fact that such perturbations induce an immediate condensation of the corresponding point-like excitations, or alternatively, that the perturbation strength grows linearly under renormalization.

\bigskip \noindent
Given a self-dual model in (2+1)d, such as the \emph{toric code}, where \emph{electric} and \emph{magnetic} excitations can be swapped, the tensor network representation breaks the duality. Indeed, at the entanglement level, one of the excitations is represented by a \emph{string-like} symmetry operator, whereas the other is described as a {point-like} operator that transforms as an irreducible representation. Depending on the way we assign these two kinds of virtual operators to the physical excitations, we arrive at two different tensor network representations. However, due to the self-duality of the model, these are equivalent up to swapping the excitations together with a Fourier transform on the virtual level.  We recover this self-duality at the boundary---in essence, the (1+1)d Ising self-duality---such that the mapping preserves the symmetry but exchanges the symmetric and symmetry-broken phases. Similarly, these two representations respond dually under uniform perturbations of the tensors that break the virtual symmetry, leading to a doping with electric or magnetic excitations, respectively, which breaks the topological order either way.

What happens to this scenario as we turn to (3+1)d? What is the entanglement symmetry structure of the (3+1)d toric code tensor network that allows us to understand its topological features in a way akin to (2+1)d? As before, we should be able to construct two tensor network representations, which satisfy virtual symmetry conditions, such that allowed operator insertions breaking these symmetries correspond to electric or magnetic physical excitations, respectively. Given that the model is no longer self-dual, such that electric and (bulk) magnetic excitations are \emph{point-like} and \emph{loop-like}, respectively, we expect these two tensor network representations to be inequivalent and characterized by distinct symmetry conditions. This raises the question, how will the difference in symmetry structure manifests itself in the study of the topological model? Furthermore, given that magnetic excitations possess a loop tension preventing them from proliferating, shall we expect the two representations to exhibit different robustness under symmetry-breaking perturbations?

The purpose of this manuscript is to construct explicitly these two dual tensor network representations of the (3+1)d toric code and study their properties: we identify their relevant virtual symmetries; we explain how these can be used to parametrize the ground space and the anyonic excitations; we construct renormalization group transformations through their action on the virtual symmetries; we discuss how these virtual symmetries are inherited by the boundary, and in which way the resulting boundary phases determine the condensation of anyons.

The first representation we construct has a \emph{global} entanglement symmetry, such that the tensor is invariant under a simultaneous symmetry action on \emph{all} entanglement degrees of freedom. In other words, every tensor satisfies a pulling through condition with respect to a virtual membrane-like operator that remains identical under blocking. Ground states and magnetic excitations are obtained from membrane-like symmetry operators, whereas point-like irreducible representations act as the dual electric particles. The symmetry is inherited at the boundary as a \emph{global} symmetry so that the entanglement spectrum exhibits symmetry-breaking phases in close analogy to that of the (2+1)d toric code, now detected by point-like order parameters and membrane-like disorder parameters.

The second representation is more interesting, as it fundamentally differs from the (2+1)d paradigm. It satisfies \emph{local} entanglement symmetries associated with every elementary loop of virtual indices. This corresponds to pulling through conditions with respect to string-like operators, such that the number of independent symmetries increases under blocking as the number of loops at the boundary of a region grows with its size. This requires in particular a more involved renormalization scheme, which we present in detail. With this representation, ground states and  electric excitations are parametrized by string-like symmetry operators, whereas magnetic excitations correspond to loops of irreducible representations. The symmetries of the tensors give rise to a \emph{gauge} symmetry at the boundary, where the behavior of the anyons now
probes the \emph{confined} and \emph{deconfined} phase of the gauge theory, respectively.

Given that the first (3+1)d representation shares a lot of characteristics with the (2+1)d ones---as illustrated, for instance, by its boundary phase diagram---one anticipates it to respond similarly to perturbations that break the entanglement symmetry. Indeed, we know that the boundary theory is unstable to such perturbations. Moreover, we expect these modifications to induce a condensation of the point-like electric excitations, which would in turn break the topological order, and similarly, that the perturbation acquire a linearly increasing strength under renormalization.
On the contrary, following the same line of argument, it is plausible that the second (3+1)d representation is stable to perturbations. Firstly, symmetry-breaking perturbations now amount to a breaking of a gauge symmetry at the boundary, which in virtue of Elitzur's theorem~\cite{PhysRevD.12.3978} suggests a stronger robustness. Secondly, such perturbations now result in a doping with loop-like magnetic fluxes, which, given that the tension favors small loops, make accidental braiding processes through them unlikely. Thirdly, as local virtual constraints rule out independent breaking of certain symmetries, the growing number of loops may increase robustness against small perturbations under renormalization. As we comment further at the end of the manuscript, it is possible to demonstrate that this second representation is indeed stable to tensor perturbations \cite{WDSV}.

\bigskip \noindent
After reviewing the definition of the toric code and different parametrizations of its ground state subspace in sec.~\ref{sec:TC}, we provide in sec.~\ref{sec:2D} a thorough account of the tensor network approach to the (2+1)d toric code  in order to introduce the relevant notions and concepts, emphasizing in particular the manifestations of the self-duality. The tensor network analysis of the (3+1)d toric code is presented in sec.~\ref{sec:3D}, where both representations and the corresponding duality mapping are discussed in detail.

%% file: _ToricCode.tex
\section{Toric code\label{sec:TC}}
\noindent
\emph{In this section, we briefly present the toric code and review some of its most notable features. }

\subsection{Lattice Hamiltonian}

\noindent
Let $\Sigma$ be a closed $d$-dimensional surface endowed with a cellulation $\Sigma_\Upsilon$. We denote the plaquettes, edges and vertices of $\Sigma_\Upsilon$  by $\mathsf{p}$, $\mathsf e$ and $\mathsf v$, respectively. To every edge $\mathsf{e} \subset \Sigma_\Upsilon$, we assign a qubit degree of freedom such that the microscopic Hilbert space of the model reads
\begin{equation}
	\mathcal{H}[\Sigma_\Upsilon] := \bigotimes_{\mathsf{e} \subset \Sigma_{\Upsilon}} \mathbb C^2\; .
\end{equation}
The lattice Hamiltonian is defined in terms of two collections of operators. To every vertex $\mathsf{v} \subset \Sigma_\Upsilon$, and to every plaquette $\mathsf{p} \subset \Sigma_\Upsilon$,  we assign operators
\begin{equation}
	\mathbb A_{\mathsf v} :=  \bigotimes_{\mathsf{e} \supset \mathsf v} Z_{\mathsf e}\q  \text{and}\q
	\mathbb B_{\mathsf p} :=  \bigotimes_{\mathsf{e} \subset \partial \mathsf p} X_{\mathsf e}\; ,
\end{equation}
respectively, where $Z$ and $X$ are the usual Pauli matrices. All these operators commute so that the Hamiltonian defined as
\begin{equation}
	\mathbb H[\Sigma_\Upsilon] := - \sum_{\mathsf{v} \subset \Sigma_\Upsilon}
	\mathbb A_{\mathsf v} - \sum_{\mathsf{p} \subset \Sigma_\Upsilon} \mathbb B_{\mathsf p} 
\end{equation}
is exactly solvable. The ground state subspace is spanned by states $|\psi \ra \in \mathcal{H}[\Sigma_\Upsilon]$ satisfying the stabilizer constraints $\mathbb A_{\mathsf v}|\psi \ra = | \psi \ra$ and $\mathbb B_{\mathsf p}|\psi \ra = |\psi \ra$, at every $\mathsf v, \mathsf p \subset \Sigma_\Upsilon$. The Hamiltonian projector explicitly reads
\begin{equation}
	\label{eq:gsProj}
	\mathbb P_{\Sigma_\Upsilon}:=
	\prod_{\mathsf v \subset \Sigma_\Upsilon}\frac{1}{2}({\rm id}+\mathbb A_{\mathsf v})\prod_{\mathsf p \subset \Sigma_\Upsilon}\frac{1}{2}({\rm id} +\mathbb B_{\mathsf p}) \; .
\end{equation}
When $\Sigma$ is chosen to be the $d$-torus $\mathbb T^d$ and  $\Sigma_\Upsilon \equiv \mathbb T^d_\boxempty$ a $d$-cubic lattice with periodic boundary conditions, the resulting model is habitually referred to as the ($d$+1)-dimensional \emph{toric code} \cite{Kitaev1997}.

\subsection{Ground states and anyonic excitations}

\noindent
Let us focus for now on the (2+1)d toric code.
Denoting by $|0 \ra$ and $| 1 \ra$ the eigenvectors of $Z$ with eigenvalues $+1$ and $-1$, respectively, we consider the basis of $\mathcal{H}[\mathbbold T^2_\boxempty] $ obtained by assigning one such vector to every edge of $\mathbbold T^2_\boxempty$. Every vector in this basis is in one-to-one correspondence with a $ \mathbb Z_2$-valued singular 1-chain in $C_1(\mathbbold T^2_\boxempty, \mathbb Z_2)$, that is an assignment of $\mathbb Z_2$ group variables to every edge of the cellulation. The stabilizer constraints $\mathbb A_{\mathsf v}| \psi \ra = | \psi \ra$ impose that an even number of $| 0 \ra$ or $| 1 \ra$ states must meet at a every vertex so that chains of $| 0 \ra$ or $|1 \ra$ states must form closed loops. These define the group of singular 1-cycles $Z_1(\mathbbold T^2_\boxempty, \mathbb Z_2)$. Since $X | 0 \ra = | 1 \ra$ and $X |1 \ra = |0 \ra$, the second set of stabilizer constraints, viz. $\mathbb B_{\mathsf p}|\psi \ra = |\psi \ra$, enforce that 1-cycles are defined up to $\mathbb Z_2$-multiplication  by 1-boundaries in $B_1(\mathbbold T^2_\boxempty, \mathbb Z_2)$, i.e. 1-cycles that bound plaquettes of $\mathbbold T^2_\boxempty$. Putting everything together, we obtain the first homology group $H_1(\mathbbold T^2_\boxempty,\mathbb Z_2) = Z_1(\mathbbold T^2_\boxempty,\mathbb Z_2)/B_1(\mathbbold T^2_\boxempty,\mathbb Z_2)$ of $\mathbbold T^2_\boxempty$ over $\mathbb Z_2$. The ground state subspace of the toric code is thus spanned by functions from homology classes in $H_1(\mathbbold T^2_\boxempty,\mathbb Z_2)$ to $\mathbb C$. But $H_1(\mathbbold T^2_\boxempty,\mathbb Z_2) = \mathbb Z_2 \oplus \mathbb Z_2$, so that the ground state degeneracy of the (2+1)d toric code is $2^2=4$ \cite{Freedman:1998sw, Kitaev:2006lla}. Naturally the same computation could be carried out in the $X$ basis instead in terms of the eigenvectors $| + \ra$ and $|- \ra$. The same reasoning applies, at the difference that the $\mathbb Z_2$-valued singular chains are now defined along the edges of the dual cellulation $\mathbbold T^2_{\boxempty^\vee}$. 

Let us now consider the (3+1)d toric code and repeat the analysis above. Starting with the $Z$ basis of  $\mathcal{H}[\mathbb T^3_\boxempty]$, the derivation is identical to the (2+1)d case so that the ground state subspace is spanned by functions from $H_1(\mathbb T^3_\boxempty,\mathbb Z_2)$ to $\mathbb C$. We now have $H_1(\mathbb T^3_\boxempty,\mathbb Z_2) = \mathbb Z_2 \oplus \mathbb Z_2 \oplus \mathbb Z_2$, and thus the ground state degeneracy of the toric code in (3+1)d is $2^3=8$. Interestingly, the computation in the $X$ basis proceeds differently. Basis vectors of $\mathcal{H}[\mathbb T^3_\boxempty]$ are now in one-to-one correspondence with $\mathbb Z_2$-valued singular 2-chains in $C_2(\mathbb T^3_{\boxempty^\vee}, \mathbb Z_2)$, i.e. assignments of $\mathbb Z_2$-group variables to every plaquette of the dual cubic lattice. States satisfying the stabilizer constraints $\mathbb B_{\mathsf p} |\psi \ra = |\psi \ra$ correspond to closed membrane configurations, i.e. 2-cycles in $Z_2(\mathbb T^3_{\boxempty^\vee},\mathbb Z_2)$. Stabilizer constraints $\mathbb A_{\mathsf v} |\psi \ra = |\psi \ra$ then yield $H_2(\mathbb T^3_{\boxempty^\vee},\mathbb Z_2)$ so that the ground state subspace is spanned by functions from classes in $H_2(\mathbb T^3_{\boxempty^\vee},\mathbb Z_2)$ to $\mathbb C$. Equivalence between the two bases is ensured by the fact that $H_2(\mathbb T^3_{\boxempty^\vee},\mathbb Z_2) = H_1(\mathbb T^3_{\boxempty},\mathbb Z_2)$. Our goal is to explore these two parametrizations of the ground state sector from a tensor network viewpoint. 

\bigskip \noindent
Before discussing  tensor network representations of the ground state subspace, let us briefly recall the excitation content of the model. By definition, excited states are obtained by violating some of the stabilizer constraints. Violations of $\mathbb A_{\mathsf v}|\psi \ra = | \psi \ra$ are referred to as \emph{electric charges}, whereas violations of $\mathbb B_{\mathsf p}|\psi \ra = | \psi \ra$ are referred to as \emph{magnetic fluxes}. Given a 1-path $\gamma$ along the edges of the direct lattice, and a ($d$--1)-path $\sigma$ along the ($d$--1)-cells of the dual lattice, excitations can be created at the boundary of the following operators
\begin{equation}
	W^X(\gamma) = \prod_{\mathsf{e} \subset \gamma} X_{\mathsf e} 
	\q \text{and} \q
	W^Z(\sigma) = \prod_{\mathsf{e}^\vee \subset \sigma} Z_{\mathsf{e}} \; .
\end{equation}
These definitions are valid both in (2+1)d and (3+1)d. Although both types of operators are \emph{string-like} in (2+1)d, they are string-like and membrane-like in (3+1)d, respectively. This naturally relates to the two parametrizations of the ground state subspace we discussed previously. Correspondingly, in (3+1)d, electric charges are point-like, whereas magnetic fluxes are loop-like.

%% file: _PEPS2D.tex
\section{Tensor network representations in (2+1)d\label{sec:2D}}

\noindent
\emph{In order to introduce the formalism and review basic notions, we  present in this section the tensor network representations of the toric code in (2+1)d.}

\subsection{Basic definitions}

\noindent
The tensor network approach consists in expressing the ground state wave functions in terms of so-called \emph{Projected Entangled-Pair States} (PEPSs). The merit of this formulation then stems from the fact that the global or topological behavior of the model, here the (2+1)d toric code, can be encoded into the local symmetries of a single PEPS tensor.

Let us start with some basic definitions. Given a set $\{A^i\}_{i=1,\ldots,n}$ of $
(\chi \times \chi)$-matrices, we call a \emph{Matrix Product State} (MPS) an element in $(\mathbb C^n)^{\otimes L}$ of the form
\begin{equation*}
	|\mpssf (A) \ra := \sum_{i_1,\ldots,i_L}{\rm tr}[A^{i_1}\cdots A^{i_L}] \,
	(| i_1 \ra \otimes \cdots \otimes | i_L \ra) \, ,
\end{equation*}
where $| i \ra$, $i \in 1 , \dots, n$, is an orthonormal basis of $\mathbb C^n$. For conciseness, we shall often use the notation $| i_1, \ldots, i_L\ra := |i_1 \ra \otimes \cdots \otimes | i_L \ra$. The integer $\chi$ is referred to as the \emph{bond dimension} of the MPS. It is convenient to define the rank-three tensor $A \in \mathbb C^n \otimes (\mathbb C^\chi)^{\otimes 2}$ as
\begin{equation*}
	A := \sum_{i = 1, \ldots,n}A^i \otimes | i \ra = \!\! \sum_{\substack{i=1,\ldots, n \\ a_1,a_2 = 1,\ldots, \chi}}
	\!\! (A^{i})_{a_1a_2}|a_1 \ra \la a_2 |  \otimes |i\ra  \, ,
\end{equation*}
where we distinguish the \emph{virtual} indices $a_1$ and $a_2$ from the \emph{physical} index $i$.
By definition the network underlying the definition of $|\mpssf(A) \ra$ is one-dimensional. Naturally, this definition can be generalized to two-dimensional networks. Given for instance a set $\{B^i\}_{i=1,\ldots,d}$ of $(\chi^{\times 4})$-tensors, we write the rank-five tensor $B \in \mathbb C^n \otimes (\mathbb C^\chi)^{\otimes 4}$ as
\begin{equation*}
	B := \!\! \sum_{\substack{i=1,\ldots,n \\ a_1,\ldots,a_4= 1,\ldots,\chi}}
	\!\! (B^i)_{a_1 a_2 a_3 a_4}|a_1,a_2 \ra \la a_3,a_4 |  \otimes |i \ra \, ,
\end{equation*} 
and depict it graphically as
\begin{equation*}
	B \equiv \!\! \sum_{\substack{i=1,\ldots,n \\ a_1,\ldots,a_4= 1,\ldots,\chi}} \!\!\! 
	\includeTikz{0}{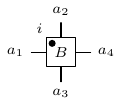}{\PEPS{0.5}} \, ,
\end{equation*}
where the \emph{black} dot, which is labelled by the physical index, stands for an additional leg sticking out of the paper plane towards the reader. Note that in the situation where the black dot is labelled by a physical index that can be written as a tensor product, we shall often write as many black dots accordingly.
We then call a \emph{Projected Entangled-Pair State} (PEPS) an element in $(\mathbb C^n)^{\otimes (L^2)}$ of the form
\begin{equation*}
	| \pepssf(B)\ra := 
	\includeTikz{0}{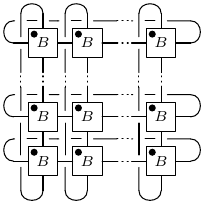}{\PEPSNet{0.5}} \, ,
\end{equation*}
where connected legs translate into tensor contractions along the corresponding indices. Generalizations, where the underlying tensor network is not a square lattice or higher-dimensional, function similarly.

In the following, we shall also need so-called \emph{Projected Entangled-Pair Operators} (PEPOs) that are the operator analogues to PEPSs. These are defined in terms of tensors of the form
\begin{equation*}
	C := \!\!\! \sum_{\substack{i,j=1,\ldots,n \\ a_1,\ldots,a_4= 1,\ldots,\chi}}
	\!\!\! (C^{ij})_{a_1 a_2 a_3 a_4}|a_1,a_2 \ra \la a_3,a_4 |  \otimes |i \ra \la j | \, .
\end{equation*} 
Graphically, we shall make use of the notation 
\begin{equation*}
	C \equiv \!\!\! \sum_{\substack{i,j=1,\ldots,n \\ a_1,a_2,a_3,a_4=1,\ldots,\chi}} \!\!\! 
	\includeTikz{0}{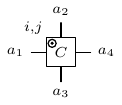}{\PEPO{0.5}} \, ,
\end{equation*}
where the symbol $\begin{tikzpicture}[scale=0.6, baseline=-0.3em] \ddott{0,0} \end{tikzpicture}$, labelled by the physical indices $i$ and $j$, stands for two additional legs sticking out of the paper plane towards and away from the reader, respectively.

\subsection{PEPS building blocks\label{sec:gsTwoD}}

\noindent
Let us now derive the PEPS representations of the (2+1)d toric code. More specifically, we shall distinguish two representations that correspond to the $Z$ and $X$ bases of the ground state subspace discussed in sec.~\ref{sec:TC}. Since we have access to the Hamiltonian projector \eqref{eq:gsProj}, ground states can be simply found by projecting any 1-chains in $C_1(\mathbbold T^2_\boxempty , \mathbb Z_2)$ onto the ground state subspace. In practice, we find it more convenient to first pick a state that manifestly satisfies one of the two sets of stabiliser constraints, and then project with respect to the second set of constraints. In other words, we pick a state that corresponds to either a 1-cycle in $Z_1(\mathbbold T^2_\boxempty, \mathbb Z_2)$ or a 1-cycle in $Z_1(\mathbbold T^2_{\boxempty^\vee}, \mathbb Z_2)$ and then apply the relevant part of the Hamiltonian projector, e.g.
\begin{align}
	\label{eq:polar1}
	| \psi \ra_{\rm g.s.} &= \prod_{\mathsf v \subset \mathbbold T^2_\boxempty} \frac{1}{2}({\rm id}+ \mathbb A_{\mathsf v}) | + ,+ , \ldots ,+ \ra
	\\
	\label{eq:polar2}
	| \psi \ra_{\rm g.s.} &= \prod_{\mathsf p \subset \mathbbold T^2_\boxempty} \frac{1}{2}({\rm id}+ \mathbb B_{\mathsf p}) | 0 ,0 , \ldots ,0 \ra \; .
\end{align}
Let us now derive the two tensor network representations that correspond to the two choices illustrated above. In order to do so, we first need to express the components of the Hamiltonian projector in terms of PEPOs. We shall describe the latter case in detail and deduce the former one by analogy.

Given a plaquette $\mathsf{p} \subset \mathbbold T^2_\boxempty$, the projector $\tfrac{1}{2}({\rm id} + \mathbb B_{\mathsf{p}})$ can be written as the contraction of two PEPO tensors as
\begin{equation*}
	\includeTikz{0}{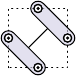}{\PEPOB{0.5}}
	\;\;\; \text{with} \;\;\; 
	\includeTikz{8}{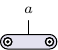}{\PEPOBHalf{0.5}}
	\equiv \frac{1}{2^\frac{1}{2}} \! \begin{cases} \, \mathbbm 1 \otimes \mathbbm 1 \!\!\! &\text{if $| a\ra=|0\ra$} \\ X \otimes X \!\!\! &\text{if $| a\ra=| 1\ra$}\end{cases}  ,
\end{equation*}
where the dotted square represents the plaquette $\mathsf{p}$. This operator needs to be applied to every plaquette of $\mathbbold T^2_\boxempty$. We choose to do so according the following pattern
\begin{equation}
	\label{eq:toricProj}
	\includeTikz{0}{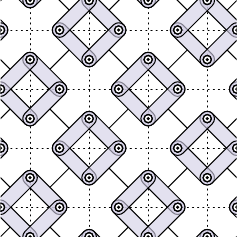}{\toricProj{0.5}} \, ,
\end{equation}
where it is understood that two PEPO tensors are contracted along one of their physical indices at every qubit location. Acting with such operators on the kinematical state $|0,0,\ldots,0\ra$ yields a ground state wave function. We are interested in the tensor $T^X_{\rm 2d}$ that results from blocking four PEPO tensors around a vertex $\mathsf{v} \subset \mathbbold T^2_\boxempty$, and acting on the state $|0,0,0,0\ra$, i.e.\footnote{The symbol $\triangleright$ indicates that the operator acts (from the left) on the microscopic state via matrix multiplication along the physical indices.}
\begin{equation*}
	\includeTikz{0}{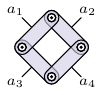}{\toricProjLocB{0.5}{1}}
	\, \triangleright \, 
	\includeTikz{0}{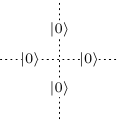}{\toricProjLocB{0.5}{2}} =:	
	\includeTikz{0}{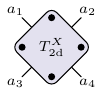}{\toricProjLocB{0.5}{3}} \; .
\end{equation*}
Going through the derivation, we find that
\begin{align}
	\nn
	T^X_{\rm 2d} &\propto  \sum_{\{a=0,1\}}  | a_1,a_2 \ra \la a_3,a_4 | 
	\\[-1em] \nn
	&\hspace{4.2em}  \otimes | a_1+a_2, a_2+a_4,a_1+a_3,a_3+a_4 \ra 
	\\
	\label{eq:expDef2DOne}
	&\equiv \sum_{\{a=0,1\}}
	\includeTikz{0}{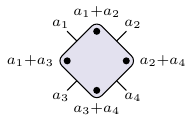}{\toricProjLocB{0.5}{4}}\; .
\end{align}
As per the nomenclature introduced previously, the indices labelled by $a_1,\ldots,a_4$ in equation \eqref{eq:expDef2DOne} are referred to as the virtual indices, while the remaining four represented as black dots are the physical indices.
Putting everything together, we obtain that $| \pepssf(T^X_{\rm 2d})\ra$ is a tensor network description of a ground state of the (2+1)d toric code. This can be confirmed graphically using the following properties
\begin{align}
	\label{eq:propHalfPEPOB}
	\includeTikz{7}{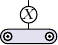}{\PEPOBHalfProp{0.5}{1}} = 
	\includeTikz{7}{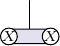}{\PEPOBHalfProp{0.5}{2}} \q , \q \includeTikz{7}{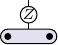}{\PEPOBHalfProp{0.5}{3}}  = 
	\includeTikz{7}{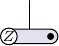}{\PEPOBHalfProp{0.5}{4}} = 
	\includeTikz{7}{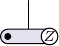}{\PEPOBHalfProp{0.5}{5}} \, .
\end{align}
Let us now investigate some properties of this PEPS tensor. First of all, let us introduce a lighter notation for $T^X_{\rm 2d}$:
\begin{equation}
	\label{eq:notationTwoDX}
	\includeTikz{0}{toricProjLocB3}{\toricProjLocB{0.5}{3}} \;\; \equiv \;\; \includeTikz{0}{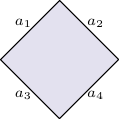}{\toricProjLocB{0.5}{5}} \, ,
\end{equation}
where physical indices are now omitted. Moreover, virtual indices are now supported on edges of the square. Inspecting the definition of $T^X_{\rm 2d}$, we notice immediately that it satisfies a $\mathbb Z_2$-symmetry condition, which in terms of the notation we have just introduced reads:
\begin{equation}
	\label{eq:pull2D}
	\includeTikz{0}{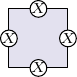}{\SymTwoDPepsOne{0.5}{2}} \; = \;\; 
	\includeTikz{0}{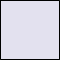}{\SymTwoDPepsOne{0.5}{1}}
	\;\; \Leftrightarrow \; 
	\includeTikz{0}{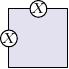}{\SymTwoDPepsOne{0.5}{3}} \; = \;\; 
	\includeTikz{0}{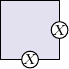}{\SymTwoDPepsOne{0.5}{4}} \, .
\end{equation}
Note that we rotated and rescaled the diagrammatic representation of $T^X_{\rm 2d}$ in the equation above for convenience.
The symmetry condition presented above is the only one, and thus it \emph{fully} specifies the support of the tensor on the virtual degrees of freedom.
Moreover, let $\mathcal{P}(T^X_{\rm 2d})$ be the tensor that maps the virtual indices of $T^X_{\rm 2d}$ to its physical ones, i.e. the tensor $T^X_{\rm 2d}$ reshaped as a matrix whose indices are the tensor products of all the virtual and physical indices of $T^X_{\rm 2d}$, respectively. Given the definition \eqref{eq:expDef2DOne}, it reads
\begin{equation*}
	\mathcal{P}(T^X_{\rm 2d}) \propto
	\!\! \sum_{\{a=0,1\}} \!\!  | a_1+a_2, a_2+a_4,a_1+a_3,a_3+a_4 \ra  \la  \{a_n\}_{n=1}^4 |  \, ,
\end{equation*}
where $\la \{a_n\}_{n=1}^4| \equiv \la a_1, \ldots, a_4|$. It follows from the definition and a shift of summation variables that 
\begin{align}
	\nn
	\mathbb E(T^X_{\rm 2d}) &:= \mathcal{P}({T^X_{\rm 2d}}^\intercal) \mathcal{P}(T^X_{\rm 2d}) 
	\\
	\nn
	&\propto
	\sum_{\substack{\{a=0,1\} \\ \{b=0,1\}}} \delta_{a_1+a_2,b_1+b_2} \, \delta_{a_2+a_4,b_2+b_4} \, \delta_{a_1+a_3,b_1+b_3}
	\\[-1.6em]
	\nn
	&\hspace{5em} \times | \{b_m\}_{m=1}^4 \ra \la \{a_n\}_{n=1}^4 |
	\\
	\nn
	&\propto \sum_{\substack{b=0,1 \\ \{a=0,1\}}} \!\! |a_1+b,a_2+b,a_3+b,a_4+b \ra \la \{a_n\}_{n=1}^4 |
	\\
	\label{eq:onSite2D}
	&\propto \frac{1}{2}
	(\mathbbm 1^{\otimes 4} + X^{\otimes 4}) \, ,
\end{align}
so that $\mathbb E(T^X_{\rm 2d})$ performs a group averaging, and hence defines a projector onto the $\mathbb Z_2$-invariant subspace. It follows that when restricting to the $\mathbb Z_2$-invariant subspace, $\mathcal{P}({T^X_{\rm 2d}}^\intercal)$ is the unitary left-inverse of $\mathcal{P}(T^X_{\rm 2d})$. By applying this unitary inverse to the physical indices of $T^X_{\rm 2d}$, we obtain the following isomorphism\footnote{Although we use the same notation for \eqref{eq:onSite2D} and \eqref{eq:isoTwoDx}, these are strictly speaking distinct objects. Whereas the matrices appearing in the definition of the former tensor are solely between virtual states, the ones appearing in the definition of the latter  are between virtual and physical states. Nevertheless, physical and virtual vector spaces happen to be isomorphic, which justifies our notation.}
\begin{align}
	T^X_{\rm 2d} \simeq 
	\label{eq:isoTwoDx}
	\frac{1}{2}(\mathbbm 1^{\otimes 4} + X^{\otimes 4}) = \frac{1}{2}\sum_{a=0,1}(X^a)^{\otimes 4} \, ,
\end{align}
which indicates that the tensor $T^X_{\rm 2d}$ is itself a \emph{projector} onto the $\mathbb Z_2$-invariant subspace. Crucially, this isomorphism renders explicit the fact that there is a one-to-one correspondence between physical and virtual degrees of freedom up to a $\mathbb Z_2$-symmetry. This implies that it is enough to focus on operations at the virtual level, hence justifying the notation \eqref{eq:notationTwoDX}. In the following, it will be very convenient to use the isomorphism \eqref{eq:isoTwoDx} in order to carry out computations.

\bigskip \noindent
The other tensor network representation, i.e. associated with \eqref{eq:polar1}, is derived following exactly the same procedure. Given a vertex $\mathsf{v} \subset \mathbbold T^2_\boxempty$, the projector $\frac{1}{2}({\rm id}+\mathbb A_\mathsf{v})$ can be written as the contraction of two PEPO tensors as
\begin{equation*}
	\includeTikz{0}{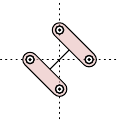}{\PEPOA{0.5}}
	\;\; \text{with} \;\;\; 
	\includeTikz{8}{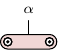}{\PEPOAHalf{0.5}} 
	\equiv \frac{1}{2^\frac{1}{2}} \! \begin{cases}\, \mathbbm 1 \otimes \mathbbm 1 \!\!\! &\text{if $| \alpha\ra=|+\ra$} \\ Z \otimes Z \!\!\! &\text{if $| \alpha\ra=| -\ra$}\end{cases}  .
\end{equation*}
The PEPS tensor then results from blocking four such PEPO tensors around a plaquette $\mathsf p \subset \mathbbold T^2_\boxempty$, and acting on the state $|+,+,+,+\ra$, i.e.
\begin{equation*}
	\includeTikz{0}{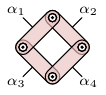}{\toricProjLocA{0.5}{1}}
	\, \triangleright \, 
	\includeTikz{0}{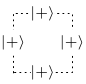}{\toricProjLocA{0.5}{2}} \, =: \, 	
	\includeTikz{0}{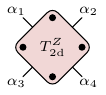}{\toricProjLocA{0.5}{3}} \, ,
\end{equation*}
so that
\begin{equation*}
	T^Z_{\rm 2d} \propto \! \sum_{\{\alpha = \pm\}} \! 
	| \alpha_1,\alpha_2 \ra \la \alpha_3, \alpha_4 |  \otimes | \alpha_1\alpha_2, \alpha_2 \alpha_4, \alpha_1 \alpha_3, \alpha_3 \alpha_4 \ra 
	\, .
\end{equation*}
We can confirm that this tensor network state is in the ground state sector of the model using the following properties
	\begin{align}
	\label{eq:propHalfPEPOA}
	\includeTikz{7}{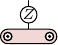}{\PEPOAHalfProp{0.5}{1}} = 
	\includeTikz{7}{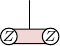}{\PEPOAHalfProp{0.5}{2}} \q , \q 
	\includeTikz{7}{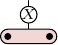}{\PEPOAHalfProp{0.5}{3}}  = 
	\includeTikz{7}{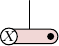}{\PEPOAHalfProp{0.5}{4}}= 
	\includeTikz{7}{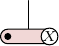}{\PEPOAHalfProp{0.5}{5}} \, .
	\end{align}
As before, we introduce a lighter notation for $T^Z_{\rm 2d}$:
\begin{equation}
	\includeTikz{0}{toricProjLocA3}{\toricProjLocA{0.5}{3}} \;\; \equiv \;\;  
	\includeTikz{0}{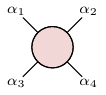}{\toricProjLocA{0.5}{5}} \, ,
\end{equation}
where physical indices are omitted. Note that this notation purposefully differs from the one of $T^X_{\rm 2d}$. Although, this distinction might seem a little bit artificial at this point, it will turn out to be very useful when generalizing to (3+1)d. Inspecting the definition of $T^Z_{\rm 2d}$, we find that it satisfies the same symmetry condition as $T^X_{\rm 2d}$ but with respect to Pauli-$Z$ operators, i.e.
\begin{equation*}
	\includeTikz{0}{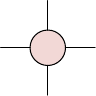}{\SymTwoDPepsTwo{0.5}{2}} \, = \, 
	\includeTikz{0}{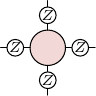}{\SymTwoDPepsTwo{0.5}{1}} 
	\; \Leftrightarrow \; 
	\includeTikz{0}{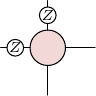}{\SymTwoDPepsTwo{0.5}{3}} \, = \, 
	\includeTikz{0}{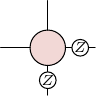}{\SymTwoDPepsTwo{0.5}{4}} \; .
\end{equation*}
Finally, repeating the previous analysis yields the isomorphism
\begin{equation*}
T_{\rm 2d}^Z \simeq \frac{1}{2} (\mathbbm 1 ^{\otimes 4} + Z^{\otimes 4}) \, ,
\end{equation*}
so that $T^Z_{\rm 2d}$ is itself a projector onto the $\mathbb Z_2$-invariant subspace.

\subsection{Ground state subspace and excitations\label{sec:gsExc2D}}
\noindent
We derived above a PEPS tensor $T^X_{\rm 2d}$ such that $| \pepssf(T^X_{\rm 2d}) \ra$ defines a ground state of the (2+1)d toric code. Since this tensor has the geometry of a square such that virtual indices are supported on edges, contraction along a virtual index is now performed for two squares sharing a common edge, and thus $| \pepssf(T^X_{\rm 2d}) \ra$ has the geometry of a \emph{square tesselation}.

As mentioned above, the tensor $T^X_{\rm 2d}$ satisfies a $\mathbb Z_2$-symmetry condition under $X^{\otimes 4}$.
Such symmetry along the virtual indices of the tensor is a crucial feature of the topological order. Indeed, it is immediate to check that this $\mathbb Z_2$-invariance is stable under concatenation of several such tensors:
\begin{equation*}
	\includeTikz{0}{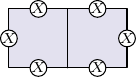}{\SymTwoDPepsOne{0.5}{5}} = 
	\includeTikz{0}{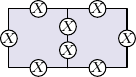}{\SymTwoDPepsOne{0.5}{6}} \stackrel{\eqref{eq:pull2D}}{=} 
	\includeTikz{0}{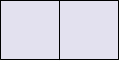}{\SymTwoDPepsOne{0.5}{7}} \, ,
\end{equation*} 
where we used the fact that $X^2 = \mathbbm 1$, and thus any loop of $X$ operators along a contractible 1-cycle of the manifold, i.e. 1-boundaries, can be removed. Similarly, any string of $X$ operators can be pulled through the PEPS tensors, as illustrated on the r.h.s of \eqref{eq:pull2D}, making the bulk of such strings locally undetectable. Interestingly, a closed loop of $X$ operators along any of the non-contractible 1-cycles does have a \emph{global} effect, namely to introduce a non-trivial flux in the wave function, hence allowing to map one ground state to another. This yields yet another way of phrasing that the ground state degeneracy of the model is $2^2$. 

Although the symmetry condition \eqref{eq:pull2D} is a salient feature of the topological order, it is not a \emph{sufficient} condition. Indeed, we can easily engineer tensors that display such symmetry conditions, but these do not necessarily parametrize the ground state of the same topological model.   This implies in particular that by simply fixing a symmetry condition for the PEPS tensors, we do not fully specify a topological phase, and thus we can \emph{a priori} explore different topological phases and the corresponding phase transitions. That being said, the fact that the tensor is itself a projector onto the $\mathbb Z_2$ invariant subspace, as seen from \eqref{eq:isoTwoDx}, is \emph{characteristic} of the topological order.

Crucially, anyonic (physical) excitations can also be studied in terms of virtual operators. Pairs of point-like electric excitations labelled by $e$ can be conveniently created at the endpoints of an open string of $X$ operators along the virtual indices of the PEPS tensors. As mentioned above, away from its endpoints, such string can be freely deformed by pulling it through the tensors. Pairs of point-like magnetic excitations labelled by $m$ can be created by inserting a pair of $Z$ operators at virtual indices of the network, which violate the local symmetries of the tensors sharing this virtual index:\footnote{Properties \eqref{eq:propHalfPEPOB} may be used to confirm that these virtual operators indeed create the expected physical excitations.}
\begin{equation}
	\includeTikz{-3}{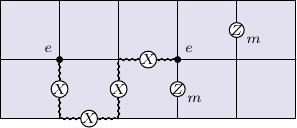}{\TwoDAnyons{0.5}} 
	\; .
\end{equation}
As before, we rotated and rescaled the diagrammatic depiction of the tensor network corresponding to \eqref{eq:toricProj} for convenience.
Notice that there is no `string' connecting the $Z$ operators, as such we could be tempted to consider a single magnetic excitation by inserting a single $Z$ operator. However, there is no operation on the physical indices that would induce such a configuration. This follows from the fact that a single plaquette operator is equal to the product of all the plaquette operators but the one considered. 
This PEPS formulation of the excitations can be used to recover the fusion rules of the model, namely $e \times e = m \times m = \epsilon \times \epsilon = 1$, $e \times m = \epsilon$, $e \times \epsilon = m $ and $m \times \epsilon = e$, 
where $\epsilon$ denotes the fermionic composite excitation.
Furthermore, both types of excitations can be moved over the network via \emph{local unitary operators} \cite{SCHUCH20102153}. 

As reviewed below, the tensor $T^X_{\rm 2d}$ encodes the renormalization group fixed point of the phase. Importantly, the virtual operators defined above still describe excitations when deforming the tensor away from the fixed point via an invertible physical operation. In the following, this will enable us to construct order parameters for detecting phase transitions.

Before pursuing our analysis, let us mention that the other tensor network representation behaves identically except that a pair of electric charges is now obtained by inserting of pair of $X$ operators, whereas a pair of magnetic fluxes can be created at the endpoints of an open string of $Z$ operators across the virtual indices. 
These two tensor network representations are related via a duality map, which we could write explicitly in terms of Hadamard matrices. In sec.~\ref{sec:mixed2D}, we shall discuss how we can define mixed tensor network representations, involving both $T^X_{\rm 2d}$ and $T^Z_{\rm 2d}$.

\subsection{Renormalization group fixed point\label{sec:RG2D}}
\noindent
Generally, \emph{gapped} phases of matter can be defined as equivalence classes of physical systems displaying a spectral gap above the ground state that persists in the thermodynamic limit. It is believed---although not proven---that such gapped phases admit in the infra-red limit an effective description in terms of \emph{topological quantum field theories}. In this terminology, the toric code constitutes a concrete realization of a gapped phase described in the infra-red limit by $\mathbb Z_2$-BF theory. 

Such definition implies that two gapped ground states belong to the same phase if they can be related by an adiabatic evolution that does not close the energy gap. More practically, such evolution may be implemented by means of \emph{local unitary transformations}. But these local transformations can typically be arranged in order to implement a (real space) \emph{renormalization} group flow, so that the task of finding equivalence classes of states with respect to these transformations boils down to finding fixed points of the corresponding flow. It is expected that such fixed points capture the long-range entanglement of the system, which is a defining feature of (intrinsic) topological order \cite{Chen:2010gda}. We shall now illustrate this remark by implementing a renormalization group flow at the level of the tensor network, and confirm that the PEPS tensor indeed describe fixed points of such a flow.

Let us investigate the renormalization of $|\pepssf(T^X_{\rm 2d})\ra$. Since we are dealing with a homogeneous tensor network, it is enough to understand the renormalization of two neighboring tensors. Let us consider the following concatenation of tensors:
\begin{equation}
	\label{eq:concat2D}
	T^X_{\rm 2d}T^X_{\rm 2d} \equiv 
	\includeTikz{0}{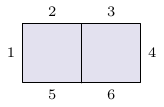}{\TwoDFlow{0.5}{0}} \, ,
\end{equation}
where the numbers identify the different edges/bonds. As mentioned above, this block of two tensors still satisfies the $\mathbb Z_2$-symmetry condition. Crucially, the isomorphism property \eqref{eq:isoTwoDx} is also stable under concatenation. This can be shown by following the same approach as before. Applying the map $\mathcal{P}((T_{\rm 2d}^XT_{\rm 2d}^X)^\intercal)$ to $T_{\rm 2d}^XT_{\rm 2d}^X$ and proceeding as in \eqref{eq:onSite2D}, we find
\begin{align}
	\nn
	T_{\rm 2d}^XT^X_{\rm 2d} &\simeq \frac{1}{4} \sum_{a_1,a_2=0,1}
	(X^{a_1})^{\otimes 3} \, {\rm tr}(X^{a_1}X^{a_2}) \,  (X^{a_2})^{\otimes 3}
	\\
	\label{eq:isoRG2D}
	&= \frac{1}{2}(\mathbbm 1^{\otimes 6} + X^{\otimes 6}) \, ,
\end{align} 
where we used the fact that ${\rm tr}(X^{a_1}X^{a_2}) = 2\delta_{a_1,a_2}$.
Using this isomorphism, let us now demonstrate that the tensor $T^X_{\rm 2d}$ is a fixed point of the renormalization group flow. In order to prove this statement, we need to show $(i)$ that the tensor $T^X_{\rm 2d}T^X_{\rm 2d}$ is equal, up to isomorphisms, to a tensor product between $T^X_{\rm 2d}$ and another term, and $(ii)$ that the second term describes purely short-range correlations so that it can be discarded as we focus on non-local properties.

In light of the isomorphism \eqref{eq:isoRG2D}, we understand that it is enough to know how to perform the desired factorization at the level of the symmetry operators. Given two neighboring bonds, we have $X^a \otimes X^a \simeq X^a \otimes \mathbbm 1$. This isomorphism can be physically implemented via the controlled NOT gate $\cX$
\begin{equation*}
	{\rm c}X_{\mathsf e_1,\mathsf e_2} : |a_1,a_2 \ra_{\mathsf e_1,\mathsf e_2} \equiv |a \ra_{\mathsf e_1} \otimes |a_2\ra_{\mathsf e_2} \mapsto | a_1 , a_1+a_2 \ra_{\mathsf e_1,\mathsf e_2} \, ,
\end{equation*}
such that 
\begin{equation}
	\label{eq:cXaction}
	{\rm c}X^{\dagger}(X^a \otimes X^a)\, {\rm c}X = X^a \otimes \mathbbm 1 \; ,
\end{equation} 
where $\mathsf e_1$ and $\mathsf e_2$ refer to the bonds/edges along which it is applied.
Considering \eqref{eq:concat2D}, we apply $\cX_{2,3}$ to the pair of physical indices $\text{\small {2,3}}$ and $(\cX_{2,3})^\dagger$ to the corresponding pair of indices belonging to the upper neighbor. Similarly, we apply $\cX_{5,6}$ and $(\cX_{5,6})^\dagger$ to the tensor $T^X_{\rm 2d}T^X_{\rm 2d}$ and its lower neighbor, respectively. By further inserting the resolutions of the identity ${\rm c}X_{2,3}({\rm c}X_{2,3})^\dagger$ and ${\rm c}X_{5,6}({\rm c}X_{5,6})^\dagger$ along the virtual indices between $T^X_{\rm 2d}T^X_{\rm 2d}$ and its neighbors, we can apply equation \eqref{eq:cXaction} twice.
We obtain that the  controlled NOT gates act as \emph{disentanglers},
which in virtue of \eqref{eq:isoRG2D} induce the isomorphism
\begin{align}
	\nn
	T_{\rm 2d}^X T^X_{\rm 2d} &\simeq T_{\rm 2d}^X \otimes \Big( \sum_{a_1,a_2=0,1} |a_1\ra \la a_1|  \otimes |a_2\ra \la a_2| \Big) 
\end{align}
where the second term on the r.h.s is a tensor product of tensors that map one-to-one the physical system to the virtual one, representing maximally entangled states at the interface between neighboring PEPS tensors.
Introducing the graphical notation
\begin{equation}
	\includeTikz{0}{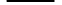}{\TwoDFlow{0.5}{5}} \equiv \sum_{a=0,1}|a \ra \la a | \, ,
\end{equation}
we can rewrite the previous isomorphism as
\begin{equation}
	T^X_{\rm 2d}T^X_{\rm 2d} \simeq 
	\;\; 
	\includeTikz{0}{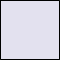}{\TwoDFlow{0.5}{2}} \; \otimes \; 
	\includeTikz{0}{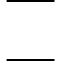}{\TwoDFlow{0.5}{4}} \, .
\end{equation}
Similarly, we find that the blocking of four neighboring $T^X_{\rm 2d}$ tensors satisfies 
\begin{equation}
	\label{eq:flow2D}
	\includeTikz{0}{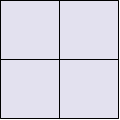}{\TwoDFlow{0.5}{1}} \;\; \simeq \;\; \includeTikz{0}{TwoDflow2}{\TwoDFlow{0.5}{2}} \; \otimes \; 
	\includeTikz{0}{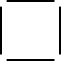}{\TwoDFlow{0.5}{3}} \, .
\end{equation}
Since we are only interested in the \emph{long-range} information, the short-range maximally entangled states can be safely discarded, hence confirming that the PEPS tensors do appear as fixed points of the renormalization group flow. The same reasoning would apply with the other tensor network representation, but the physical operations would be expressed in terms of the controlled gates $(H \otimes H)^\dagger \, {\rm c}X \,(H \otimes H)$, with $H$ the \emph{Hadamard} matrix.

Before concluding this part, let us comment briefly about the entanglement entropy. Given a connected and topologically trivial region $\Omega$ of the tensor network, the corresponding reduced density matrix has a degenerate spectrum. Moreover, we count $|\mathbb Z_2|^{|\partial \Omega| -1}$ dominant singular values, where $|\partial \Omega|$ is the length of the boundary, as defined by the number of tensor bonds along the boundary. The `$-1$' in the previous formula amounts to the overall $\mathbb Z_2$-symmetry condition of $\Omega$. Considering the region defined by the configuration on the left-hand side of \eqref{eq:flow2D}, we know that the corresponding entanglement entropy is equal to $7{\rm log}(2)$. Since each  maximally entangled state on the right-hand side contributes by a factor ${\rm log}(2)$, we check that the isomorphism preserves the entanglement entropy as expected.

\subsection{Transfer operator and edge physics}

\noindent
Let us now briefly review how the topological order of the (2+1)d toric code can be analysed from the point of view of the \emph{transfer operator} \cite{PhysRevLett.111.090501,haegeman2015shadows}, which turns out to be a precious tool when performing numerical simulations. Let us first consider $N \times N'$ copies of the PEPS tensor $T^X_{\rm 2d}$ and contract them on a cylinder $\mathbbold S^1 \times \mathbbold I$ with $\mathbbold I \equiv [0,1]$ such that the cylinder  has circumference $N$ and length $N'$. The resulting tensor possesses two sets of open virtual indices, at which we can assign boundary conditions $| \psi_L \ra$ and $| \psi_R \ra$. There is a natural action of $X^{\otimes N}$ on both sides of the cylinder. Furthermore, it follows from the symmetry condition of $T^X_{\rm 2d}$ that the tensor network on the cylinder remains invariant under $X^{\otimes N} \otimes X^{\otimes N}$.

Recall that the four degenerate ground states on the torus can be obtained by inserting strings of $X^g$ operators, $g \in \{0,1\}$, along both non-contractible 1-cycles. An alternative basis of the ground state subspace is obtained by inserting a string of $X^g$ operators along one of the non-contractible 1-cycles, and a string of $\mathbb P_{\rho}:= \frac{1}{2}(\mathbbm 1 + \rho X)$ operators, $\rho = \pm 1$, along the other one. The operators $\mathbb P_{+}$ and $\mathbb P_{-}$ correspond to the projectors onto the even and odd sectors, respectively. Going back to the cylinder, we distinguish  four (minimally-entangled) topological sectors, which we identify in a similar way. On the one hand, we have the possibility of inserting a string of $X^g$ operators along the cylinder. On the other hand, we have the possibility of inserting a string of $\mathbb P_{\rho}$ operators around the cylinder. But using the symmetry conditions of the individual PEPS tensors, such a string can be moved to the boundary, at which point it can act on one of the boundary conditions, projecting it to the even or the odd sector. The symmetry under $X^{\otimes N} \otimes X^{\otimes N}$ then enforces that both boundary conditions must be in the same sector.  Putting everything together, we obtain that the topological sectors on the cylinder are identified by the possibility of inserting a string of $X^g$ operators along the cylinder and a choice of boundary conditions identified by an irreducible representation $\rho = \pm 1 : a \mapsto (\pm 1)^a$ of $\mathbb Z_2$. These topological sectors correspond to the four quasi-particles of the model and can be depicted as 
\begin{equation*}
	\la \psi_\rho | \!\!\!\!\!\!\!\! 
	\includeTikz{0}{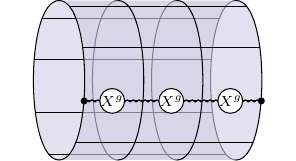}{\transferTwoD{0.5}} \!\!\!\!\!\!\!\! | \psi_\rho \ra \, .
\end{equation*}
These are by construction in one-to-one correspondence with the four degenerate ground states of the model on the torus. Recall that these four quasi-particles can be created by acting on the ground state subspace with a string of $X^g$ operators along a path $\gamma$ with endpoints  $Z_\rho \equiv |0\ra\la 0|+\rho|1\ra\la1|$ that transform like an irreducible representation  of $\mathbb Z_2$. We denote by $\mathcal R_{g,\rho}(\gamma)$ the combined operator and by $|g,\rho \ra \equiv | \mathcal R_{g,\rho}(\gamma)(\pepssf)\ra$ the corresponding excited state.

Let us now consider the eight-valent tensor $\mathbb E(T_{\rm 2d}^X)$ defined in \eqref{eq:onSite2D}. The symmetry of the tensor $T_{\rm 2d}^X$ induces the same $\mathbb Z_2$-symmetry independently in the `bra' and `ket' layers of $\mathbb E(T_{\rm 2d}^X)$. We consider $N$ copies of $\mathbb E(T_{\rm 2d}^X)$ and contract them on a ring. The resulting matrix is the so-called \emph{transfer operator} denoted by $\mathbb T(T_{\rm 2d}^X)$. The symmetry conditions on $\mathbb E (T_{\rm 2d}^X)$ implies $[\mathbb T(T^X_{\rm 2d}), X^{\otimes N} \otimesbk \mathbbm 1 ^{\otimes N}] = [\mathbb T(T^X_{\rm 2d}), \mathbbm 1^{\otimes N} \otimesbk  X^{\otimes N}] = 0$, 
where we introduced the notation $\otimesbk$ to indicate that the tensor product is with respect to  the bra and ket layers of $\mathbb T(T_{\rm 2d}^X)$. Therefore, $\mathbb T(T^X_{\rm 2d})$ carries a representation $(\rho ,\rho')$ of $\mathbb Z_2 \times \mathbb Z_2$. Moreover, we have as before the possibility of inserting a string of $X^g$ operators  between neighboring tensors at a given position in the bra and/or ket layers. As such,  the transfer operator decomposes \emph{a priori} into 16 blocks denoted by $\mathbb T^{g,\rho}_{g',\rho'}(T_{\rm 2d}^X)$.

From the explicit definition of $\mathbb E(T_{\rm 2d}^X)$ and using the fact that ${\rm tr}(X^gX^{g'}) = 2 \delta_{g,g'}$, we obtain $\mathbb T^{g}_{g'}(T_{\rm 2d}^X) = \frac{1}{2}(\mathbbm 1^{\otimes N}\otimes \mathbbm 1^{\otimes N} + X^{\otimes N}\otimes X^{\otimes N}) \, \delta_{g,g'}$.
For $g=g'$, this transfer operator has two degenerate fixed points, which are found to be spanned by the product states $\mathbbm 1 ^{\otimes N}$ and $X^{\otimes N}$. Overall, we obtain  four degenerate fixed points that correspond to the four non-vanishing blocks $\mathbb T^{g,\rho}_{g,\rho}(T^X_{\rm 2d})$. We thus find that the number of degenerate fixed points of the transfer operator equals the ground state degeneracy of the Hamiltonian, as expected.\footnote{Note that we should in principle consider left and right fixed points of the transfer operator, but these are the same since $\mathbb T(T^X_{\rm 2d})$ is Hermitian.}

The doubly degenerate fixed point described above is invariant under the action of $X^{\otimes N}\otimesbk X^{\otimes N}$, but breaks the symmetry of the transfer operator with respect to $\mathbbm 1^{\otimes N} \otimesbk X^{\otimes N}$ or $X^{\otimes N} \otimesbk \mathbbm 1^{\otimes N}$. In other words, the symmetry group $\mathbb Z_2 \times \mathbb Z_2$ of the transfer operator is broken down to its diagonal subgroup ${\rm diag}(\mathbb Z_2 \times \mathbb Z_2) \cong \mathbb Z_2$ at the fixed points.
As a matter of fact, acting with either $X^{\otimes N}\otimesbk \mathbbm 1^{\otimes N}$ or $\mathbbm 1^{\otimes N} \otimesbk X^{\otimes N}$ has the effect of mapping one fixed point to the other. It turns out that this symmetry structure of the fixed points is characteristic of the renormalization group fixed point. We mentioned earlier that the virtual symmetry satisfied by the tensors is not in one-to-one correspondence with the topological order. We can thus ask, which phases can we encounter as we move away from the renormalization group fixed point within the space of $\mathbb Z_2$ symmetric tensors? This question can be answered by monitoring the symmetry pattern of the transfer operator fixed points \cite{haegeman2015shadows, PhysRevB.95.235119}. In order to understand this mechanism, we must first emphasize what the fixed point structure of the transfer operator indicates regarding the excitation content of the theory.  The transfer operator effectively computes the overlap between topological sectors on the cylinders. The fixed point structure we found, or equivalently the fact that only the blocks $\mathbb T^{a,\rho}_{a,\rho}(T^X_{\rm 2d})$ are non-vanishing, implies that 
\begin{equation*}
	\la g',\rho' | g , \rho \ra = 
	\begin{cases}\,1  \!\!\! &\text{if $g=g'$ and $\rho=\rho'$} \\ \, 0 \!\!\! &\text{otherwise}\end{cases}  
\end{equation*}
in the thermodynamic limit.
This indicates that all the quasi-particle states are orthogonal among themselves, and in particular all the non-trivial excited states are orthogonal to the vacuum state $| 0, + \ra$.

As we deform the model away from the renormalization group fixed point, these virtual operators still encode physical excitations. Nevertheless, as we keep increasing the deformations, we may encounter a phase transition, at which point a given virtual operator might encode a different physical excitation. In particular, the corresponding physical excitation might now be identified with the vacuum sector. As alluded earlier, these condensates are associated with different symmetry breaking patterns in the fixed point sector.

Starting from our tensor network ansatz for the (2+1)d toric code, we distinguish \emph{three} different symmetry patterns in the fixed point sector. The first scenario is the one that we already described, where the symmetry group is ${\rm diag}(\mathbb Z_2 \times \mathbb Z_2)$. The second possibility corresponds to breaking all the symmetries, i.e. the symmetry with respect to $X^{\otimes N} \otimesbk X^{\otimes N}$ is also broken, which can be detected via the local order parameter $Z \otimesbk \mathbbm 1$ since $XZ=-ZX$. The fixed points are of the form $(|a\ra \la a' |)^{\otimes N}$ in this case. The third possibility corresponds to having fixed points that are also invariant under $X^{\otimes  N} \otimesbk \mathbbm 1 ^{\otimes N}$ and $\mathbbm 1 ^{\otimes N} \otimesbk X^{\otimes N}$, which can be probed via the local order parameter $Z \otimesbk Z$, so that the symmetry group is $\mathbb Z_2 \times \mathbb Z_2$. This scenario arises when applying the projector $(\mathbbm 1 + X)^{\otimes 4}$ to the physical indices so that the (unique) fixed point of the transfer operator is equal to $(| + \ra \la + |)^{\otimes N}$.

In order to make the relation between these symmetry patterns and the condensation of the bulk excitations explicit, it is convenient to rephrase our analysis on the half-infinite plane instead of the cylinder, by making the circumference of the cylinder infinite and cutting it open. At this point, it is possible to use the symmetry conditions of the PEPS tensors in order to pull through the string of $X^g$ operators as follows:
\begin{equation*}
	\includeTikz{0}{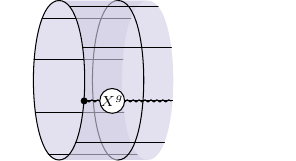}{\transferTwoDBis{0.5}{1}}
	\hspace{-5.5em} \to \;  
	\includeTikz{0}{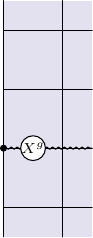}{\transferTwoDBis{0.5}{2}} \; \xrightarrow{\eqref{eq:pull2D}} \;  
	\includeTikz{0}{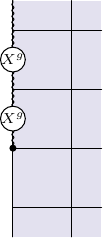}{\transferTwoDBis{0.5}{3}} \, ,
\end{equation*}
such that one of the endpoints of the string is sent to infinity. Consequently, excited states with a non-trivial electric charge $e$ now correspond to 0d domain wall excitations of the transfer operator. Crucially, we can evaluate the overlap $\la \mathcal R_{g',\rho'}(\gamma)(\pepssf) |  \mathcal R_{g,\rho}(\gamma) (\pepssf) \ra$
of excited states as the expectation value 
\begin{equation*}
	\la \mathcal{R}_{g',\rho'}^\dagger \otimesbk \mathcal R_{g,\rho} \ra_{\rm f.p.} \equiv \la {\rm f.p.} | \mathcal{R}_{g',\rho'}^\dagger \otimesbk \mathcal R_{g,\rho} |{\rm f.p.}\ra
\end{equation*}
of the corresponding operators between the fixed points $| {\rm f.p.} \ra$ of the transfer operator, where we assume $\la {\rm f.p.}| {\rm f.p.} \ra =1$. We obtain that if the magnetic fluxes condense into the ground state such that $\la 0, \rho' | 0 , \rho \ra 
\neq 0$ for $\rho \neq \rho'$, then the symmetry with respect to  $X^\infty \otimesbk X^\infty$ must be broken. Indeed, as long as the symmetry is satisfied, i.e. $(X^\infty \otimesbk X^\infty) | {\rm f.p} \ra = | {\rm f.p.} \ra$, we have 
\begin{align}
	\la Z  \ra_\mathsf{PEPS} &= \la Z \otimesbk \mathbbm 1 \ra_{\rm f.p.} = \la {\rm f.p.}| Z \otimesbk \mathbbm 1 | {\rm f.p.} \ra 
	\\ 
	\nn &= \la {\rm f.p.}| X^\infty Z \otimesbk X^\infty | {\rm f.p.} \ra  
	\\
	\nn
	&= 
	-\la {\rm f.p.}| ZX^\infty  \otimesbk X^\infty | {\rm f.p.} \ra 
	= - \la Z \otimesbk \mathbbm 1 \ra_{\rm f.p.} = 0 \, .
\end{align}
Similarly, as long as the $\mathbb Z_2 \times \mathbb Z_2$ symmetry is preserved, the electric excitations condense into the ground states such that $\la g , + | g', + \ra \neq 0 $ for $g \neq g'$.  In summary, we have the following correspondence between symmetry structures of the fixed point of the transfer operator and phases in the bulk:
\begin{align*}
	\setlength{\tabcolsep}{1em}
	\begin{tabularx}{0.98\columnwidth}{ccc}
		Fixed point $\mathbb T(T^X_{\rm 2d})$ & Phase & Condensate
		\\
		\midrule
		$\mathbb Z_2 \times \mathbb Z_2$ & trivial & electric
		\\
		${\rm diag}(\mathbb Z_2 \times \mathbb Z_2) \cong \mathbb Z_2$ & toric code & --
		\\ broken & trivial & magnetic
	\end{tabularx}.
\end{align*}
Carrying out the same analysis using the other tensor network representation would yield the same relation between symmetry breaking and condensation except that the symmetry broken phase would now be associated with the condensation of the electric excitations, and vice versa.

A related question pertains to the existence of boundary Hamiltonians encoding the dynamics of the virtual degrees of freedom.
Let us consider an open tensor network cut out of $| \pepssf(T^X_{\rm 2d}) \ra$ such that the virtual indices along the boundary are left uncontracted. Setting a value for each uncontracted virtual index amounts to defining an entanglement state on the boundary. The PEPS then defines a linear map from the one-dimensional Hilbert space of boundary states to a subspace of physical states.  Using the one-to-one correspondence between virtual and physical degrees of freedom along the boundary, parent Hamiltonians for the edge states can be engineered. But the virtual symmetry on the boundary that is inherited from the bulk tensors imposes a global parity constraint. This implies that the boundary is anomalous since only parity-preserving Hamiltonian can be defined \cite{yang2014edge,PhysRevB.83.245134}. 

The boundary theory is protected against symmetry breaking under local perturbations by the topological order in the bulk, unless a phase transition takes place. The phase diagram of such a one-dimensional model is in bijection with the \emph{gapped boundary} conditions of the (2+1)d toric code. In (2+1)d, we distinguish two such gapped boundary conditions, characterized by the condensation of the electric or magnetic excitations, respectively. By tuning parameters at the boundary, we could drive the toric code to either condensate. Going from one condensate to the other, our previous analysis suggests that the system will undergo a (second order) phase transition involving  a $\mathbb Z_2$-symmetry breaking, which can be checked to be in the same universality class as the Ising model \cite{yang2014edge,PhysRevB.83.245134, PhysRevLett.114.076401, PhysRevResearch.1.033054, chen2019topological, lichtman2020bulk}.
The phase transition between the two gapped boundaries, which are identified by a choice of condensate, closes the gap giving rise to a gapless edge at criticality.

\subsection{Intertwining MPO and duality on the boundary \label{sec:mixed2D}}

\noindent
We derived above two tensor network representations for the (2+1)d toric code. These two representations turn out to display the same $\mathbb Z_2$-symmetry. As we shall see later, this situation is specific to (2+1)d where the toric code is \emph{self-dual}. So far we have been considering each representation separately, raising the question as to whether these can be combined in order to define \emph{mixed} tensor network representations of the ground state subspace. 

Let us consider a  patch of the square lattice with the following choice of $1$-chain in $C_1(\mathbbold T^2_\boxempty, \mathbb Z_2)$:
\begin{equation*}
	\includeTikz{0}{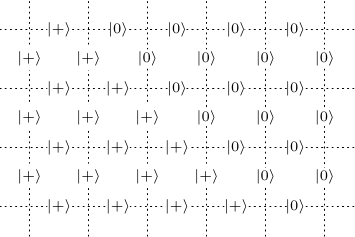}{\mixedTwoD{0.5}{1}} \, 
\end{equation*}
where we identify a diagonal strip of vertex and plaquette operators where both $|+\ra$ and $|0\ra$ states meet. Away from this diagonal strip, we fall within one of the two configurations considered earlier so that the stabilizer constraints can be enforced by means of the same PEPO operators, i.e.
\begin{equation*}
	\includeTikz{0}{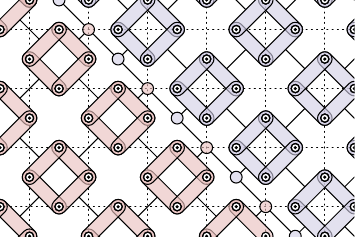}{\mixedTwoD{0.5}{2}} \, ,
\end{equation*}
where we introduced the following complex-valued tensors along the diagonal strip:
\begin{align}
	\begin{split}
	\label{eq:deltaXZ}
	\includeTikz{11}{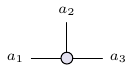}{\deltaT{0.5}{1}} 
	&\equiv 
	\begin{cases} 
		 1 \!\!\! &\text{if $ a_1=a_2=a_3$} 
		\\ 
		0 \!\!\! &\text{otherwise}
	\end{cases} \, , 
	\\ 
	\includeTikz{-11}{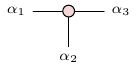}{\deltaT{0.5}{11}} \! &\equiv
		\begin{cases} 
	 1 \!\!\! &\text{if $ \alpha_1 =\alpha_2 = \alpha_3$} 
	\\ 
	0 \!\!\! &\text{otherwise}
	\end{cases} \, ,
	\end{split}
\end{align}
denoted by $\delta_Z$ and $\delta_X$ tensors, respectively.
It is understood in the definitions above that $\{a=0,1\}$ and $\{\alpha=\pm \}$ label eigenstates of $Z$ and $X$, respectively, so that the following symmetry conditions are satisfied:
\begin{align}
		\label{eq:propDeltaZ}
		\includeTikz{6.7}{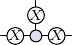}{\deltaT{0.5}{2}} = 
		\includeTikz{6.7}{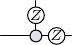}{\deltaT{0.5}{3}} = 
		\includeTikz{6.7}{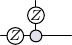}{\deltaT{0.5}{4}} = 
		\includeTikz{6.7}{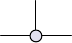}{\deltaT{0.5}{0}} \, ,
		\\[0.2em]
		\label{eq:propDeltaX}
		\includeTikz{-6.7}{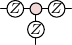}{\deltaT{0.5}{12}} = 
		\includeTikz{-6.7}{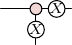}{\deltaT{0.5}{13}} = 
		\includeTikz{-6.7}{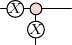}{\deltaT{0.5}{14}}  = 
		\includeTikz{-6.7}{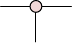}{\deltaT{0.5}{10}} \, .
\end{align}
Contracting these tensors along a line as depicted above, we obtain a so-called Matrix Product Operator (MPO), which is the operator analogue of an MPS. This MPO plays the role of an intertwiner between the two PEPS representations. We shall now demonstrate that this MPO, together with its neighboring PEPOs, does enforce the stabilizer constraints at the vertices and plaquettes along the diagonal strip. We shall focus on the plaquette constraints and deduce the vertex ones by duality. We need to confirm the relation:
\begin{equation}
	\includeTikz{0}{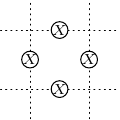}{\interMPOSym{0.5}{0}} \triangleright 
	\includeTikz{0}{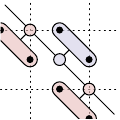}{\interMPOSym{0.5}{11}} = 
	\includeTikz{0}{interMPOSym11}{\interMPOSym{0.5}{11}}, 
\end{equation}
where
\begin{equation}
	\includeTikz{0}{interMPOSym11}{\interMPOSym{0.5}{11}} := \includeTikz{0}{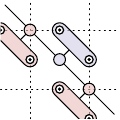}{\interMPOSym{0.5}{1}} \triangleright 
	\includeTikz{0}{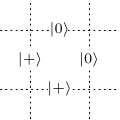}{\interMPOSym{0.5}{-1}} \, . 
\end{equation}
Using the symmetry conditions satisfied by the delta tensors introduced above, together with the symmetry conditions satisfied by PEPS tensors appearing above, one has
\begin{align}
	&
	\includeTikz{0}{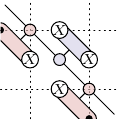}{\interMPOSym{0.5}{2}} \stackrel{(\ref{eq:propHalfPEPOB},\ref{eq:propHalfPEPOA})}{=} \; \includeTikz{0}{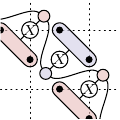}{\interMPOSym{0.5}{3}}
	\\ \nn
	& \q \stackrel{\eqref{eq:propDeltaX}}{=} \; 
	\includeTikz{0}{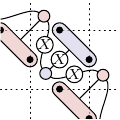}{\interMPOSym{0.5}{4}}
	\! \stackrel{\eqref{eq:propDeltaZ}}{=} \; 	\includeTikz{0}{interMPOSym11}{\interMPOSym{0.5}{11}} \, ,
\end{align}
as required.
The fact that the stabilizer constraints at the vertices are enforced follows by duality. Putting everything together, we obtain that the following PEPS is a (mixed) tensor network representation of the ground state subspace:\footnote{Notice that we started from a very specific configuration of microscopic states, with the purpose of deriving the intertwining MPO and study its properties. Following the same idea, we could derive more general mixed representations, but due to the geometry of the problem, additional PEPS tensors would typically have to be defined.}
\begin{equation*}
	\includeTikz{0}{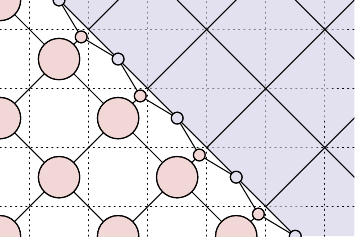}{\mixedTwoD{0.5}{3}} \, .
\end{equation*}
Moreover, the MPO behaves as expected with respect to excitation-creating operators. Indeed, we can check that it maps a string of $X$ operators on the $T^X_{\rm 2d}$-side of the diagonal strip to a pair of $X$ operators on the $T^Z_{\rm 2d}$-side:
\begin{align}
	\label{eq:mappingString}
	&
	\includeTikz{0}{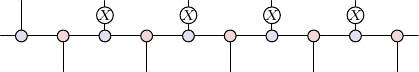}{\mixedTwoDMPO{0.5}{0}} 
	\\ \nn
	 =\; &
	\includeTikz{0}{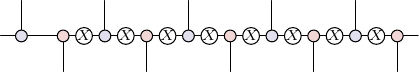}{\mixedTwoDMPO{0.5}{1}} 
	\\ \nn
	 = \; &
	\includeTikz{0}{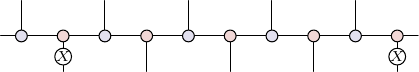}{\mixedTwoDMPO{0.5}{2}} \; .
\end{align}
Similarly, a string of $Z$ operators on the $T^Z_{\rm 2d}$-side would be mapped to a pair of $Z$-operators on the $T^X_{\rm 2d}$-side. 

It is worth emphasizing that the MPO considered in this section does \emph{not} represent a topological defect. Indeed, it merely acts as an `intertwiner' between different tensor network representations, and as such preserves excitations that go through it. By adding some Hadamard matrices to its definition, we could turn this MPO into a topological duality defect, which would perform the permutation of anyons $e \leftrightarrow m$ \cite{bridgeman2017tensor,Williamson:2017uzx}. In order to make the distinction between both scenarios explicit, we briefly present such duality defect in app.~\ref{sec:app_dual2D}. But,  instead of deriving the corresponding MPO, we design new PEPS tensors that parametrize the ground state subspace along the topological defect. 

Finally, we explained earlier that given a state $| \pepssf(T^X_{\rm 2d})\ra$ on an open manifold, we could define an Ising-like Hamiltonian on the one-dimensional Hilbert space of virtual indices that are left uncontracted. Applying the intertwining MPO that we have just defined would map the model to its dual. Indeed, the intertwining MPO performs in particular the identifications $Z_iZ_{i+1} \leftrightarrow Z^\vee_{i+1/2}$ and $Z^\vee_{i-1/2}Z^\vee_{i+1/2} \leftrightarrow Z_i$. \emph{Self-duality} of the (1+1)d Ising model is then reflected in the fact that both tensor network representations have the same symmetry structure.

%% file: _PEPS3D.tex
\section{Tensor network representations in (3+1)d\label{sec:3D}}
\noindent
\emph{In this section, we define and analyze two canonical tensor network representations of the (3+1)d toric code, which are the higher-dimensional analogues of the ones defined in the previous section. Some features of the (3+1)d model are similar to (2+1)d, in which case we shall be brief and invoke the results of the previous section.}

\subsection{PEPS building blocks}

\noindent
We distinguish two tensor network representations, which are obtained following the same approach as in sec.~\ref{sec:gsTwoD} for the (2+1)d model. Given a plaquette $\mathsf{p} \subset \mathbb T^3_\boxempty$, the projector $\frac{1}{2}({\rm id} + \mathbb B_\mathsf{p})$ is the same as before, and so is its tensor network representation in terms of PEPO tensors. These PEPO tensors need to be applied to every plaquette $\mathsf p \subset \mathbb T^3_\boxempty$. We do so according to a pattern akin to (2+1)d such that PEPOs are organized by blocks around every other vertex of the lattice.
Acting with such operators on the kinematical state $| 0,0,\ldots 0 \ra$ yields a ground state wave function for the (3+1)d toric code. We are interested in the tensor $T^X_{\rm 3d}$ that results from blocking twelve PEPO tensors around a vertex $\mathsf v \subset \mathbb T^3_\boxempty$, and acting on the state $|0 \ra^{\otimes 6}$, i.e.
\begin{align}
	\nn
	&
	\includeTikz{0}{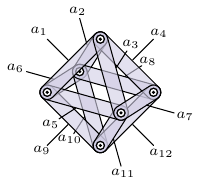}{\ThreeToricProjLocB{0.5}{1}} \hspace{-1em}
	\triangleright \, 
	\includeTikz{0}{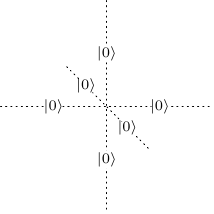}{\ThreeToricProjLocB{0.5}{2}}
	\\[-2em]
	\label{eq:defT3DxRaw}
	& \q\;\;\;\; =:	
	\includeTikz{0}{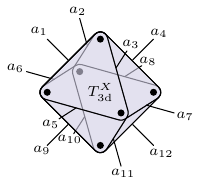}{\ThreeToricProjLocB{0.5}{3}} \; .
\end{align}
Before writing explicitly what the tensor $T^X_{\rm 3d}$ is, let us introduce right away a different notation that is reminiscent of the one used in (2+1)d for $T^X_{\rm 2d}$:
\begin{equation}
	\label{eq:notationThreeDX}
	\!\!
	\includeTikz{0}{ThreeToricProjLocB3}{\ThreeToricProjLocB{0.5}{3}} \!\!\!\!
	\equiv \;\; 
	\includeTikz{0}{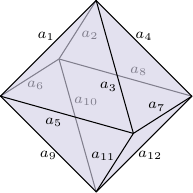}{\ThreeToricProjLocB{0.45}{4}} \,,
\end{equation}
where physical indices are omitted and virtual indices are supported on the edges of the \emph{octahedron}. With this notation in mind, we have the explicit formula:
\begin{align}
	\label{eq:defT3Dx}
	T^X_{\rm 3d}  \propto \! 
	\sum_{\{a=0,1\}}\bigotimes_{\mathsf e \in \{1,\ldots,6\}}
	\!\! |a_{\mathsf e}\ra \!\!
	\bigotimes_{\mathsf e \in \{7,\ldots,12\}} \!\! \la a_{\mathsf e}| \,  \bigotimes_{\mathsf v \subset \tet} \big| \sum_{\mathsf e \supset \mathsf v} a_{\mathsf e} \big \ra
	\, ,
\end{align}
where the labels $\mathsf e, \mathsf v \subset \btet$ now refer to edges and vertices of the underlying octahedron, respectively. Putting everything together, we obtain that $| \pepssf(T^X_{\rm 3d})\ra$ is a tensor network description of a ground state of the (3+1)d toric code.

Let us now investigate some properties of this PEPS tensor. By inspecting definition \eqref{eq:defT3Dx}, or alternatively using the properties \eqref{eq:propHalfPEPOB}, we notice that the tensor remains invariant under the following virtual $\mathbb Z_2$-actions:
\begin{align*}
	&
	\includeTikz{0}{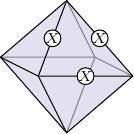}{\symThreeDPepsOne{0.32}{1}} \! = \! 
	\includeTikz{0}{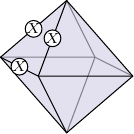}{\symThreeDPepsOne{0.32}{2}} \! = \! 
	\includeTikz{0}{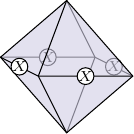}{\symThreeDPepsOne{0.32}{3}} 
	\\[-1.5em]
	&\q\;\;\;\!\!\! =
	\includeTikz{0}{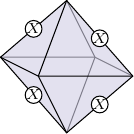}{\symThreeDPepsOne{0.32}{4}} \! = \ldots = \! 
	\includeTikz{0}{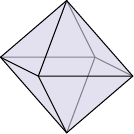}{\symThreeDPepsOne{0.32}{5}} \, ,
\end{align*}
or equivalently
\begin{align*}
	\includeTikz{0}{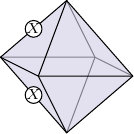}{\symThreeDPepsOne{0.32}{6}} \! = \! 
	\includeTikz{0}{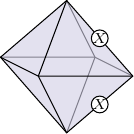}{\symThreeDPepsOne{0.32}{7}} \! = \ldots = \!   
	\includeTikz{0}{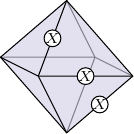}{\symThreeDPepsOne{0.32}{8}} \, .
\end{align*}
More precisely, we have a symmetry condition for every closed loop of $X$ operators along the virtual indices of $T^X_{\rm 3d}$. For instance, every loop around  a triangular face of the octahedron yields a symmetry condition. But we notice that the product of all such operators is equal to the identity map, so that we distinguish $8-1=7$ independent symmetry conditions. This redundancy implicitly follows from the fact that plaquette operators along 2-boundaries of $\mathbb T^3_\boxempty$ multiply to the identity. More generally, generating sets of symmetry conditions are in one-to-one correspondence with generating sets of loops along the edges of the octahedron.

Note that the tensor $T^X_{\rm 3d}$ does not display any additional symmetry, hence the ones displayed above fully specify its support on the virtual degrees of freedom. Moreover, let $\mathcal{P}(T^X_{\rm 3d})$ be the tensor that maps the virtual indices of $T^X_{\rm 3d}$ to its physical ones. Repeating the same manipulations as in equation \eqref{eq:onSite2D} yields
\begin{align}
	\label{eq:oneSiteThreeD}
	\mathbb{E}(T^X_{\rm 3d})
	&:= 
	\mathcal{P}({T^X_{\rm 3d}}^\intercal)\mathcal{P}(T^X_{\rm 3d})
	\\[0.2em]
	\nn
	&\propto \frac{1}{2^7} \sum_{\{a=0,1\}} \bigotimes_{\mathsf e \subset \tet}\Big(\prod_{\substack{\triangle \supset \mathsf e\\ \triangle \neq \triangle_0}}X_{\mathsf e}^{a_\triangle} \Big)
	\\
	\nn
	&\propto \frac{1}{2^7}\prod_{\substack{\triangle \subset \tet \\ \triangle \neq \triangle_0}}\Big(\sum_{a_\triangle}\bigotimes_{\mathsf e \subset \triangle}X_{\mathsf e}^{a_{\triangle}} \Big) \, 	 ,
\end{align}
where $\triangle_0$ is a fixed triangle in the octahedron. 
Dissecting this formula, we understand that $\mathbb{E}(T^X_{\rm 3d})$ is a projector onto the subspace of states that are invariant under the symmetry actions at every triangular face of the octahedron. More generally, we have
\begin{align}
	\label{eq:EmapThreeDx}
	\mathbb{E}(T^X_{\rm 3d}) \propto
	\frac{1}{2^{|\mathcal L(\tet)|}}\prod_{\ell \in \mathcal L(\tet)}\Big( \sum_{a_\ell=0,1} \bigotimes_{\mathsf e \subset \ell} X_{\mathsf e}^{a_\ell}\Big)\, ,
\end{align}
where $\mathcal{L}(\btet)$ is a generating set of loops of virtual indices on the octahedron, and the tensor product is over the edges/indices $\mathsf e$ included in the loop $\ell$. When restricting to the invariant subspace, we obtain that $\mathcal{P}({T^X_{\rm 3d}}^\intercal)$ is the unitary left-inverse of $\mathcal{P}({T^X_{\rm 3d}})$. Acting with this map on the physical vector space of $T^X_{\rm 3d}$, we obtain the following isomorphic PEPS tensor:
\begin{align}
	\label{eq:isoThreeDx}
	T^X_{\rm 3d} \simeq \!  
	 \prod_{\ell \in \mathcal L(\tet)} \! \mathbb P_{+,\ell} \q {\rm with} \q
	\mathbb P_{+,\ell} := \frac{1}{2} \sum_{a_\ell=0,1} \bigotimes_{\mathsf e \subset \ell} X_{\mathsf e}^{a_\ell} \, .
\end{align} 
Notice that we can choose any generating set of loops in order to write this isomorphism. Going from one choice to another simply amounts to a relabelling of summation variables. These isomorphisms make explicit the fact that there is a one-to-one correspondence between physical and virtual degrees of freedom up to a (local) $\mathbb Z_2$-symmetry at every triangle. As in (2+1)d, this implies that it is enough to focus on operations at the virtual level, hence justifying the simplified notation \eqref{eq:notationThreeDX}.

The fact that $T^X_{\rm 3d}$ is isomorphic to a product of projectors may be surprising since the tensor, as initially defined in \eqref{eq:defT3Dx}, does not have the same number of virtual and physical indices. We already discussed that we distinguish seven independent symmetry conditions along the virtual indices of $T^X_{\rm 3d}$ so that there are only five independent virtual degrees of freedom. But, $T^X_{\rm 3d}$ is defined by projecting a microscopic state centered around a vertex at which the stabilizer constraint is satisfied. Therefore, only $6-1=5$ physical indices are independent. Virtual and physical indices having the same bond dimension, this confirms that the virtual and physical vector spaces are isomorphic.

\bigskip \noindent
Let us now construct the second tensor network representation. Interestingly, because we are now dealing with a cubic cellulation of the three-torus, the vertex operators defer from the (2+1)d ones. Indeed, they now involve six qubit degrees of freedom instead of four. This implies that new PEPO tensors must be defined.

Given a vertex $\mathsf{v} \subset \mathbb T^3_\boxempty$, the projector $\frac{1}{2}({\rm id} + \mathbb A_\mathsf{v})$ can be written as the following contraction of two PEPO tensors:
\begin{gather*}
	\includeTikz{0}{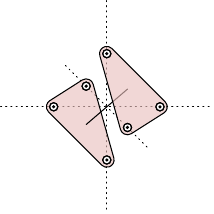}{\PEPOAThree{0.5}}
	\\[-0.5em]
	\text{with} \;\;\; 
	\includeTikz{0}{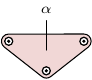}{\PEPOAThreeHalf{0.5}} 
	\equiv \frac{1}{2^\frac{1}{2}} \! \begin{cases} \, \mathbbm 1 \otimes \mathbbm 1 \otimes \mathbbm 1 \!\!\! &\text{if $| \alpha\ra=|+\ra$} \\ Z \otimes Z \otimes Z \!\!\! &\text{if $| \alpha\ra=| -\ra$}\end{cases}  .
\end{gather*}
This operator needs to be applied to every vertex of $\mathbb T^3_\boxempty$. We do so according to pattern akin to (2+1)d such that PEPO tensors are organized by blocks around every other cube of the lattice. The PEPS tensor then results from blocking eight such PEPO tensors around a cube $\mathsf{c} \subset \mathbb T^3_\boxempty$, and acting on the state $| + \ra^{\otimes 12}$, i.e.
\begin{align*}
	&
	\includeTikz{0}{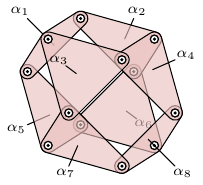}{\ThreeToricProjLocA{0.5}{1}}
	\triangleright \;  
	\includeTikz{0}{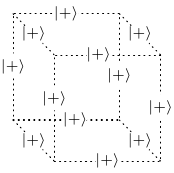}{\ThreeToricProjLocA{0.5}{2}}
	\\[-0.5em]
	& \q\;\; =: 
	\includeTikz{0}{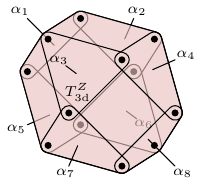}{\ThreeToricProjLocA{0.5}{3}}  \, .
\end{align*}
We notice immediately that both the number of virtual and physical indices differ from the other representation.
Going through the derivation, we find the following explicit formula
\begin{align}
	\label{eq:defT3Dz}
	T^Z_{\rm 3d} \propto \!
	\sum_{\{\alpha=\pm\}}\bigotimes_{\mathsf v \in \{1,\ldots,4\}} \!\!
	 |\alpha_{\mathsf v}\ra 
	\bigotimes_{\mathsf v \in \{5,\ldots,8\}} \!\!  \la \alpha_{\mathsf v}| \, \bigotimes_{\mathsf e} \big| \prod_{\mathsf v \subset \partial \mathsf e} \alpha_{\mathsf v} \big \ra
	\, ,
\end{align}
where the labels $\mathsf v,\mathsf e \subset \mathbbold T^3_{\boxempty}$ here refer to vertices and edges of the original cubic lattice, respectively.
Henceforth, we shall use the following simplified notation
\begin{equation*}
	\includeTikz{0}{ThreeToricProjLocA3}{\ThreeToricProjLocA{0.5}{3}} 
	\equiv 
	\includeTikz{0}{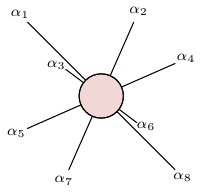}{\ThreeToricProjLocA{0.5}{4}} \, ,
\end{equation*}
that mimics the one of $T^Z_{\rm 2d}$. Using the properties of the PEPO tensors, we find that the tensor $T^Z_{\rm 3d}$ satisfies the following symmetry property
\begin{align*}
	\includeTikz{0}{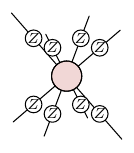}{\symThreeDPepsTwo{0.32}{1}} &= 
	\includeTikz{0}{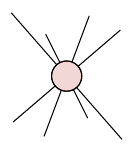}{\symThreeDPepsTwo{0.32}{2}}  \, ,
	\\[-0.5em]
	{\rm or} \;\; 
	\includeTikz{0}{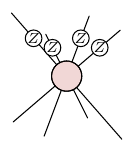}{\symThreeDPepsTwo{0.32}{3}} &= 
	\includeTikz{0}{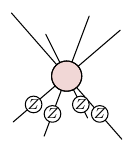}{\symThreeDPepsTwo{0.32}{4}}  \, .
\end{align*}
The tensor does not display any additional symmetry condition so that its support on the virtual degrees of freedom is fully specified by the one above. Following the same steps as in (2+1)d, we obtain the following isomorphism:
\begin{equation}
	\label{eq:isoThreeDZ}
	T^Z_{\rm 3d} \simeq \frac{1}{2}(\mathbbm 1^{\otimes 8} + Z^{\otimes 8}) \, ,
\end{equation}
which indicates that there is a one-to-one correspondence between physical and virtual degrees of freedom up to a (global) $\mathbb Z_2$-symmetry. As for the other representation, we can check this is compatible with the dimension of the virtual and physical vector spaces of $T^Z_{\rm 3d}$, as initially defined in \eqref{eq:defT3Dz}. Since there is a unique symmetry condition, we have $8-1=7$ independent virtual degrees of freedom. But, $T^Z_{\rm 3d}$ is defined by projecting a kinematical state centred around a cube such that the stabilizer constraints are satisfied at every plaquette. Since the plaquette operators around a cube multiply to the identity, one stabilizer constraint is redundant, and thus $12-(6-1)=7$ physical indices are independent, as expected.

\subsection{Ground state subspace and excitations\label{sec:excOpThreeD}}

\noindent
We derived above two tensors $T^X_{\rm 3d}$ and $T^Z_{\rm 3d}$ such that both $| \pepssf(T^X_{\rm 3d}) \ra$ and $| \pepssf(T^Z_{\rm 3d}) \ra$ define ground states of the (3+1)d toric code. These tensors are the three-dimensional analogues of $T^X_{\rm 2d}$ and $T^Z_{\rm 2d}$, respectively. In sec.~\ref{sec:2D}, we explained that in (2+1)d, both tensors have the same virtual symmetry---up to a change of basis---so that they yield the same parametrization of the ground state subspace. We shall now explain that this is no longer the case in (3+1)d, as anticipated from our discussion in sec.~\ref{sec:TC}.

Let us first focus on the parametrization of the ground state subspace provided by $T^X_{\rm 3d}$. Recall that $T^X_{\rm 3d}$ has the geometry of an octahedron such that contractions along virtual indices are performed for two octahedra sharing a common edge, and thus the tensor network we obtain has the geometry of a \emph{tetrahedral-octahedral honeycomb}:
\begin{equation*}
	\includeTikz{0}{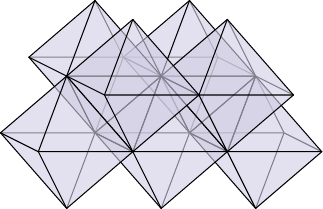}{\tetNetwork{0.32}{1}} \, .
\end{equation*}
As explained in detailed above, $T^X_{\rm 3d}$ satisfies as many symmetry conditions as the number of independent loops of virtual indices (or edges) on the octahedron. It is immediate to check that the same statement holds when considering the concatenation of several such tensors. But, we now find that the number of symmetry conditions is \emph{extensive}.

Although we now have several symmetry conditions per PEPS tensor, these are of the same nature as those appearing in (2+1)d. Consequently, this tensor network representation yields a parametrization of the ground state subspace that is similar to the (2+1)d one. Any closed loop of $X$ operators can be removed using these symmetry conditions, unless it goes along one of the three non-contractible 1-cycles in $\mathbb T^3_\boxempty$, in which case it introduces a non-trivial flux in the wave function, mapping one ground state to another. This confirms the eight-fold degeneracy of the ground state. As usual, the symmetry conditions satisfied by $T^X_{\rm 3d}$ are a crucial feature of the topological order, but these are not enough to fully characterize it. This implies in particular that by working within the space of tensors satisfying such symmetry conditions, we may be able to explore different topological phases. The fact that the tensor is itself a projector onto the invariant subspace at every closed loop is however \emph{characteristic} of the (3+1)d toric code.

Analogously to (2+1)d, point-like electric charges are created at the endpoints of an open string of $X$ operators, the symmetry conditions making the bulk of such strings undetectable, whereas loop-like flux excitations are obtained by inserting loops of $Z$ operators on the virtual indices that break some of the symmetry conditions. Note that these loops of $Z$ operators are not along the one-skeleton of the tetrahedral-octahedral honeycomb but rather its dual, e.g.
\begin{equation}
	\label{eq:excTetra}
	\includeTikz{0}{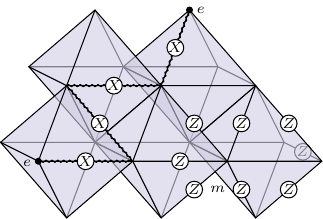}{\tetNetwork{0.32}{2}} \, .
\end{equation}
Let us explain this scenario more carefully by emphasizing the interplay between physical and virtual degrees of freedom. Let us consider a single tensor:
\begin{equation*}
	\includeTikz{0}{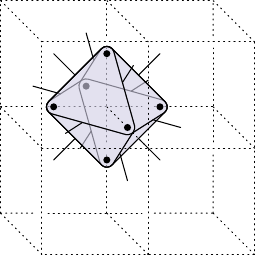}{\loopFlux{0.5}{1}} \; ,
\end{equation*}
where we also represented the cellulation $\mathbb T^3_{\boxempty}$.
Recall that a loop-like excitation is created at the boundary of a membrane of $Z$ operators along the dual lattice. Let us consider the smallest type of membrane operator, which intersects a single edge of the direct lattice. We apply such membrane operator to the configuration above so that a single $Z$ operator acts on the state in the middle. It follows from properties \eqref{eq:propHalfPEPOB} that
\begin{equation*}
	\includeTikz{0}{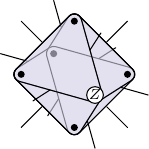}{\loopFlux{0.5}{3}} = 	\includeTikz{0}{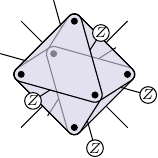}{\loopFlux{0.5}{4}} \; ,
\end{equation*}
which yields the corresponding operator on the virtual level. We can now confirm that such an excitation corresponds to the violation of the stabilizer constraints at the four plaquettes surrounding the edge considered. Combining such operators, we can obtain a configuration like the one depicted in \eqref{eq:excTetra}. Note that because of the redundancies between the virtual symmetry conditions,  only virtual operators resulting from the procedure described above may be inserted, otherwise the tensor network would contract to zero. Notice further that it is not possible to violate a single symmetry condition for a given octahedron.

\bigskip \noindent
As anticipated in sec.~\ref{sec:TC}, the parametrization of the ground state subspace yielded by the PEPS tensor $T^Z_{\rm 3d}$ is noticeably different. First of all, it follows from the construction and the choice of notation that $| \mathsf{PEPS}(T^Z_{\rm 3d}) \ra$  has the geometry of a \emph{body-centered cubic} lattice. Moreover, as in (2+1)d, $T^Z_{\rm 3d}$ displays a unique symmetry condition. However, this symmetry condition is of a different nature. Indeed, due to the geometry of the tensor, we now have that any closed surface of $Z$ operators along a  contractible 2-cycle of the manifold can be removed. Equivalently, given a membrane of $Z$ operators, it can be pulled through the PEPS tensors making it locally undetectable. The ground state subspace is thus parametrized in terms of closed membranes going along one of the three non-contractible 2-cycles of the three-torus, which have a global effect that maps one ground state to another. 

Furthermore, string-like flux excitations are now created at the boundary of an open membrane of $Z$ operators, while pairs of point-like charge excitations can be created by inserting locally $X$ operators, e.g. 
\begin{equation*}
	\includeTikz{0}{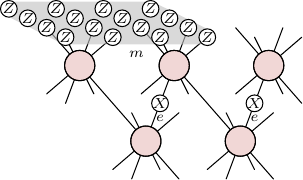}{\cubNetwork{0.32}{1}} \; .
\end{equation*}
Remark that, using the symmetry conditions of the PEPS tensors, we can pull the bulk of the open membrane operator through one of the point-like charge excitations. This process is equivalent to the braiding between the point-like charge and the loop-like flux. The condition $XZ=-ZX$ yields the phase associated with this process.

As in (2+1)d, the virtual operators we have just defined still encode physical excitations as we deform the tensors $T^{X}_{\rm 3d}$ and $T^Z_{\rm 3d}$, which, as we demonstrate below, define renormalization group fixed points.

\subsection{Renormalization group fixed point}
\noindent
Let us now derive the renormalization group flow in the space of tensors, with respect to which the PEPS tensors derived above are fixed points. As before, the tensor network being homogeneous, it is enough to understand the renormalization of a few neighboring tensors.

Let us start with the PEPS representation provided by the tensor $T^Z_{\rm 3d}$. Since $T^Z_{\rm 3d}$ satisfies a unique symmetry condition, regardless of the fact that it is with respect to a (closed) membrane operator instead of a string-like one, the renormalization of $|\mathsf{PEPS}(T^Z_{\rm 3d})\ra$ proceeds analogously to (2+1)d. Indeed, the isomorphism property \eqref{eq:isoThreeDZ} can be shown, following exactly the same steps as in (2+1)d, to be stable under concatenation, and the physical operations implementing the renormalization can be expressed in terms of $(H \otimes H)^\dagger \, {\rm c}X \,(H \otimes H)$ gates. The only difference is that due to the geometry of the underlying network, it is now necessary to act on six pairs of physical indices. Correspondingly, more maximally entangled states at the interface between neighboring tensors will be discarded.

The $T^X_{\rm 3d}$ representation has a much more interesting behavior upon renormalization. We consider the following concatenation of two neighboring tensors
\begin{equation}
	\label{eq:flowThreeD}
	T^X_{\rm 3d} T^X_{\rm 3d} \equiv \hspace{-2em} 
	\includeTikz{0}{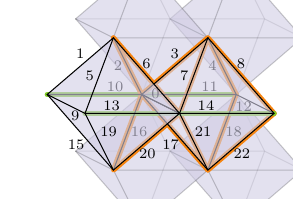}{\ThreeDFlow{0.32}{0}} \; ,
\end{equation}
where the numbers are used to identify the different edge/indices of the block of two tensors. We need to imagine that every pair of neighboring tensors in the same direction are blocked together. But this blocking can be done following different patterns. We choose a \emph{zig-zag} pattern such that the faded tensors above represent two neighboring blocks of tensors. More precisely, the block $T^X_{\rm 3d} T^X_{\rm 3d}$ of tensors we consider is contracted to its upper-back-right neighbor along the virtual indices $\snum{\{2\}}$, $\snum{\{3\}}$ and $\snum{\{4\}}$. Similarly, it is contracted to its lower-back-right neighbors along the virtual indices $\snum{\{16\}}$, $\snum{\{17\}}$ and $\snum{\{18\}}$. The remaining neighboring blocks are not represented but for instance the upper-back-left neighbor is contracted to $T^X_{\rm 3d}T^X_{\rm 3d}$ along the virtual index $\snum{\{1\}}$. The front neighboring blocks follow the same pattern. The asymmetry in the contraction pattern is partly responsible for making the renormalization flow of this representation more exotic.  

Let us study the symmetry conditions satisfied by the block of tensors depicted in \eqref{eq:flowThreeD}. As we mentioned earlier, the number of symmetries is now extensive and grows as we consider blocks of more and more tensors. However, independent symmetries can always be identified via a set of independent loops along the virtual indices. For the configuration at issue, we distinguish 13 such symmetries. We consider the generating set of symmetry conditions defined as
\begin{align*}
	\setlength{\tabcolsep}{0.137em}
	\begin{tabularx}{\columnwidth}{l|cccccccccccccccccccccc}
		{} & ${\sss 1}$  & ${\sss 2}$ & ${\sss 3}$ & ${\sss 4}$ & ${\sss 5}$ & ${\sss 6}$ & ${\sss 7}$ & ${\sss 8}$ & ${\sss 9}$ & ${\sss 10}$ & ${\sss 11}$ & ${\sss 12}$ & ${\sss 13}$ & ${\sss 14}$ & ${\sss 15}$ & ${\sss 16}$ & ${\sss 17}$ & ${\sss 18}$ & ${\sss 19}$ & ${\sss 20}$ & ${\sss 21}$ & ${\sss 22}$
		\\%
		\midrule
		${\sss a_1}$ & ${\sss X}$ & \marktopleft{c1}${\sss X}$ & ${\sss X}$ & ${\sss X}$ & -- & \marktopleft{c2}-- & -- & -- & -- & \marktopleft{c5}${\sss X}$ & ${\sss X}$ & -- & \marktopleft{c6}-- & -- & -- & \marktopleft{c3}-- & -- & -- & -- & \marktopleft{c4}-- & -- & -- 
		\\[-0.2em]
		${\sss a_2}$ & -- & -- & -- & -- & ${\sss X}$ & ${\sss X}$ & ${\sss X}$ & ${\sss X}$ & -- & -- & -- & -- & ${\sss X}$ & ${\sss X}$ & -- & -- & -- & -- & -- & -- & -- & --
		\\[-0.2em]
		${\sss a_3}$ & -- & -- & -- & -- & -- & -- & -- & -- & -- &  ${\sss X}$ & ${\sss X}$ & -- & -- & -- & ${\sss X}$ & ${\sss X}$ & ${\sss X}$ & ${\sss X}$ & -- & -- & -- & --
		\\[-0.2em]
		${\sss a_4}$ & -- & -- & -- & -- & -- & -- & -- & -- & -- & -- & -- & -- & ${\sss X}$ & ${\sss X}$ & -- & -- & -- & -- & ${\sss X}$ & ${\sss X}$ & ${\sss X}$ & ${\sss X}$
		\\[-0.2em]
		${\sss a_5}$ & -- & ${\sss X}$ & ${\sss X}$ & ${\sss X}$ & -- & ${\sss X}$ & ${\sss X}$ & ${\sss X}$ & -- & -- & -- & ${\sss X}$ & -- & -- & -- & -- & -- & -- & -- & -- & -- & --
		\\[-0.2em]
		${\sss a_6}$ & -- & -- & -- & -- & -- & -- & -- & -- & -- & -- & -- & ${\sss X}$ & -- & -- & -- & ${\sss X}$ & ${\sss X}$ & ${\sss X}$ & -- & ${\sss X}$ & ${\sss X}$ & ${\sss X}$ 
		\\[-0.2em]
		${\sss a_7}$ & ${\sss X}$ & -- & -- & -- & ${\sss X}$ & -- & -- & -- & ${\sss X}$ & -- & -- & -- & -- & -- & -- & -- & -- & -- & -- & -- & -- & --
		\\[-0.2em]
		${\sss a_8}$ & -- & ${\sss X}$ & ${\sss X}$ & -- & -- & ${\sss X}$ & ${\sss X}$ & -- & -- & -- & -- & -- & -- & -- & -- & -- & -- & -- & -- & -- & -- & --
		\\[-0.2em]
		${\sss a_9}$ & -- & -- & -- & -- & -- & -- & -- & -- & -- & -- & -- & -- & -- & -- & -- & ${\sss X}$ & ${\sss X}$ & -- & -- & ${\sss X}$ & ${\sss X}$ & --
		\\[-0.2em]
		${\sss a_{10}}$ & -- & -- & ${\sss X}$ & ${\sss X}$ & -- & -- & -- & -- & -- &  -- & ${\sss X}$  & -- &  -- & -- & -- & -- & -- & -- & -- & -- & -- & --
		\\[-0.2em]
		${\sss a_{11}}$ & -- & -- & -- & -- & -- & -- & -- & -- & -- & -- & ${\sss X}$ & -- & -- & -- & -- & -- & ${\sss X}$ & ${\sss X}$ & -- & -- & -- & --
		\\[-0.2em]
		${\sss a_{12}}$ & -- & -- & -- & -- & -- & -- & ${\sss X}$ & ${\sss X}$ & -- & -- & -- & -- & -- & ${\sss X}$ & -- & -- & -- & -- & -- & -- & -- & --
		\\[-0.2em]
		${\sss a_{13}}$ & -- & -- & -- & --\markbottomright{c1}{BurntOrange} & -- & -- & -- & --\markbottomright{c2}{BurntOrange} & -- & -- & --\markbottomright{c5}{LimeGreen} & -- & -- & ${\sss X}$\markbottomright{c6}{LimeGreen} & -- & -- & -- & --\markbottomright{c3}{BurntOrange} & -- & -- & ${\sss X}$ & ${\sss X}\markbottomright{c4}{BurntOrange}$
	\end{tabularx}
\end{align*}
with every horizontal line identified with a variable $a_n$ corresponding to one of the 13 symmetry conditions, such that an `$X$' indicates that we act with the Pauli $X$ operator on the virtual index indicated by the column. For instance, the second line in the table corresponds to the symmetry condition
\begin{align*}
	\includeTikz{0}{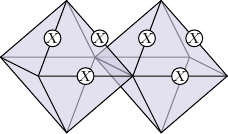}{\exampleSymFlow{0.32}} \! = \! \includeTikz{0}{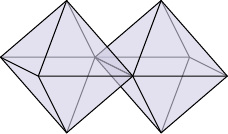}{\ThreeDFlow{0.32}{1}} \, .
\end{align*}
The nomenclature employed here is consistent with that of equation \eqref{eq:exUnitaryMap} below.
The set of loops associated with these symmetry conditions is denoted by $\mathcal{L}(\bttet)$. For convenience, we highlighted sets of bonds that satisfy the same pattern of symmetry conditions, which reflects our choice of concatenation pattern. The choice of colours is in concordance with \eqref{eq:flowThreeD}.

Crucially, the isomorphism property \eqref{eq:isoThreeDx} is also stable under concatenation. More precisely, we can find an isomorphism of the form \eqref{eq:isoThreeDx} for $T^X_{\rm 3d}T^X_{\rm 3d}$ associated with the set of symmetry conditions described above. In order to show this, let us start by applying the isomorphism property to each tensor in \eqref{eq:flowThreeD}, where the sets of independent loops are provided by seven of the eight triangles, respectively. More specifically, we choose to express the l.h.s and r.h.s tensors, respectively, as
\begin{align*}
	T^X_{\rm 3d} {\sss |}_{\rm l.h.s}&\simeq
	\frac{1}{2}
	\Big(\! \prod_{\substack{\triangle \subset \tet \\ \{0\} \not\subset \triangle }}
	\mathbb P_{+,\triangle} \Big)
	\sum_{a =0,1}\bigotimes_{\mathsf e \subset \{0,2,6\}}X^a_{\mathsf e}
	\\
	T^X_{\rm 3d} {\sss |}_{\rm r.h.s}&\simeq
	\frac{1}{2}
	\Big(\! \prod_{\substack{\triangle \subset \tet \\ \{0\} \not\subset \triangle }}
	\mathbb P_{+,\triangle} \Big)
	 \sum_{a =0,1}  \bigotimes_{\mathsf e \subset \{0,3,7\}}X^a_{\mathsf e} \, ,
\end{align*}
where `$\snum{\{0\}}$' in the products above refers to the bond along which the concatenation is performed, so that the products are over all the triangles in the octahedra that do not contain the edge $\snum{\{0\}}$.
Given these isomorphisms, let us apply the map $\mathcal{P}((T_{\rm 3d}^XT_{\rm 3d}^X)^\intercal)$ to the physical space of $T^X_{\rm 3d}T^X_{\rm 3d}$. We find 
\begin{align}
	\nn
	T^X_{\rm 3d}T^X_{\rm 3d} &\simeq 
	\frac{1}{4}\Big(\prod_{\triangle \subset \ttet} \mathbb P_{+,\triangle}\Big)
	\\
	\nn
	&  \q\, \times  \sum_{a_1,a_2}  (X^{a_1}_2 \otimes X^{a_1}_6)\,  {\rm tr}(X^{a_1}_0 X^{a_2}_0) \,
	(X^{a_2}_3 \otimes X_7^{a_2}) 
	\\
	&=
	\frac{1}{2}\Big( \!\! \prod_{\triangle \subset \ttet} \!\! \mathbb P_{+,\triangle}\Big)\! \sum_{a}\, \bigotimes_{\mathsf e \subset \{2,3,6,7\}} \!\! X^a_{\mathsf e} \; ,
\end{align}
where we used  the fact that ${\rm tr}(X^{a_1}X^{a_2}) = 2\delta_{a_1,a_2}$. It is understood in the formulae above that the virtual bond along the edge $\snum{\{0\}}$ is summed over so that the triangles $\triangle$ containing $\snum{\{0\}}$ are not included in $\bttet$. Moreover, notice that the redundant symmetry condition for $\bttet$ is associated with the plaquette labelled by $\snum{\{16,17,20,21\}}$, and as such it is not enforced in the formula above. A simple relabelling of the summation variables finally yields
\begin{equation}
	\label{eq:isoConcatenation}
	T^X_{\rm 3d}T^X_{\rm 3d} \simeq \prod_{\ell \in \mathcal L(\ttet)} \mathbb P_{+,\ell} \, . 
\end{equation}
Using this isomorphism, let us now demonstrate that $T^X_{\rm 3d}$ is a fixed point of the renormalization group flow. Recall that in order to prove such statement,  we have two requirements, which we repeat here for convenience: $(i)$ Show that the tensor $T^X_{\rm 3d}T^X_{\rm 3d}$ is equal, up to isomorphisms, to a tensor product between $T^X_{\rm 3d}$ and another term, and $(ii)$ that the second term describes purely short-range correlations so that it can be discarded as we focus on non-local properties.

As in (2+1)d, we wish to implement isomorphism $(i)$ via a unitary physical operation. 
Let us consider the following product of disentangling maps
\begin{align*}
	\mathcal{U}_{\rm ren.}
	:=
	\; & \mathcal{U}_{2,3,4}\otimes \mathcal{U}_{6,7,8} \otimes \mathcal{U}_{16,17,18} \otimes \mathcal{U}_{20,21,22}
	\\
	& \!\! \otimes {\rm c}X_{10,11} \otimes {\rm c}X_{13,14} \, ,
\end{align*}
where we defined $\mathcal{U}_{\mathsf e_1,\mathsf e_2,\mathsf e_3}  =  {\rm c}X_{\mathsf e_3,\mathsf e_1} {\rm c}X_{\mathsf e_1,\mathsf e_2} {\rm c}X_{\mathsf e_2,\mathsf e_3}$
such that
\begin{align*}
	\mathcal U_{\mathsf e_1, \mathsf e_2, \mathsf e_3}\,: \, 
	&|a_1,a_2,a_3 \ra_{\mathsf e_1,\mathsf e_2,\mathsf e_3}  \\
	&\mapsto
	|a_1+a_2+a_3,a_1+a_2,a_2+a_3\ra_{\mathsf e_1,\mathsf e_2,\mathsf e_3} \, .
\end{align*}
This map $\mathcal{U}_{{\rm ren.}}$ is applied to the physical indices of $T^X_{\rm 3d}T^X_{\rm 3d}$, whereas the map $\mathcal U_{\rm ren.}^\dagger$ is applied to the corresponding physical indices of the neighboring tensors.  In virtue of the isomorphism \eqref{eq:isoConcatenation}, it is enough to check that this physical operation performs the desired factorization at the level of the projectors $\mathbb P_{+,\ell}$, for every $\ell \in \mathcal{L}(\bttet)$. This will then induce the corresponding factorization at the level of the tensors. Since we can insert the resolution of the identity $\mathcal U_{\rm ren.}^\dagger\mathcal U_{\rm ren.} = \mathbbm 1$ between $T^X_{\rm 3d}T^X_{\rm 3d}$ and its surrounding neighbors along the virtual indices, we simply need to confirm that the map $\mathcal U_{\rm ren.}$ satisfies
\begin{align}
	\label{eq:UglobalAction}
	&\mathcal{U}_{\rm ren.}
	\Big(\! \prod_{\ell \in \mathcal{L}(\ttet)} \!\! \mathbb P_{+,\ell}\Big) \,
	\mathcal{U}^\dagger_{\rm ren.} 
	= 
	\prod_{\ell \in \mathcal{L}(\tet)} \! \mathbb P_{+,\ell} \prod_{\ell' \in \mathcal{L}({\rm corr.})} \!\!\! \mathbb P_{+,\ell'} \,   , 
\end{align}
where $\mathcal{L}(\rm corr.)$ refers to a set of collections of legs for which there is a symmetry. Although we use the same notation, note that elements in $\mathcal{L}(\rm corr.)$ do not typically correspond to loops along the virtual indices anymore. For instance, one has
\begin{align}
	\nn
	&\mathcal U_{2,3,4}  (X_2^{a_1+a_5+a_8}\otimes X_3^{a_1+a_5+a_8+a_{10}} \otimes X_4^{a_1+a_5+a_{10}})  \mathcal U_{2,3,4}^\dagger 
	\\
	\label{eq:exUnitaryMap} 
	& \q = X_2^{a_1+a_5} \otimes X_3^{a_{10}} \otimes X_{4}^{a_8} \, ,
\end{align}
for all $a_1,a_5,a_8,a_{10} = 0,1$, such that the nomenclature is the one of the previous table. Using our notation, this can be rewritten more visually as follows:\footnote{Turning these tables of symmetry operators into $\mathbb Z_2$-valued matrices, we can compute the action of the maps $\mathcal U$ via matrix multiplication.}
\begin{equation}
	\setlength{\tabcolsep}{0.137em}
	\mathcal U_{2,3,4} : \;  
	\begin{tabularx}{0.185\columnwidth}{l|ccc}
		{} & ${\sss 2}$ & ${\sss 3}$ & ${\sss 4}$
		\\
		\midrule
		${\sss a_1}$ & ${\sss X}$ & ${\sss X}$ & ${\sss X}$
		\\
		${\sss a_5}$ & ${\sss X}$ & ${\sss X}$ & ${\sss X}$
		\\
		${\sss a_8}$ & ${\sss X}$ & ${\sss X}$ & --
		\\
		${\sss a_{10}}$ & -- & ${\sss X}$ & ${\sss X}$
	\end{tabularx}
	\; \mapsto \; 
	\begin{tabularx}{0.185\columnwidth}{l|ccc}
		{} & ${\sss 2}$ & ${\sss 3}$ & ${\sss 4}$
		\\
		\midrule
		${\sss a_1}$ & ${\sss X}$ & -- & --
		\\
		${\sss a_5}$ & ${\sss X}$ & -- & --
		\\
		${\sss a_8}$ & -- & -- & ${\sss X}$
		\\
		${\sss a_{10}}$ & -- & ${\sss X}$ & --
	\end{tabularx} \, .
\end{equation}
Performing this computation for every linear map entering the definition of $\mathcal U_{\rm ren.}$ yields the l.h.s of \eqref{eq:UglobalAction}. Instead of writing the result explicitly, we shall indicate for every symmetry condition as depicted in the previous table what it is mapped to upon application of $\mathcal{U}_{\rm ren.}$:
\begin{align*}
	\setlength{\tabcolsep}{0.137em}
	\begin{tabularx}{\columnwidth}{l|cccccccccccccccccccccc}
		{} & ${\sss 1}$  & ${\sss 2}$ & ${\sss 5}$ & ${\sss 6}$ & ${\sss 9}$ & ${\sss 10}$ & ${\sss 12}$ & ${\sss 13}$ & ${\sss 15}$ & ${\sss 16}$ & ${\sss 19}$ & ${\sss 20}$ & ${\sss 3}$ & ${\sss 4}$ & ${\sss 7}$ & ${\sss 8}$ & ${\sss 11}$ & ${\sss 14}$ & ${\sss 17}$ & ${\sss 18}$ & ${\sss 21}$ & ${\sss 22}$
		\\%
		\midrule
		${\sss a_1}$ & ${\sss X}$ & ${\sss X}$ & -- & -- & -- & ${\sss X}$ & -- & -- & -- & -- & -- & -- & -- & -- & -- & -- & -- & -- & -- & -- & -- & -- 
		\\[-0.2em]
		${\sss a_2}$ & -- & -- & ${\sss X}$ & ${\sss X}$ & -- & -- & -- & ${\sss X}$ & -- & -- & -- & -- & -- & -- & -- & -- & -- & -- & -- & -- & -- & --
		\\[-0.2em]
		${\sss a_3}$ & -- & -- & -- & -- & -- & ${\sss X}$ & -- & -- & ${\sss X}$ &  ${\sss X}$ & -- & -- & -- & -- & -- & -- & -- & -- & -- & -- & -- & --
		\\[-0.2em]
		${\sss a_4}$ & -- & -- & -- & -- & -- & -- & -- & ${\sss X}$ & -- & -- & ${\sss X}$ & ${\sss X}$ & -- & -- & -- & -- & -- & -- & -- & -- & -- & --
		\\[-0.2em]
		${\sss a_5}$ & -- & ${\sss X}$ & -- & ${\sss X}$ & -- & -- & ${\sss X}$ & -- & -- & -- & -- & -- & -- & -- & -- & -- & -- & -- & -- & -- & -- & --
		\\[-0.2em]
		${\sss a_6}$ & -- & -- & -- & -- & -- & -- & ${\sss X}$ & -- & -- & ${\sss X}$ & -- & ${\sss X}$ & -- & -- & -- & --& -- & -- & -- & -- & -- & -- 
		\\[-0.2em]
		${\sss a_7}$ & ${\sss X}$ & -- & ${\sss X}$ & -- & ${\sss X}$ & -- & -- & -- & -- & -- & -- & -- & -- & -- & -- & -- & -- & -- & -- & -- & -- & --
		\\[-0.2em]
		${\sss a_8}$ & -- & -- & -- & -- & -- & -- & -- & -- & -- & -- & -- & -- & -- & ${\sss X}$ & -- & ${\sss X}$ & -- & -- & -- & -- & -- & --
		\\[-0.2em]
		${\sss a_9}$ & -- & -- & -- & -- & -- & -- & -- & -- & -- & -- & -- & -- & -- & -- & -- & -- & -- & -- & -- & ${\sss X}$ & -- & ${\sss X}$
		\\[-0.2em]
		${\sss a_{10}}$ & -- & -- & -- & -- & -- & -- & -- & -- & -- &  -- & -- & -- & ${\sss X}$ & -- & -- & -- & ${\sss X}$ & -- & -- & -- & -- & --
		\\[-0.2em]
		${\sss a_{11}}$ & -- & -- & -- & -- & -- & -- & -- & -- & -- & -- & -- & -- & -- & -- & -- & -- & ${\sss X}$ & -- & ${\sss X}$ & -- & -- & --
		\\[-0.2em]
		${\sss a_{12}}$ & -- & -- & -- & -- & -- & -- & -- & -- & -- & -- & -- & -- & -- & -- & ${\sss X}$ & -- & -- & ${\sss X}$ & -- & -- & -- & --
		\\[-0.2em]
		${\sss a_{13}}$ & -- & -- & -- & -- & -- & -- & -- & -- & -- & -- & -- & -- & -- & -- & -- & -- & -- & ${\sss X}$ & -- & -- & ${\sss X}$ & --
	\end{tabularx}
\end{align*}
such that the nomenclature is the same as above.
Note that we preserved the order of the rows so that the symmetry condition written in the $n^{\rm th}$ row of the previous table is mapped to the symmetry condition written in the $n^{\rm th}$ row of this new table, but columns have been reorganized for convenience. 

We distinguish immediately two disjoint sets of symmetry conditions. The first set contains 7 symmetries associated with 12 edges, whereas the second one contains 6 symmetries associated with 10 edges. Furthermore, by inspecting the structure of the symmetry conditions in the first set, we identify that these precisely correspond to the symmetries of a single tensor $T^X_{\rm 3d}$ with respect to a set of independent loops $\mathcal{L}(\btet)$. In virtue of the isomorphism \eqref{eq:isoConcatenation} computed above, this factorization at the level of the symmetry operators induces the following isomorphism of tensors
\begin{align}
	&
	\includeTikz{0}{ThreeDFlow1}{\ThreeDFlow{0.32}{1}} 
	\\[-1.5em] \nn
	& \q\, \simeq 
	\includeTikz{0}{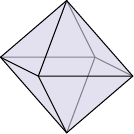}{\ThreeDFlow{0.32}{2}} \otimes \includeTikz{0}{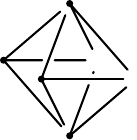}{\ThreeDFlow{0.32}{3}}  ,
\end{align}
where the second tensor on the right-hand side is defined as the tensor product
\begin{equation}
	\label{eq:GHZThreeD}
	\includeTikz{0}{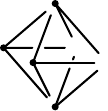}{\ThreeDFlow{0.25}{3}} 
	\equiv \Bigg( 
	\includeTikz{0}{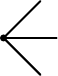}{\ThreeDFlow{0.25}{4}}  \Bigg)^{\otimes 2} \otimes \Bigg(\; 
	\includeTikz{0}{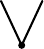}{\ThreeDFlow{0.25}{5}} \; \Bigg)^{\otimes 2} \, , 
\end{equation} 
such that sets of connected bold lines in the tensor product above represent PEPS tensors that act as projectors onto the subspace of states satisfying the following symmetry conditions:
\begin{equation*}
	 \includeTikz{0}{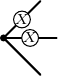}{\symCorrFlow{0.25}{1}}\! =
	 \includeTikz{0}{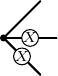}{\symCorrFlow{0.25}{2}} \! = 
	 \includeTikz{0}{ThreeDFlow4}{\ThreeDFlow{0.25}{4}} \q {\rm and} \q 
	 \includeTikz{0}{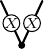}{\symCorrFlow{0.25}{3}} = 
	 \includeTikz{0}{ThreeDFlow5}{\ThreeDFlow{0.25}{5}} \, .
\end{equation*} 
More explicitly, we have
\begin{equation*}
	\includeTikz{0}{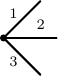}{\ThreeDFlow{0.25}{7}} \equiv  \frac{1}{4}\sum_{a_1,a_2=0,1} X_1^{a_1} \otimes X_2^{a_1+a_2} \otimes X_3^{a_2} \, . 
\end{equation*}
The set of collection of legs for which such symmetry conditions occur was the one notated via $\mathcal{L}(\rm corr.)$ in \eqref{eq:UglobalAction}. Note that two such PEPS tensors are contracted along a virtual index when they share a common edge/bond, in a way akin to the $T^X_{\rm 3d}$ tensors.

In (2+1)d, the analogues of the states depicted above corresponded to maximally entangled states at the interface between two neighboring PEPS tensors. As such these could easily be contracted away. The same does not immediately happen in (3+1)d. Indeed, we must confirm that these tensors represent purely short-range correlations. In other words, we must ensure that these do not combine in such way as to form long-range chains. In order to verify this condition, it is necessary to consider the renormalization of a larger block of tensors. More specifically, we should repeat the analysis carried out above for the pairs of tensors that are adjacent to the one considered. By construction, we apply the same disentangling map to a given block and its neighbors. But the asymmetry in the contraction pattern implies that the tensors will not transform strictly the same way, slightly changing the geometry of the resulting tensors. Nevertheless, it is not necessary to repeat the previous analysis as the result can be inferred by noticing that the roles of the legs, along which a given contraction is performed, are mirrored. In order to confirm that the correlations remain short-range, it is enough to consider the isomorphism of the following block of tensors:
\begin{align}
	\label{eq:flow3DGlobal}
	&
	\includeTikz{0}{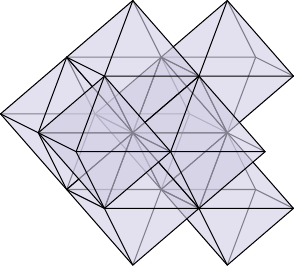}{\correlationNetwork{0.32}{1}}
	\\[-1.9em] \nn
	& \q\, \simeq 
	\includeTikz{0}{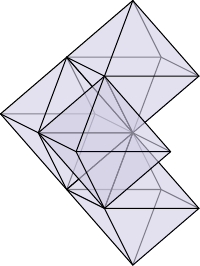}{\correlationNetwork{0.32}{2}} \otimes 
	\includeTikz{0}{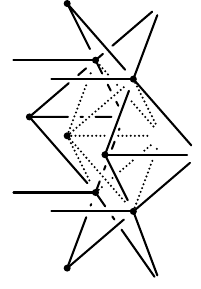}{\correlationNetwork{0.32}{3}}  ,
\end{align}
where the bonds represented by dotted bold lines happen to be summed over. In the middle of the second term in the tensor product, we notice a five-valent tensor that is traced over, which is the contraction of the same four tensors appearing in \eqref{eq:GHZThreeD} that result from the individual factorization of the four pairs of tensors on the l.h.s. of \eqref{eq:flow3DGlobal}. By considering a larger block of tensors, we would notice that the remaining tensors always get contracted away following a similar pattern.  This confirms that they only represent purely short-range information and can thus be safely discarded as we are interested in the long-range information. Hence, the PEPS tensor $T^X_{\rm 3d}$ indeed appears as the fixed point of a renormalization group flow.

As a consistency check, let us to confirm that the entanglement entropy is preserved throughout this process. We consider the region $\Omega$ defined by the block of tensors represented on the l.h.s of \eqref{eq:flow3DGlobal}. The corresponding reduced density matrix has a degenerate spectrum with $|\mathbb Z_2|^{|\partial \Omega| - |\mathcal{L}(\Omega)|}$ dominant singular values, where $|\partial \Omega|$ is the area of the boundary, as defined by the number of uncontracted bonds, and $|\mathcal L(\Omega)|$ the number of independent loops along the virtual indices on the boundary.\footnote{The same scaling  is obtained when performing the computation in terms of basis states that diagonalize the plaquette operators. In this basis, the entanglement entropy for a connected region $\Omega$ reads $S(\Omega) = {\rm log} |G| / (|G_\Omega||G_{\bar \Omega}|)$, where $G$ is the group generated by all the plaquette operators on $\mathbb T^3_\boxempty$, while $G_\Omega$ and $G_{\bar \Omega}$ are the subgroups generated by the plaquette operators within $\Omega$ and its complement, respectively. Denoting by $\boxempty_{0}$ the number of vertices in $\mathbb T^3_\boxempty$, we find $|G|=2^{2(\boxempty_0 - 1)}$. This follows from: $(i)$ there are $3\boxempty_0$ plaquettes in $\mathbb T^3_\boxempty$, $(ii)$ the product of all the plaquettes operators around any cube or along any of three non-contractible 2-cycles is the identity, $(iii)$ the product of all the plaquette operators around every cube but one is the same as the product of the operators around this cube. Following similar rules, we compute $|G_\Omega|$  and $|G_{\bar \Omega}|$ so that $S(\Omega) = (\boxempty_2^{\partial \Omega} - \boxempty_3^{\partial \Omega} + 1){\rm log}(2)$, where $\boxempty_2^{\partial \Omega}$ and $\boxempty_3^{\partial \Omega}$ denote the number of plaquettes and cubes shared by $\Omega$ and its complement, respectively \cite{PhysRevB.78.155120}.} We thus find that the entanglement entropy associated with region $\Omega$ is $23{\rm log}(2)$. On the other hand, the entanglement entropy of the first term on the r.h.s. of \eqref{eq:flow3DGlobal} is equal to $15{\rm log}(2)$. Furthermore, each tensor in the second term contributes  by a factor of ${\rm log}(2)$ since the number of independent symmetry conditions subtracted from the number of bonds is always equal to one. We count eight such tensors (nine minus the one that is traced over), and thus the entanglement entropy is preserved, as expected.

\subsection{Transfer operator}

\noindent
We explained in the previous section how the (2+1)d toric code could be analysed from the point of view of the transfer operator. More specifically, we showed that the number of degenerate fixed points of the transfer operator equals the ground state degeneracy of the Hamiltonian, and discussed how the symmetry structure of the fixed points reflects the topological order. Furthermore, we emphasized how for both representations the same two trivial phases could be obtained by condensing either type of bulk excitations, which was reflected in the symmetry structure of the fixed point sector. We shall now examine how these statements generalize to three dimensions.

Let us first focus on the PEPS representation in terms of the tensor $T^Z_{\rm 3d}$. We consider several copies of $T^Z_{\rm 3d}$ that we contract on the manifold $\mathbbold T^2 \times \mathbbold I$. It follows from the symmetry condition satisfied by $T^Z_{\rm 3d}$ that the tensor network remains invariant under the simultaneous action of two membranes of $Z^g$ operators at both ends of $\mathbbold T^2 \times \mathbbold I$. Mimicking  the (2+1)d scenario, we identify eight minimally entangled topological sectors on $\mathbbold T^2 \times \mathbbold I$. These can be distinguished by: $(i)$ the possibility of inserting two membranes of $Z^g$ and $Z^h$ operators as depicted below: 
\begin{align*}
	\includeTikz{0}{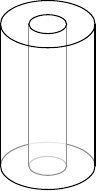}{\threeCylinder{0.8}{1}} \, , \q
	\includeTikz{0}{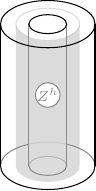}{\threeCylinder{0.8}{2}} \, , \q
	\includeTikz{0}{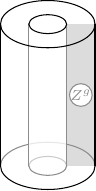}{\threeCylinder{0.8}{3}} \, , \q
	\includeTikz{0}{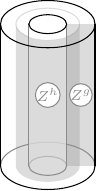}{\threeCylinder{0.8}{4}} \, .
\end{align*}
where we represented $\mathbbold T^2 \times \mathbbold I$ as a hollow cylinder whose inner and outer cylinders are identified, and $(ii)$ the possibility of inserting a torus-like membrane of $\mathbb P_\rho:= \frac{1}{2}(\mathbbm 1 + \rho Z)$ operators across the $\mathbbold I \equiv [0,1]$ direction. Using the symmetry conditions of the PEPS tensors, the latter membrane can be moved to the boundary where it projects  the boundary conditions onto the even or odd sector, respectively. The global parity constraint ensures that both boundary conditions are in the same sector. Putting everything together, we obtain that states on $\mathbbold T^2 \times \mathbbold I$ are parametrized by the possibility of inserting two membranes of $Z$ operators and a choice of boundary conditions labelled by an irreducible representation of the group $\mathbb Z_2$ generated by the torus-like membrane of $Z$ operators. By construction, these topological sectors are in bijection with the ground states on the three-torus $\mathbbold T^3$.

Recall that the (3+1)d toric code Hamiltonian yields electric and magnetic excitations that are point-like and loop-like, respectively. The topological sectors on $\mathbbold T^2 \times \mathbbold I$ described above classify a special configuration of two magnetic fluxes and an electric charge. More precisely, we should interpret these topological sectors as a loop-like flux excitation to which a charge is attached, while being threaded by another flux. Loop-like excitations can then only fuse and braid if they are threaded by the same flux, giving rise in particular to the so-called \emph{three-loop braiding} statistics \cite{Baez:2006un, Wang:2014oya, putrov2016braiding, Wang:2014xba, Chan:2017eov, Bullivant:2018pju,  Wan:2019oyr}. Furthermore, for more general models where the input group could be any finite group, the threading flux would have the effect of constraining the magnetic flux and electric charge quantum numbers of the loop-like excitation \cite{PhysRevB.91.075114, Wang:2014oya,Delcamp:2017pcw, Bullivant:2019fmk}. 

The analysis of the transfer operator proceeds very similarly to (2+1)d. We define $\mathbb T(T^Z_{\rm 3d})$ by contracting several copies of the tensor $\mathbb E(T^Z_{\rm 3d})\propto \frac{1}{2}(\mathbbm 1 ^{\otimes 8}+ Z^{\otimes 8})$ on a single slice of $\mathbbold T^2 \times \mathbbold I$. The symmetry condition satisfied by $T^Z_{\rm 3d}$ induces the same global $\mathbb Z_2$-symmetry in the bra and ket layers of $\mathbb E(T^Z_{\rm 3d})$. It follows that the transfer operator $\mathbb T(T^Z_{\rm 3d})$ commutes with a torus-like membrane of $Z$ operators in both the bra and ket layers. As such, it is labelled by a representation $(\rho,\rho')$ of $\mathbb Z_2 \times \mathbb Z_2$ that decomposes the symmetry action. Moreover, as depicted above, we have the possibility of inserting cylinder-like membranes of $Z^g$ and $Z^h$ operators along two directions, in both the bra and the ket layers. Therefore, the transfer operator decomposes \emph{a priori} into 64 blocks denoted by $\mathbb T^{g,h,\rho}_{g',h',\rho'}(T^Z_{\rm 3d})$. From the definition of $T^Z_{\rm 3d}$, we obtain that only the blocks satisfying $g=g'$ and $h=h'$ are non-vanishing, in which case the transfer operator has two degenerate fixed points labelled by $\rho=\rho'=\pm 1$. The eight degenerate ground states of the model are therefore in one-to-one correspondence with the eight non-vanishing blocks $\mathbb T^{g,h,\rho}_{g,h,\rho}(T^Z_{\rm 3d})$ of the transfer operator. The fact that only these blocks are non-vanishing indicates that none of the excitations is condensed.

\bigskip \noindent
Let us now study the tensor network representation in terms of $T^X_{\rm 3d}$. We consider several copies of  $T^X_{\rm 3d}$ that we contract on the manifold $\mathbbold T^2 \times \mathbbold I$. A \emph{double layer} of the resulting translational invariant operator can be depicted as\footnote{Because of the geometry of the tensor network, the translation invariance is with respect to two layers of tensors.}
\begin{equation}
	\label{eq:threeCylinder}
	\includeTikz{0}{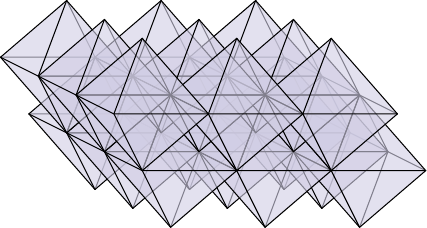}{\transferThreeD{0.32}{1}} 
\end{equation}
with periodic boundary conditions such that opposite horizontal edges are identified. We should keep in mind that the same pattern of tensors can be repeated both horizontally and vertically. As usual, this tensor network inherits from $T^X_{\rm 3d}$ some symmetry conditions so that any loop of $X$ operators along any contractible 1-cycle can be traced away. Seen from the top, the uncontracted bonds of the tensor network under consideration form a square lattice, i.e.
\begin{equation}
	\label{eq:ProjThreeCylinder}
	\includeTikz{0}{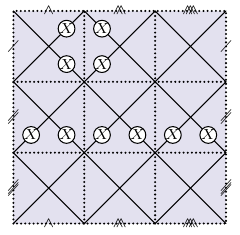}{\transferThreeD{0.24}{2}} \; ,
\end{equation}
where the densely dotted lines represent bonds that are contracted. Henceforth, we shall denote this effective square lattice formed by the upper uncontracted bonds of the network by $\mathbbold T^2_{\diam}$. For every plaquette of $\mathbbold T^2_{\diam}$, there is a $\mathbb Z_2$-symmetry condition under the action of a loop of $X$ operators as depicted for instance at the top of \eqref{eq:ProjThreeCylinder}. Loops of $X$ operators along one of the two non-contractible 1-cycles, as depicted at the bottom of \eqref{eq:ProjThreeCylinder}, have a global effect. Naturally, such operators can be moved across $\mathbbold T^2_{\diam}$ by using the symmetry conditions at every plaquette. Note that seen from the bottom, the uncontracted bonds of the tensor network \eqref{eq:threeCylinder} form an analogous effective square lattice, on which the same properties hold. It follows from the symmetry conditions in the bulk of the network that acting with two loops of $X$ operators along a given non-contractible 1-cycle at the top and at the bottom of \eqref{eq:threeCylinder} is a symmetry of the system. 

We still identify eight topological sectors on this manifold, which can be distinguished by: $(i)$ the possibility of inserting a string of $X^g$ operators `vertically' along $\mathbbold T^2 \times \mathbbold I$, and $(ii)$ the possibility of inserting two loops of $\mathbb P_{\rho}:= \frac{1}{2}(\mathbbm 1 + \rho X)$ operators along both non-contractible 1-cycles, which project the boundary conditions onto the even or odd sectors with respect to the non-contractible loops of $X$ operators. These eight topological sectors represent the same configuration of quasi-particles as before.

Let us now consider the transfer operator $\mathbb T(T^X_{\rm 3d})$ obtained by contracting  copies of the tensor $\mathbb E(T^X_{\rm 3d})$ according to the pattern depicted in \eqref{eq:threeCylinder}, i.e. two `horizontal' layers of  tensors $\mathbb E(T^X_{\rm 3d})$ with periodic boundary conditions. Henceforth, we shall refer to the underlying tesselation as $\mathbbold T^2_{\diam} \times \mathbbold I$. The symmetry conditions discussed in the previous paragraphs are now independently satisfied in the bra and ket layers of $\mathbb T(T^X_{\rm 3d})$. It follows from the definition of $\mathbb E(T^X_{\rm 3d})$ that, when working in the $| \pm \ra$ basis, bra and ket layers of the transfer operator can be identified. As such, we can couple them by contracting each pair of virtual indices in the bra and ket layers with the three-valent tensor $\delta_{X}$ defined in \eqref{eq:deltaXZ}. Let us work this operation out explicitly. First, let us rewrite $\mathbb E(T^X_{\rm 3d})$ as
\begin{align}
	\nn
	\mathbb E(T^X_{\rm 3d}) &= \frac{1}{2^{|\mathcal L(\tet)|}}\prod_{\ell \in \mathcal L(\tet)}\Big( \sum_{a_\ell=0,1} \bigotimes_{\mathsf e \subset \ell} X_{\mathsf e}^{a_\ell}\Big)
	\\
	\nn
	&= \frac{1}{2^{|\mathcal L(\tet)|}} 
	\sum_{\{a\}}
	\bigotimes_{\mathsf e \subset \tet}
	\Big(\prod_{\substack{\ell \in \mathcal L(\tet) \\ \ell \supset \mathsf e}}X_{\mathsf e}^{a_\ell}\Big)
	\\
	&=
	\frac{1}{2^{|\mathcal L(\tet)|}} 
	\sum_{\{a\}}
	\bigotimes_{\mathsf e \subset \tet}
	X_{\mathsf e}^{a_{\mathsf e}}
	\, ,
\end{align}
where we introduced the notation 
\begin{equation}
	a_\mathsf{e}:= \sum_{\substack{\mathsf \ell \in \mathcal L(\tet)  \\ \ell \supset \mathsf e}}a_\ell \; .
\end{equation} 
We then define the new $\mathbb C$-valued tensor
\begin{align}
	{\rm diag}(\mathbb E)(T^X_{\rm 3d}) &:=
	\frac{1}{2^{|\mathcal L(\tet)|}} 
	\sum_{\{a\}}
	\bigotimes_{\mathsf e \subset \tet}
	\includeTikz{0}{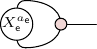}{\deltaTBis{0.5}{0}}  \; ,
\end{align}
where we made use for $\delta_X$ of the graphical notation  introduced in \eqref{eq:deltaXZ}. It follows from 
\begin{equation}
	\includeTikz{0}{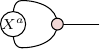}{\deltaTBis{0.5}{-1}} = 
	\includeTikz{0}{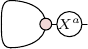}{\deltaTBis{0.5}{-2}} = |+\ra + (-1)^a |-\ra 
\end{equation}
that 
${\rm diag}(\mathbb E)(T^X_{\rm 3d})$ boils down to
\begin{align}
	\nn
	{\rm diag}(\mathbb E)(T^X_{\rm 3d}) &=
	\frac{1}{2^{|\mathcal L(\tet)|}} 
	\sum_{\{a\}}
	\bigotimes_{\mathsf e \subset \tet}
	\Big(\sum_{\alpha_\mathsf{e}=\pm }\alpha_\mathsf{e}^{a_\mathsf{e}}| \alpha_{\mathsf e}\ra \Big)
	\\
	\nn
	&= 
	\frac{1}{2^{|\mathcal L(\tet)|}} 
	\sum_{\{a\}}
	\sum_{\{\alpha\}}
	\bigotimes_{\mathsf e \subset \tet}\Big( \prod_{\substack{\ell \in \mathcal L(\tet) \\ \ell \supset \mathsf e }}\alpha_{\mathsf e}^{a_\ell}\Big)| \alpha_{\mathsf e}\ra 
	\\
	\nn
	&=
	\frac{1}{2^{|\mathcal L(\tet)|}} 
	\sum_{\{\alpha\}}
	\! \Big( \! \bigotimes_{\mathsf e \subset \tet} |\alpha_{\mathsf e} \ra \! \Big)
	\!\!\! \prod_{\ell \in \mathcal L(\tet)} \!\!\!
	\Big(\sum_{a_\ell}\prod_{\mathsf e \subset \ell}\alpha_{\mathsf e}^{a_\ell}
	 \Big)
	 \\
	 \label{eq:onSiteBF}
	 &= \sum_{\{\alpha=\pm1\}}\Big(\prod_{\ell \in \mathcal L(\tet)} \!\! \delta(\alpha_\ell) \Big)\bigotimes_{\mathsf e \subset \tet}|\alpha_{\mathsf e}\ra \, ,
\end{align}
where we introduced the notation
\begin{equation}
	\alpha_\ell := \prod_{\mathsf e \subset \ell}\alpha_{\mathsf e} \, .
\end{equation}
We can now construct the transfer operator ${\rm diag}(\mathbb T)(T^X_{\rm 3d})$ obtained by contracting several copies of the on-site term ${\rm diag}(\mathbb E)(T^X_{\rm 3d})$ according to the pattern depicted in \eqref{eq:threeCylinder}. We shall demonstrate that it corresponds to the ground state projector of the (2+1)d toric code on $\mathbbold T^2_{\diam}$. In order to be able to construct ${\rm diag}(\mathbb T)(T^X_{\rm 3d})$ from ${\rm diag}(\mathbb E)(T^X_{\rm 3d})$, we need to specify whether the bonds correspond to vector or co-vectors, which boils down to choosing an orientation convention for the edges of the underlying octahedron. Naturally, opposite edges must have opposite conventions and we choose the four top edges to correspond to co-vectors. It follows that the transfer operator reads 
\begin{equation*}
	{\rm diag}(\mathbb T)(T^X_{\rm 3d}) \propto \sum_{\{\alpha\}} 
	\Big( \!\! \prod_{\ell \in \mathcal L(\mathbbold T^2_{\diamsss}\times \mathbbold I)} \!\!\!\! \delta(\alpha_{\mathsf \ell})\Big) 
	\!\!\! \bigotimes_{\mathsf e \subset \mathbbold T^2_{\diamsss} \times \{0\}}  \!\!\!\! |\alpha_\mathsf{e}\ra \!\!\! \bigotimes_{\mathsf e \subset \mathbbold T_{\diamsss}^2 \times \{1\}} \!\!\!\! \la \alpha_\mathsf{e} | 
\end{equation*}
and amounts to the map that the $\mathbb Z_2$-BF theory assigns to the cobordism $\mathbbold T_{\diam}^2 \times \mathbbold I$. 

Let $\mathcal H[\mathbbold T^2_{\diam}]=\bigotimes_{\mathsf e \subset \mathbbold T^2_{\diamsss}} \mathbb C[\mathbb Z_2]$ be the  microscopic Hilbert space associated with the top uncontracted indices of \eqref{eq:threeCylinder}. Since the transfer operator includes a delta function for every plaquette $\mathsf p \subset \mathbbold T^2_{\diam}$, it projects out every state $| \psi \ra \in \mathcal H [\mathbbold T^2_{\diam}]$  that does not fulfil the stabilizer constraints $\mathbb B_\mathsf{p}|\psi \ra = |\psi \ra$, where $\mathbb B_\mathsf{p}$ was defined in sec.~\ref{sec:TC}. In other words, the transfer operator acts on plaquettes as $\prod_{\mathsf p \subset \mathbbold T^2_{\diamsss}}\frac{1}{2}({\rm id}+\mathbb B_\mathsf{p})$.

Let us now show how the transfer operator acts on the vertices of $\mathbbold T^2_{\diam}$. Considering the upper horizontal layer of tensors in ${\rm diag}(\mathbb T)(T^X_{\rm 3d})$, we realize that a given tensor in this layer can be thought as a map from the Hilbert space of states defined by the top uncontracted indices to that of the bottom indices, whose action is given by
\begin{align*}
	\Bigg |  
	\includeTikz{0}{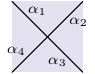}{\onSiteTransfer{0.24}{3}}  \Bigg\ra \mapsto
	\frac{1}{2}\sum_{\kappa = \pm 1 }\Bigg| \! 
	\includeTikz{0}{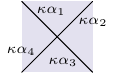}{\onSiteTransfer{0.24}{4}} \!\!\Bigg\ra \, ,
\end{align*}
i.e. this single tensor acts on a given state in $\mathbb C[\mathbb Z_2]^{\otimes 4} \subset \mathcal H[\mathbbold T^2_{\diam}] $ as $\frac{1}{2}({\rm id}+ \mathbb A_\mathsf{v})$, with $\mathsf{v}$ the top vertex of the octahedron. Inspecting carefully the contraction pattern of \eqref{eq:threeCylinder}, we notice that the upper horizontal layer of the transfer operator includes one such projector for every other vertex of $\mathbb T^2_{\diam}$. The lower horizontal layer of tensors then performs the same projection on the remaining vertices. Combining our results, we find that
\begin{equation}
	{\rm diag}(\mathbb T)(T^X_{\rm 3d}) = \prod_{\mathsf v \subset \mathbbold T^2_{\diamsss}}\frac{1}{2}({\rm id}+\mathbb A_\mathsf v) \prod_{\mathsf p \subset \mathbbold T^2_{\diamsss}}\frac{1}{2}({\rm id}+\mathbb B_\mathsf p) \, ,
\end{equation}
i.e. it is equal to the Hamiltonian projector of the (2+1)d toric code. It follows immediately that the transfer operator has four degenerate fixed points. Finally, since we have the possibility of inserting a string of $X^g$ operators along the transfer operator, we have twice as many fixed points in total. Putting everything together, this confirms the eight-fold degeneracy of the fixed point sector.

\bigskip \noindent
Let us now explore what are the different phases that can be encoded within the space of symmetric tensors, a given phase being encoded into the fixed point structure of the transfer operator. Practically speaking, we can think of these symmetric tensors as being variationally optimized wave functions for a perturbed version of the original Hamiltonian, a typical perturbation being the insertion of a (sufficiently strong) uniform magnetic field. In (2+1)d, the different phases we encountered were characterized by different symmetry patterns of the fixed point sector of the transfer operator. Phase transitions then involved symmetry breaking patterns, which could be detected using local order parameters. Crucially, these local order parameters turned out to be related to excitation-creating operators, drawing a correspondence between the phase diagram in the fixed point of the transfer operator and the condensation of excitations. We shall now derive the analogue of these results for our tensor network representations of the (3+1)d toric code.

As in (2+1)d, the fixed points of the transfer operator in the $T^Z_{\rm 3d}$ representation break the $\mathbb Z_2 \times \mathbb Z_2$-symmetry of the transfer operator down to its diagonal subgroup ${\rm diag}(\mathbb Z_2 \times \mathbb Z_2) \cong \mathbb Z_2$. Furthermore, we distinguish two other symmetry patterns which correspond to the condensation of the electric charges or the magnetic fluxes, respectively. Making this last statement more explicit requires emphasizing the relation between symmetry breaking and expectation values of excitation-creating operators. To do so, we shall consider the transfer operator associated with the half-infinite space, generalizing what we did in (2+1)d, so that expectation values can be conveniently evaluated between the fixed points of the transfer operator.  In this context, excitations with a non-trivial magnetic flux correspond to 1d domain wall excitations of the transfer operator. Condensing such loop-like flux excitations thus requires the fixed points to be symmetric under $\mathbb Z_2 \times \mathbb Z_2$. Conversely, if the point-like charges are condensed, then all the symmetry conditions must be broken. Overall, the $T^Z_{\rm 3d}$ representation features the same correspondence as in (2+1)d between symmetry structure of the fixed points and condensation of the excitations.

What about the $T^X_{\rm 3d}$ representation? Given that we now have `local' virtual symmetries and the fact that both excitation-creating operators are one-dimensional instead of zero- and two-dimensional, respectively, as for the other representation, we do not expect the previous analysis to apply here. We have already found that in this representation, the (3+1)d toric code phase is characterized by the fact that the fixed point sector of the transfer operator ${\rm diag}(\mathbb T)(T^X_{\rm 3d})$ corresponds to the ground state subspace of the (2+1)d toric code on $\mathbbold T^2_{\diam}$. Using a slightly abusive notation, we shall thus refer to this phase as the ${\rm diag}(\tc_{\rm 2d} \times \tc_{\rm 2d})\cong \tc_{\rm 2d}$ one.
Relaxing the identification between bra and ket layers, we obtain a fixed point sector that describes two independent copies of the (2+1)d toric code phase in the bra and ket layers, respectively. The corresponding phase is denoted by $\tc_{\rm 2d} \times \tc_{\rm 2d} \equiv \tc_{\rm 2d}^{\times 2}$. Assuming the system has been deformed towards such a phase, let us now examine what it means for the excitations. Let us consider the operator
\begin{equation}	
	\mathcal W_\rho(\sigma):= \prod_{\mathsf e^\vee \subset \sigma}(Z_\rho)_{\mathsf e}
\end{equation}
that creates a loop-like flux excitation, where $Z_\rho= | 0 \ra \la 0 | + \rho | 1 \ra \la 1 |$ and $\sigma$ is a closed loop along the dual of the effective square lattice $\mathbbold T^2_{\diam}$. As before, we can compute $\la \mathcal W_\rho (\sigma)\ra_\mathsf{PEPS}$ as the expectation value $\la \mathcal W_\rho(\sigma)\otimesbk {\rm id}\ra_{\rm f.p.}$ between the fixed points of the transfer operator associated with the half-infinite space. Recall that we use the notation $\otimesbk$ to indicate that the tensor product is between the bra and ket layers of the transfer operator. But  $\la \mathcal{W}_\rho(\sigma)\ra_{\rm TC_{\rm 2d}} \neq 0$, and thus $\la \mathcal W_\rho(\sigma)\otimesbk {\rm id}\ra_{\rm f.p.} \neq 0$, which implies that the loop-like flux excitations of the (3+1)d toric code are condensed. We thus identify this phase as the magnetic condensate. Furthermore, considering the electric operator
\begin{equation}
	\mathcal R_{g}(\gamma) = \prod_{\mathsf e \subset \gamma}X_\mathsf e^g \, ,
\end{equation}
where $\gamma$ is path along the edges of $\mathbbold T^2_{\diam}$, we find that $\la \mathcal R_{g}(\gamma)(\pepssf) |  \mathcal R_{g}(\gamma)(\pepssf) \ra = \la \mathcal{R}_{g}(\gamma)^\dagger \otimesbk \mathcal R_{g}(\gamma) \ra_{\rm f.p.} =0$ since bra and ket layers of the fixed point correspond to independent copies of $\tc_{\rm 2d}$. This implies that the electric charge excitations are confined.

Let us now suppose that we drive the system towards a third phase such that the bra and ket layers of the transfer operators are still identified, but instead of projecting onto the ground state sector the (2+1)d toric code, it projects onto that of the topologically trivial phase. In this scenario, the fixed point of the transfer operator at the renormalization group fixed point is unique and given by $|+,\ldots, +\ra$. 
As long as bra and ket layers of the transfer operator are identified, we can express $\la \mathcal R_g(\gamma)\ra_\mathsf{PEPS}$ between the fixed points of ${\rm diag}(\mathbb T)(T^X_{\rm 3d})$ as
\begin{align}
	\nn
	&\bigg\la \bigg(
	\includeTikz{0}{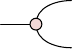}{\deltaTBis{0.5}{1}}
	\, \bigg)^{\!\!\otimes \infty}
	\!\!\! \mathcal R_g(\gamma)\otimesbk {\rm id} \, \bigg(\, 
	\includeTikz{0}{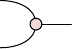}{\deltaTBis{0.5}{2}}
	\bigg)^{\!\! \otimes \infty} \bigg\ra_{\rm f.p.}
	\\
	\nn
	& \q = \la +, \ldots, +|\mathcal R_g(\gamma) | +, \ldots + \ra
	\\
	& \q =\la + , \ldots, + | +, \ldots , + \ra =1 \, ,
\end{align}
where we used the fact that
\begin{align}
	\includeTikz{0}{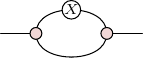}{\deltaTBis{0.5}{3}} = X 
\end{align} 
between the first and second line. This indicates that electric excitations must be condensed. Similarly, since
\begin{align}
	\includeTikz{0}{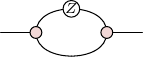}{\deltaTBis{0.5}{4}} = 0 \, ,
\end{align} 
it is not possible for loop-like excitations to condense as long as both layers of the transfer operator are identified. We thus identify this phase with the electric condensate. In summary, we have the following correspondence between the fixed point sector of the transfer operator and the phase in the bulk:\footnote{Alternatively, we can identify the $\mathbb Z_2$ topologically ordered phase $\text{\scriptsize TC}_{\rm 2d}$ with a 1-form $\mathbb Z_2$ symmetry spontaneously broken phase, where the Wilson loop $\mathcal W_\rho(\sigma)$ plays the role of a generalized order parameter \cite{gaiotto2015generalized,Ji:2019jhk}.}
\begin{align*}
	\setlength{\tabcolsep}{0.83em}
	\begin{tabularx}{0.98\columnwidth}{cccc}
	F.p. $\mathbb T(T^X_{\rm 3d})$ & F.p. $\mathbb T(T^Z_{\rm 3d})$ &  Phase & Condensate
	\\
	\midrule
	$\tc_{\rm 2d}^{\times 2}$ & $\mathbb Z_2^2$ &  trivial & magnetic
	\\
	${\rm diag}(\tc_{\rm 2d}^{\times 2}) $ & ${\rm diag}(\mathbb Z_2^2)$ & $\tc_{\rm 3d}$ & --
	\\${\rm diag}({\rm triv.}^{\times 2})$  & broken & trivial & electric
	\end{tabularx} ,
\end{align*}
where we also included for comparison what we obtained with the other representation. Crucially, the same bulk phases manifest themselves differently in the fixed point sector of the transfer operator depending on the representation we use.

As in (2+1)d, the electric and the magnetic condensates provide two gapped boundary conditions for the (3+1)d toric code. Note however that these do not exhaust all the possible choices of gapped boundaries. Indeed, given the magnetic condensate, we have for instance the possibility to further append a topological state in the \emph{double semion} phase along the boundary, which is defined in terms of the group $\mathbb Z_2$ and the non-trivial cohomology class in $H^3(\mathbb Z_2, {\rm U}(1))$ \cite{wang2017symmetric, wang2018gapped, Bullivant:2020xhy}.

\subsection{Edge physics}

\noindent
Rephrasing the previous arguments, we shall now argue that the duality between the two representations for the bulk phase induces a duality relation at the boundary.

Let us first consider the tensor network state $|\pepssf(T^Z_{\rm 3d})\ra$ with open virtual indices. Similarly to (2+1)d, we can define a Hamiltonian at the edge of the tensor network that is invariant under an anomalous global $\mathbb Z_2$ symmetry condition. Given that the charge and flux operators play the role of order and disorder parameters at the boundary, respectively, driving the system from one condensate to the other along the edge, we expect a phase transition in the same universality class as that of the (2+1)d Ising model, such that the electric and the magnetic condensates are identified with the ferromagnetic and the paramagnetic phases, respectively \cite{Ji:2019jhk, kong2020algebraic}.

Let us now consider the tensor network state $|\pepssf (T^X_{\rm 3d}) \ra$ with open virtual indices.  Given the virtual symmetry conditions satisfied by $T^X_{\rm 3d}$, we find that the tensor network now hosts $\mathbb Z_2$ gauge theories at the boundary, where it is understood that the \emph{Gau{\ss} constraint} is enforced at every plaquette of the boundary lattice. In this context, the operators $\mathcal W_\rho(\sigma)$ generating the loop-like flux excitations correspond to \emph{Wilson} operators for the boundary theory, whereas the string operators $\mathcal R_{g}(\gamma)$ are identified with the so-called \emph{monopole} operators. When all the gauge fields fluctuations are cancelled, the boundary $\mathbb Z_2$ gauge theory boils down to the (2+1)d toric code. Slightly driving the system away from the renormalization group fixed point amounts to inserting small fluctuations of the gauge field so that the expectation value of $\mathcal W_\rho(\sigma)$ obeys a perimeter law, i.e. the energy cost grows according to the length of $\sigma$.  In this case, the ground states of the boundary topological order are no longer degenerate but separated by a gap that decays exponentially with some characteristic dimension of the boundary manifold. This corresponds to the so-called \emph{deconfined} phase of the $\mathbb Z_2$ gauge theory \cite{Sachdev_2018, Wegner:1984qt, RevModPhys.51.659, Fisher2004}. The expectation value of $\mathcal W_\rho(\sigma)$ being non-zero up to local counter-terms that depend on the geometry, this translates into the condensation of the flux excitations for the bulk theory. Conversely, when the fluctuations of the gauge field dominate, the Wilson loop operator obeys an area law, so that the expectation value vanishes when taken to be infinitely large. This is the confining phase of the $\mathbb Z_2$ gauge theory. Furthermore, this phase corresponds to a condensate of monopole excitations \cite{Sachdev_2018, Wegner:1984qt, RevModPhys.51.659, Fisher2004}, which amounts to a condensation of the electric excitations for the bulk toric code.

By tuning parameters at the edge, we can drive the (3+1)d toric code from one gapped boundary condition to another, each arising from the condensation of the point-like charges or the loop-like fluxes. In light of the dictionary spelt out in the previous paragraph between the virtual operators of the tensor network state $|\pepssf (T^X_{\rm 3d}) \ra$ and that of the boundary gauge theory, we expect the corresponding phase transition to be in the universality class of the (2+1)d $\mathbb Z_2$ gauge theory.

Comparing the two descriptions of the boundary phase diagram of the (3+1)d toric code encoded by the $T^X_{\rm 3d}$ and $T^Z_{\rm 3d}$ representations, respectively, we recover the well-known duality between the (2+1)d Ising model and the (2+1)d $\mathbb Z_2$ gauge theory \cite{doi:10.1063/1.1665530,RevModPhys.51.659,fradkin2013field,Fisher2004}. More specifically, the symmetry broken phase of the Ising model corresponds to the confining phase of the gauge theory, such that the order parameter and the 1d domain wall operator of the former are mapped to the \emph{monopole} and the Wilson loop operators of the latter, whereas the symmetry preserving phase is the analogue of the deconfined phase.

\subsection{Intertwining PEPO and duality on the boundary}

\noindent
Mimicking the (2+1)d study, we shall now define a PEPO that intertwines the two tensor network representations defined in this section. In particular, this intertwining PEPO is required to map the operators of one representation to that of the other. As a consequence of the analysis above, we shall find that this PEPO also performs the duality mapping at the boundary. Given $| \pepssf(T^X_{\rm 3d}) \ra$, let us consider for instance a diagonal slice that supports a loop-like flux excitation:
\begin{equation}
	\label{eq:threeDSlice}
	\includeTikz{0}{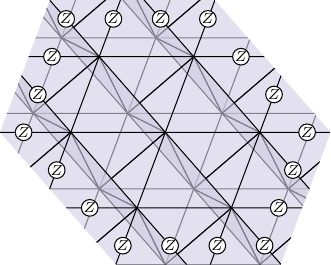}{\mixedThreeD{0.32}{1}} \, .
\end{equation}
In the other tensor network representation, the same excitation would be obtained at the boundary of a membrane of $Z$-operators. The PEPO that intertwines the two representations is obtained as the contraction of $\delta_Z$ and $\delta_X$ tensors according to the following pattern:
\begin{equation*}
	\includeTikz{0}{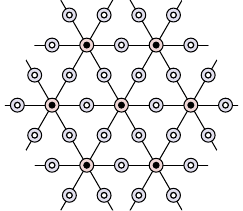}{\mixedThreeDPEPO{0.235}{1}} \, ,
\end{equation*}
where $\begin{tikzpicture}[scale=1.1,baseline=-0.3em]
\node[circle, fill = black, inner sep=1.2pt] at (0,0) {};
\end{tikzpicture}$ and $\begin{tikzpicture}[scale=0.9, baseline=-0.3em]
\node[circle, fill = black, inner sep=1.3pt] at (0,0) {};
\node[circle, fill = white, inner sep=0.8pt] at (0,0) {};
\end{tikzpicture}$ indicate physical indices sticking out of the paper plane towards and away from the reader, respectively. Let us now  check that this PEPO maps a membrane of $Z$ operators on one side to a loop of $Z$ operators on the other side:
\begin{equation*}
	\includeTikz{0}{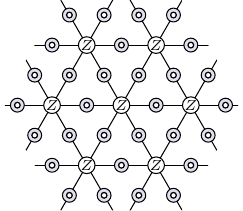}{\mixedThreeDPEPO{0.235}{2}} \!  = \!  \includeTikz{0}{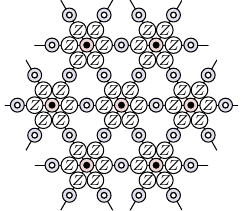}{\mixedThreeDPEPO{0.235}{3}} \! ,
\end{equation*}
where we used the symmetry conditions \eqref{eq:propDeltaX} of the $\delta_X$ tensors. Using now the symmetry conditions \eqref{eq:propDeltaZ} of the $\delta_Z$ tensors, we find
\begin{equation*}
	\includeTikz{0}{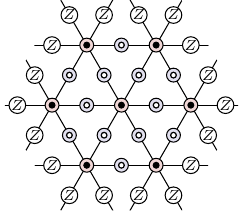}{\mixedThreeDPEPO{0.235}{4}} \! ,
\end{equation*}
which is precisely the operator that was  depicted in \eqref{eq:threeDSlice}. Proceeding exactly as in \eqref{eq:mappingString}, we find that a string of $X$ operators on one side is mapped to a pair of $X$ operators on the other side. As expected, this intertwiner PEPO is also the map that sends the (2+1)d Ising model to its dual $\mathbb Z_2$ gauge theory \cite{RevModPhys.52.453}.

%% file: _Discussion.tex
\section{Discussion}

\noindent
We have shown that the toric code admits two canonical tensor network representations that differ in the order according to which the two families of stabilizer constraints are enforced. In (2+1)d, both representations are equivalent and display the same form of virtual symmetry, and as such they encode the properties of the model in a similar fashion. In contrast, in (3+1)d, the two representations are inequivalent and are characterized by distinct symmetry conditions: whereas one representation displays a unique global virtual symmetry condition with respect to a membrane-like operator, the other one displays several local virtual constraints with respect to string-like operators. The manifestations of these different symmetry structures was  the subject of study of this manuscript. 

Although we focused exclusively on the toric code, we could have almost as easily described the case where the group $\mathbb Z_2$ is replaced by any finite \emph{abelian} group. More generally, it was established in~\cite{SCHUCH20102153, PhysRevLett.111.090501, BUERSCHAPER2014447, csahinouglu2014characterizing, BULTINCK2017183, Bultinck_2017, Williamson:2017uzx} that in (2+1)d topological order can be characterized by the ability of pulling MPOs throughout the tensor network along the virtual degrees of freedom. For the (2+1)d toric code, the strings of Pauli operators play the role of these MPOs. Moreover, closed MPOs can be shown to satisfy algebras, \emph{Morita} equivalence classes of which yields a classifying tool for non chiral topological phases in (2+1)d. It follows that there exist as many tensor network representations of a given phase as
the number of representatives in the corresponding Morita class~\cite{Lootens:2020mso}. This tensor network classification encompasses in particular \emph{string net} models \cite{Turaev:1992hq, Barrett:1993ab, Levin:2004mi}, whose input data are \emph{spherical fusion categories}. For these models, the algebra satisfied by the MPOs is the fusion algebra of the category and the different tensors in play are found to be expressible in terms of the basis elements of the associator associated with this algebra. In this terminology, the fact that both tensor network representations of the (2+1)d toric code have the same form of virtual symmetry follows from the fact that the corresponding closed MPOs are identified with the simple objects of categories that are \emph{monoidally} equivalent.

A similar classification in terms of tensor networks is expected to hold in (3+1)d, and our study of the toric code reveals certain features that should persist in the more general case. In (3+1)d, string net models are replaced by \emph{membrane net} models whose input data are so-called \emph{spherical fusion 2-categories} \cite{douglas2018fusion}. In this scenario, PEPS tensors should now evaluate to some 10j-symbols that characterize the monoidal \emph{pentagonator} of the 2-category. Our study suggests that there should be two distinguishable ways of generalizing the (2+1)d construction in terms of  pulling-through conditions with respect to membrane-like or string-like PEPOs, respectively. Closed PEPOs should in turn define two different category theoretical-like structures, which we expect to be related via some higher categorical version of Morita equivalence. 

Let us finally touch upon a concrete application of the ideas presented in this paper. We mentioned in the introduction that, given their distinct symmetry properties, we naturally expect the two tensor network representations we constructed to respond differently under arbitrary perturbations. It turns out that---in close analogy with the (2+1)d scenario, where tensor network representations of topological order are unstable to perturbations breaking the entanglement symmetry \cite{PhysRevB.82.165119,PhysRevB.98.125112}---the $T^Z_{\rm 3d}$ representation can be shown to be unstable. The underlying mechanism is the same as in (2+1)d, namely that operator insertions that break the entanglement symmetry lead to a condensation of the point-like electric excitations, which in turn drives the system to the trivial phase. However, the $T^X_{\rm 3d}$ representation happens to be \emph{stable} to arbitrary perturbations, including those that break the entanglement symmetry. We proposed some heuristic arguments in favor of this novel feature in the introduction, but this can be rigorously demonstrated by showing that the phases encoded by the perturbed tensor network map to those of a specific classical statistical model with a non-zero finite temperature phase transition \cite{WDSV}.

%% file: _Appendix.tex
\section{Duality defects\label{sec:app_dual2D}}
\noindent
We derived in the main text two tensor network representations of the (2+1)d toric code. Both representations display the same $\mathbb Z_2$-symmetry, which implicitly follows from the self-duality of the model.
The duality can be further probed by inserting topological defects lines that shift the lattice and modify accordingly some of the terms in the Hamiltonian \cite{PhysRevB.86.161107,Kitaev_2012}. It is well-known that this defect has the effect of swapping electric and magnetic point-like excitations. We could realize such a topological defect by means of an MPO that is reminiscent of the one introduced in sec.~\ref{sec:mixed2D}. Instead, we shall introduce a new PEPS tensor that parametrizes the ground state subspace along the defect.

Let us consider the square lattice with a horizontal topological defect along which qubit degrees of freedom have been removed:
\begin{equation*}
	\includeTikz{0}{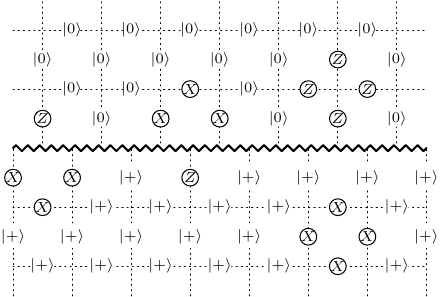}{\defectTwoD{0.5}{1}} \, ,
\end{equation*}
where we illustrated how the plaquette terms are modified in the vicinity of the topological defect. We also depicted a specific choice of 1-chain in $C_1(\mathbb T^2_\boxempty, \mathbb Z_2)$. This choice of 1-chain is such that all the stabilizer constraints with respect to the vertex operators are satisfied in the upper-half, whereas all the stabilizer constraints with respect to the plaquette operators are verified in the lower-half. Consequently, it remains to enforce the plaquette constraints above the defect line, including the ones modified due to the presence of the defect, as well as the vertex constraints below the defect line. These constraints are enforced by means of the PEPO operators introduced in the main text, with the exception of the ones in the vicinity of the defect line that require new tensors:
\begin{equation*}
	\includeTikz{0}{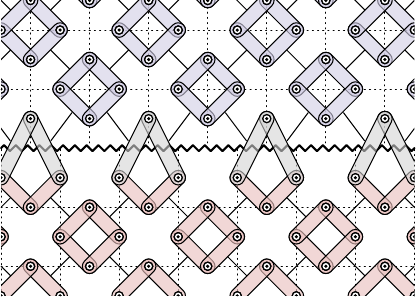}{\defectTwoD{0.5}{2}} \, ,
\end{equation*}
where we introduced the new type of PEPO tensors:
\begin{align}
	\includeTikz{0}{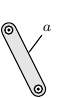}{\PEPOCHalf{0.5}{1}} 
	&\equiv \; \frac{1}{2^\frac{1}{2}} \! \begin{cases}\, \mathbbm 1 \otimes \mathbbm 1 \!\!\! &\text{if $| a\ra=|0\ra$} \\ X \otimes Z \!\!\! &\text{if $| a\ra=| 1\ra$}\end{cases} ,
	\\
	\includeTikz{0}{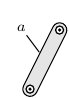}{\PEPOCHalf{0.5}{2}} 
	&\equiv \; \frac{1}{2^\frac{1}{2}} \! \begin{cases}\, \mathbbm 1 \otimes \mathbbm 1 \!\!\! &\text{if $| a\ra=|0\ra$} \\ Z \otimes X \!\!\! &\text{if $| a\ra=| 1\ra$}\end{cases} .
\end{align}
We then define a new type of PEPS tensors that results from blocking four PEPO tensors across the defect line, and acting on the state $|+,+,+,0\ra$, i.e. 
\begin{equation*}
	\includeTikz{0}{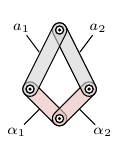}{\toricProjLocC{0.5}{1}} \!  \triangleright \, 
	\includeTikz{0}{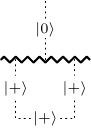}{\toricProjLocC{0.5}{2}} \, =: \, 	
	\includeTikz{0}{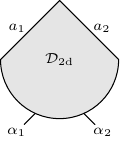}{\toricProjLocC{0.5}{3}} \; ,
\end{equation*}
where the notation descends from the one used to define $T^X_{\rm 2d}$ and $T^Z_{\rm 2d}$.
Going through the derivation, we find
\begin{align*}
	\mathcal{D}_{\rm 2d} \propto \!
	\sum_{\substack{\{a = 0,1\} \\ \{\alpha= \pm\}}} 
	\, &|a_1,a_2 \ra \la \alpha_1, \alpha_2 | \\[-1.6em]  & \otimes |(-1)^{a_1} \alpha_1, \alpha_1 \alpha_2, (-1)^{a_2} \alpha_2, a_1+a_2 \ra \, .
\end{align*}
Putting everything together, we obtain the following PEPS obtained as the contraction of tensors $T^X_{\rm 2d}$, $T^Z_{\rm 2d}$ and $\mathcal{D}_{\rm 2d}$:
\begin{equation*}
	\includeTikz{0}{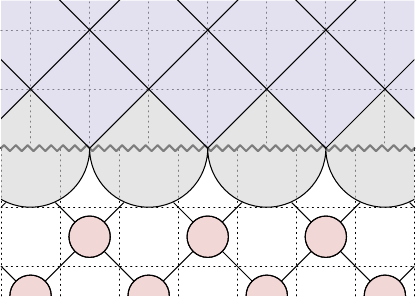}{\defectTwoD{0.5}{3}} \; .
\end{equation*}
By inspecting the defining formula of $\mathcal{D}_{\rm 2d}$, we find that it satisfies the following \emph{mixed} $\mathbb Z_2$-symmetry:
\begin{equation*}
	\includeTikz{0}{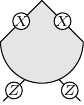}{\symTwoDPepsThree{0.35}{1}}\; = \;
	\includeTikz{0}{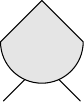}{\symTwoDPepsThree{0.35}{2}} \; \Leftrightarrow \;
	\includeTikz{0}{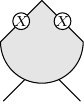}{\symTwoDPepsThree{0.35}{3}} \; = \; 
	\includeTikz{0}{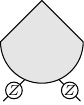}{\symTwoDPepsThree{0.35}{4}} \, .
\end{equation*}
Crucially, this $\mathbb Z_2$-symmetry implements the electric-magnetic duality taking place at the defect line. Indeed, consider a closed loop of $X$ operators along a 1-boundary above the defect line. Using the $\mathbb Z_2$-invariance of $T^X_{\rm 2d}$, this loop can be freely deformed and moved towards the defect line, at which point the symmetry of the $\mathcal{D}_{\rm 2d}$ tensors can be used to push the loop of operators through the defect line. In the process of going through the defect line, operators are swapped so that we end up with a closed loop of $Z$ operators. Considering open string of operators, the same mechanism shows that electric excitations are transformed into magnetic excitations while crossing the defect line, and vice versa. Indeed, take an open string of $X$ operators in the upper half, at the endpoints of which a pair of electric charge excitations is created. One of the excitations can be moved towards the defect line by means of local unitary transformations. When moving the excitation across the defect line, the string of operators must now be made out of $Z$ matrices so that it can still be freely deformed away from its endpoints, ensuring that it remains locally undetectable. 

Following a similar approach, we could consider dislocations in (3+1)d that probe the duality between the toric code model and the so-called \emph{2-form model} \cite{gaiotto2015generalized,kapustin2014coupling, Delcamp2019}. This model, which has a 2-form $\mathbb Z_2$ gauge theory interpretation, is characterized by loop-like electric excitations and point-like magnetic excitations. In this context, the duality defect would swap a magnetic excitation of the toric code with an electric excitation of the 2-form model, which are both loop-like, and similarly for the point-like ones.